\documentclass[12pt]{article}
\usepackage{hyperref}
\usepackage{sint_modi,cite}
\usepackage{amsmath,amssymb}
\usepackage{epsfig}
\usepackage{amssymb}

\usepackage{amssymb}
\usepackage{epsfig}

\def\a{\alpha}
\def\b{\beta}
\def\g{\gamma}

\def\D{\Delta}
\def\d{\delta}
\def\e{\varepsilon}

\def\s{\sigma}
\def\S{\Sigma}

\def\beq{\begin{equation}}
\def\eeq{\end{equation}}
\def\beqn{\begin{eqnarray}}
\def\eeqn{\end{eqnarray}}
\def\ba{\begin{eqnarray}}
\def\ea{\end{eqnarray}}

\def\m{{\tt -}}

\def\xprim2bar{\overline{x}^{\prime\prime}}

\def\beq{\begin{equation}}
\def\eeq{\end{equation}}
\def\tr{{\bf tr}}

\newcommand{\beqa}{\begin{eqnarray}}
\newcommand{\eeqa}{\end{eqnarray}}

\let\a=\alpha   \let\b=\beta   \let\g=\gamma   \let\d=\delta
\let\e=\epsilon         
        \let\m=\mu
\let\n=\nu                 
\let\s=\sigma        

  \let\D=\Delta   
         \let\S=\Sigma  

\newcommand{\Imag}{{\rm Im}\,}
\newcommand{\Real}{{\rm Re}\,}
\newcommand{\re}{{\rm e}}
\newcommand{\rd}{{\rm d}}
\newcommand{\rD}{{\rm D}}
\newcommand{\Seff}{S_{\rm eff}}

\newcommand{\oV}{\overline V}
\newcommand{\oH}{\overline H}
\newcommand{\oh}{\overline h}
\newcommand{\ov}{\overline v}

\let\a=\alpha   \let\b=\beta   \let\g=\gamma   \let\d=\delta
\let\e=\epsilon         
        \let\m=\mu
\let\n=\nu                 
\let\s=\sigma        

  \let\D=\Delta   
         \let\S=\Sigma  

\newcommand{\be}{\begin{equation}}
\newcommand{\ee}{\end{equation}}
\newcommand{\bea}{\begin{eqnarray}}
\newcommand{\eea}{\end{eqnarray}}

\usepackage{graphics}
\usepackage{graphicx}
\def\tr{{\rm tr}}
\newcommand{\eq}[1]{Eq.~(\ref{#1})}
\newcommand{\fig}[1]{Fig.~\ref{#1}}

\def\A5{(A_5)_{\rm lat}}
\def\thintablerule{\hrule height0.4pt}
\begin{document}

\rightline{WUB/09-05}

\vskip 1.5cm
\centerline{\LARGE Mean-Field Gauge Interactions in Five Dimensions I.}
\centerline{\LARGE The Torus}

\vskip 2 cm
\centerline{\large Nikos Irges and Francesco Knechtli}
\vskip1ex
\centerline{\it Department of Physics, Bergische Universit\"at Wuppertal}
\centerline{\it Gaussstr. 20, D-42119 Wuppertal, Germany}
\centerline{\it e-mail: irges, knechtli@physik.uni-wuppertal.de}
\vskip 1.5 true cm
\thintablerule
\vskip 2.0ex
\leftline{\bf Abstract}
We consider the lattice regularization of a five dimensional $SU(2)$ gauge
theory with periodic boundary conditions.
We determine a consistent mean-field background and perform computations 
of various observables originating from fluctuations around this background.
Our aim is to extract the properties of the system in regimes
of its phase diagram where it seems to be in a dimensionally reduced state.
Within the mean-field theory we establish the existence of a second order
phase transition at finite value of the gauge coupling for anisotropy
parameter less than one, where there is evidence for dimensional reduction.

\vskip 1.0ex\noindent
\vskip 2.0ex
\thintablerule

\vskip-0.2cm
\newpage

\section{Introduction}

It is possible that our world has more than four space-time dimensions.
There are different ways that extra dimensions could leave a trace
depending, among other things, on which of the fundamental forces we choose to look at.
We will concentrate here on the gauge interactions.
Leaving to the side for the moment the fermionic sector
we present here our investigation of the dynamics of extra dimensional pure
gauge theories with focus on dimensional reduction.
More specifically, we would like to propose a scheme where this phenomenon
can be analyzed analytically, far from the perturbative regime.
 
The phase diagram of a five dimensional $SU(N)$ gauge theory in infinite 
four dimensional spatial volume
is parametrized by the two dimensionless couplings
$\b = 2N/(g_5^2\Lambda )$ and $N_5=2 \pi R \Lambda$, with 
$g_5$ the five dimensional bare coupling,
$R$ the physical length parametrizing the size of the fifth 
dimension and $\Lambda$ a cut-off
\footnote{Later we will be interested in anisotropic spaces where a third
parameter, $\g$, will appear.}. 
The domain of weak coupling perturbation theory is the vicinity of the
"trivial" point $\b, N_5 \to \infty$ of the phase diagram
which is approached as one removes the cutoff.
From the formal point of view, up to now
most analytical investigations of higher dimensional theories have been carried
out in this domain, which however comes with two caveats: 
as one approaches the fixed point, physics is 
governed by the triviality of the coupling and as one 
tries to enter in the interior of the phase
diagram cut-off effects dominate.
Far from the trivial point, in the small $\b$ regime, 
the system finds itself in a confined phase below a critical value $\b_c$ of order one.
The regime $\b_c < \b << \infty$ where neither perturbation theory 
nor the strong coupling expansion is useful, up to now, is analytically 
basically unexplored.

The success of the Standard Model (SM) in explaining experimental data requires, 
after introducing a new ingredient, a natural way to sufficiently hide it.
As far as higher dimensional theories are concerned this would mean either 
that the extra dimensions are very small or
that there is a four dimensional slice in the higher dimensional space where at least 
part of the physics is localized. 
From the phenomenological point of view the former situation
has been also investigated by standard methods,
mainly with a combination of Kaluza-Klein theory and weak coupling perturbation theory
\cite{Quiros:2003gg}. 
In this approach the size of the extra dimension(s)
can be tuned to small values and the delicate issue is to make it small enough so that
it does not lead to a contradiction with the data but large 
enough so that it can have a sizable effect.
In the localization scheme, the extra dimension can be, in principle,
as large as any other dimension (or even larger) but a 
dynamical mechanism is necessary to implement it.
Unlike for gravity, for gauge fields a classical localization mechanism is not
known and perturbation theory does not seem to help in this respect. 
Consequently in most model building approaches when a certain field needs to be
localized (for some phenomenological reason) one just assumes that it is.
Nevertheless, a few non-perturbative quantum mechanisms of gauge field 
localization were invented in the past
\cite{Fu:1983ei,NPdimred}.  
In these, lattice Monte Carlo simulations have been proven a crucial tool.
Despite all efforts, from the five dimensional point of view a quantum
field theoretical approach to 
the localization of non-abelian gauge fields has been elusive.

This state of affairs calls for a complementary tool to the perturbative computations
and the numerical simulations, 
one which would enable us to probe the system analytically 
up to its phase transition. 
The method known to work in the domain of $\b$ of order one
is the mean-field approximation \cite{Drouffe:1983fv,ZJbook}. 
This approach has been used before to locate the 
critical value of $\b_c$ for various gauge theories on toroidal geometries 
(see \cite{Drouffe:1983fv} and references therein) 
and in \cite{Knechtli:2005dw} it was used to locate $\b_c$ for an orbifold geometry. 
In the 80's there was a considerable effort to define an
expansion in the fluctuations around the saddle-point (or zero'th order)
approximation \cite{Ruhl:1982er}. 
The free energy was indeed computed to
a high order in order to locate any phase transitions
very precisely and hopefully predict their order.
However, extensive observables other than the free energy were never 
computed sytematically especially for theories of large dimensionality.

In fact, it is believed that the expansion around the mean-field background
becomes a better and better approximation to the non-perturbative 
behavior of the system as the number of space-time dimensions increases so 
it seems that it is just tailored for our
purposes. In addition, we need a regulator that maintains a 
finite cutoff while preserving gauge invariance. 
The computational scheme will be therefore a lattice regularization 
with lattice spacing $a=1/\Lambda$
in the mean-field approximation to zero'th order. 
At higher orders we will perturb away from the
mean-field background by corrections that indeed go as one over 
the number of space-time dimensions.
For the basic definitions and conditions for a consistent 
formulation of a  lattice non-abelian gauge theory in five 
dimensions as well as some related Monte Carlo results see \cite{Irges:2004gy,Irges:2006hg}. 

In section 2 we describe the mean-field method applied to gauge 
theories and show how to compute 
local and global observables using fluctuations around the background. 
In section 3 we turn to the definition of lattice gauge 
theories around the mean-field background and concentrate on $SU(2)$.
In section 4 we derive the free energy, the static potential and the mass
formulae for the $SU(2)$ lattice gauge theory in five dimensions. 
In section 5 we discuss numerical results for these quantities and determine
the phase diagram. In section 6 we give our conclusions.
 
\section{Gauge theories in and around the mean-field background}

\subsection{The zero'th order approximation}
  
The partition function of an $SU(N)$ gauge theory on a Euclidean lattice is
\be
Z = \int \rD U \re^{-S_G[U]} \,,
\ee
where $S_G[U]$ is the Wilson plaquette action defined in terms of oriented
plaquettes $U(p)$. A plaquette is the product of four links around an
elementary square on the lattice. Links are generically denoted by $U$.
The mean-field computation
starts by inserting an identity in the path integral, using the Fourier representation of the
$\delta$ function
\be
\delta(f(x)) = \int_{-i\infty}^{+i\infty}\frac{\rd\alpha(x)}{2\pi i}
\re^{-\alpha(x)f(x)}.
\ee
This allows to integrate out the $SU(N)$ 
variables $U$ in favor of a set of
unconstrained variables $V$ and the set of Lagrange multipliers 
$H$ that implement the constraint $U=V$:
\bea
Z & = & \int \rD U \int \rD V \, \delta(V-U) \re^{-S_G[U]} 
\label{deltaconst} \\
& = & \int \rD U \int \rD V \int \rD H \, \re^{(1/N)\Real {\rm tr} \{H(U-V)\}} 
\re^{-S_G[V]} \,.
\eea
An effective mean-field action $\Seff[V,H]$ is defined using
\be
\re^{-u(H)} = \int \rD U \, \re^{(1/N)\Real\tr\{UH\}}\label{ueff}
\ee
in terms of which we can express conveniently the partition function as
\be
Z = \int \rD V \int \rD H \, \re^{-\Seff[V,H]} \,,\quad
\Seff = S_G[V] + u(H) + (1/N)\Real {\rm tr} \{HV\} \,. \label{effaction}
\ee
The mean-field or zero'th order approximation amounts to finding the minimum
of the effective action when
\bea
H\longrightarrow \bar{H}\mathbf{1} \,,&
V\longrightarrow \bar{V}\mathbf{1} \,,&
\Seff[\bar{V},\bar{H}]\;\mbox{=minimal} \,. \label{saddlepoint}
\eea
Taking derivatives of eq. (\ref{effaction}) with respect to $V$ and $H$ one obtains
the equations
\bea
{\overline V}&=&-\frac{\partial u}{\partial H}\Biggr|_{{\overline H}}\nonumber\\
{\overline H}&=&-\frac{\partial S_G[V]}{\partial V} \Biggr|_{\overline V}\label{mf2}
\eea
which determine the mean-field background. It is a set of 
non-linear coupled algebraic equations
that can be solved numerically by an iterative method.

The free energy per lattice site is defined as
\be
F = -\frac{1}{\cal N}\ln (Z) \,,
\ee
where $\cal N$ is the number of lattice sites.
At 0th order it is simply
\be
F^{(0)} = -\frac{1}{\cal N}\ln (Z[{\overline V},{\overline H}]) = 
\frac{S_{\rm eff}[{\overline V},{\overline H}]}{\cal N}.
\ee

\subsection{First order corrections}

Gaussian fluctuations are defined by setting
\bea
H = \bar{H} + h \;& \mbox{and}\; & V = \bar{V} + v 
\eea
and Taylor expanding the effective action to second order in the derivatives
\footnote{The first order correction to the mean-field approximation in the
  fluctuations is second order with respect to the derivative expansion.}
\be
\Seff = \Seff[\bar{V},\bar{H}] +
\frac{1}{2}\left(\left.\frac{\delta^2\Seff}{\delta H^2}\right|_{\oV,\oH}h^2 +
2\left.\frac{\delta^2\Seff}{\delta H\delta V}\right|_{\oV,\oH}hv +
\left.\frac{\delta^2\Seff}{\delta V^2}\right|_{\oV,\oH}v^2\right).
\ee
We define the quadratic pieces of the propagator 
\bea
&& \left.\frac{\delta^2\Seff}{\delta H^2}\right|_{\oV,\oH}h^2 = 
h_iK^{(hh)}_{ij}h_j = h^T K^{(hh)} h \,,\\
&&  \left.\frac{\delta^2\Seff}{{\delta V}{\delta H}}
\right|_{\oV,\oH}v h = v_iK^{(vh)}_{ij}h_j = v^T K^{(vh)} h \,,\\
 &&\left.\frac{\delta^2\Seff}{{\delta V^2}}\right|_{\oV,\oH}v^2 = 
v_iK^{(vv)}_{ij}v_j = v^T K^{(vv)} v \,,
\eea
in terms of which we can express the part of the effective action quadratic in the
fluctuations as
\be
S^{(2)}[v,h]=\frac{1}{2}\left(
h^T K^{(hh)} h + 2v^T K^{(vh)} h + v^T K^{(vv)} v\right) \,. \label{seffquadratic}
\ee
The integral of the fluctuations
\be
z = \int \rD v \int \rD h \, \re^{-S^{(2)}[v,h]} 
\ee
is a Gaussian integral and it can be easily performed to give
\be
z= \frac{(2\pi)^{|h|/2}(2\pi)^{|v|/2}}
         {\sqrt{\det(K^{(hh)}K)}}\,,
\ee
with $|h|$ and $|v|$ the dimensionalities of the $h$ and $v$ parameters
respectively and
\be 
K=-K^{(vh)}{K^{(hh)}}^{-1}K^{(vh)}+K^{(vv)} \,. \label{Kmatrix}
\ee
Therefore to first order
\be
Z^{(1)}=Z[{\overline V},{\overline H}]\cdot z  = 
e^{-S_{\rm eff}[{\overline V},{\overline H}]}\cdot z \label{Z1l}
\ee 
The typical quantity of interest is
\bea
\langle {\cal O} \rangle & = &
\frac{1}{Z} \int \rD U \, {\cal O}[U] \re^{-S_G[U]} \\
& = & \frac{1}{Z} \int \rD V \int \rD H \, {\cal O}[V] \re^{-\Seff[V,H]},
\eea
with ${\cal O}$ a gauge invariant operator.
Note that the $\delta$-function constraint in \eq{deltaconst} has been used
to replace ${\cal O}[U]$ with ${\cal O}[V]$. 
The purpose of this step is to replace a constrained $SU(N)$ path integral by a path integral over
unconstrained matrix valued complex numbers. The number of degrees of freedom apparently
increases but in fact the relevant constraint is encoded in
$K^{(hh)}$.

The observable has a Taylor expansion of the form
\be
{\cal O}[V] = {\cal O}[\oV] + 
\left.\frac{\delta{\cal O}}{\delta V}\right|_{\oV} v +
\frac{1}{2}\left.\frac{\delta^2{\cal O}}{\delta V^2}\right|_{\oV} v^2 + \ldots
\ee
The combined expansion of observable and action to the same order is 
\bea
\langle {\cal O} \rangle & = &
\frac{1}{Z}\int \rD v \int \rD h \, \left({\cal O}[\oV]+
\frac{1}{2}\left.\frac{\delta^2{\cal O}}{\delta V^2}\right|_{\oV} v^2\right)
\re^{-\left(\Seff[\oV,\oH] + S^{(2)}[v,h]\right)} \\
& = & {\cal O}[\oV] +
\frac{1}{2}\frac{\delta^2{\cal O}}{\delta V^2}\Biggr|_{\oV}
\frac{1}{z}\int \rD v \int \rD h \, v^2 \re^{-S^{(2)}[v,h]}
\,.
\eea
The link two point function can be integrated to
\be
<v_iv_j>=\frac{1}{z}\int \rD v \int \rD h \, v_i v_j 
\re^{-S^{(2)}[v,h]}= (K^{-1})_{ij}\, ,\label{vh}
\ee
where $K$ is the matrix defined in \eq{Kmatrix}.
The indices $i$ and $j$ denote collectively the indices on which the links depend.
The physical observable expectation value to first order can be then expressed as
\be
<{\cal O}> =
{\cal O}[{\overline V}] + \frac{1}{2}{\rm tr} 
\left\{\frac{\delta^2{\cal O}}{\delta V^2}\Biggr|_{\oV} K^{-1}\right\} \,. \label{correction}
\ee
At first order we define the free energy
from \eq{Z1l} as
\be
F^{(1)} = F^{(0)} - \frac{1}{\cal N}\ln(z)
= F^{(0)} + \frac{1}{2{\cal N}}\ln
\Bigl[\det(K^{(hh)}K)\Delta_{\rm FP}^{-2}
\Bigr] \,, \label{fe1def}
\ee
where we have dropped an irrelevant additive constant and $\Delta_{\rm FP}$ is
the Faddeev-Popov determinant that appears after fixing the gauge.

\subsubsection{Phase transitions}
\label{PhTr}

At zero'th order
in the confined phase, the free energy $F$ has a
minimum at $\overline H =0$ (that typically implies ${\overline V}=0$) 
which dominates over any other
minima located at any other finite values of $\overline H$. 
As $\b$ increases, a second local minimum
develops and the critical value of $\b$, which indicates when 
the system enters in the deconfined phase,
is reached when this second local minimum becomes of equal height as the
minimum at $\overline H=0$.
In the deconfined phase, the local minimum at $\overline H\ne 0$ turns into a 
global minimum. Typically, between the two minima the free energy 
develops a local maximum. 
 
An alternative way to locate the phase transition is to look at 
the zero'th order mean-field conditions Eqs. (\ref{mf2}) and define the
deconfined phase wherever the numerical iterative method converges to a 
non-trivial solution. 
In general however we would like to emphasize that
this is a safe method only as long as one is sure that the solution 
found corresponds to a global minimum of the free energy. 
Also we point out that the numerical 
value of $\b_c$ depends quite a bit on wether we fix the gauge or
not (this is an option at zero'th order), 
wether we use the iterative method or the 
direct minimization of the free energy, but not its
existence. The numerical value has no physical meaning as far as we can tell. 
The only thing that has a physical meaning is if we are 
``near'' or ``far'' from the phase transition and whether the phase we are
sitting in is the global minum of the free energy or not.

To first order it is however necessary to fix the gauge. 
In gauges which require ghosts, a divergence in the free energy 
develops due to the presence of the Faddeev-Popov determinant that vanishes
when the background is taken to zero. If the divergence is regulated
(i.e. dropped) then the phase transition can be still located at the value of   
$\b$ below which the global minimum is at ${\overline V}=0$. If it is kept then 
the global minimum of the free energy has a discontinuous jump from a larger
value to a much smaller but non-zero value. The value of $\b$ at which this
jump takes place signals the phase transition.

Our last comment on the issue of the phase transition concerns its degree. 
In \cite{Creutz:1979dw} it was established and in \cite{Drouffe:1983fv} 
confirmed that on the isotropic lattice it is a strong, first order phase transition. 

\subsubsection{Extracting masses at first order}

Let us take a generic, gauge invariant, time dependent 
operator or "observable" ${\cal O} (t)$
and its 2-point function ${\cal O} (t_0+t){\cal O} (t_0) $.
In order to extract the mass associated with ${\cal O} (t)$, we need the 
connected correlator
\bea
C (t) &=& <{\cal O} (t_0+t){\cal O} (t_0)> -
<{\cal O} (t_0+t)> <{\cal O} (t_0)>\nonumber\\
&=& C^{(0)} (t) + C^{(1)} (t) + \cdots \label{2p-correlator}
\eea
where the $C^{(0)} (t)$ and $C^{(1)} (t)$ can be identified from
\bea
<{\cal O} (t_0+t){\cal O} (t_0)> &=& {\cal O}^{(0)} (t_0+t){\cal O}^{(0)}
(t_0) + \frac{1}{2} {\rm tr} 
\left\{ \frac{\d^2 ({\cal O} (t_0+t){\cal O} (t_0))}{\d^2 v } K^{-1}\right\} +
\cdots \nonumber\\
\eea
Since $C^{(0)} (t)=0$ and because time independent contributions drop
out from $C^{(1)}(t)$, at first order we have
\be
 C^{(1)} (t) = \frac{1}{2} {\rm tr} 
\left\{ \frac{\d^{(1,1)}({\cal O} (t_0+t){\cal O} (t_0))}{\d^2 v } K^{-1}\right\} 
= \frac{1}{2} {\rm tr} 
\left\{ \frac{{\tilde \d}^{(1,1)}({\cal O} (t_0+t){\cal O} (t_0))}{\d^2 v } {\tilde K}^{-1}\right\} 
\label{Higgsmass1}
\ee
where the notation $\d^{(1,1)}$ means one derivative acting on each 
of the ${\cal O}(t_0+t)$ and ${\cal O}(t_0)$ operators.
In the second part of the equation tilded quantities are the Fourier
transforms of the corresponding untilded quantities.

To extract the scalar mass we observe that a gauge invariant correlator admits an expansion 
in terms of the energy eigenvalues of the states it contains as
\be
C(t) = \sum_\lambda c_\lambda e^{- E_\lambda t},
\ee
where $E_0=m\,, \;\; E_1=m^*\,, \cdots$ and therefore we have that
\be
m \simeq \lim_{t\to \infty} \ln \frac{C^{(1)} (t)}{C^{(1)} (t-1)}.
\ee

\subsubsection{The static potential}

The static potential can be obtained from a Wilson loop ${\cal O}_W$ extending in the time
direction and in one spatial dimension. In the infinite time limit the
expectation value of the loop is related to the static potential as
\be
t\to \infty : \hskip .5cm {\rm e}^{-V t} \simeq \; <{\cal O}_W>  .
\ee
This immediately implies that in the mean-field background the potential 
is constant
\be
{\cal O}_W[\overline V] = N\ov_0^{2(r+t)}.
\ee
To obtain the first order correction, one must compute a gauge boson exchange between the two
time-like legs of the loop plus the self energy and tadpole diagrams.
All other possible exchanges of gauge bosons vanish in
the infinite time limit. 
The static potential is the quantity that 
will decide whether the system is in a dimensionally reduced state or not.

The static potential can be extracted to first order form
\be
V = -\lim_{t\to \infty} \frac{1}{t} \log {({\cal O}_W[\overline V])}-
\lim_{t\to \infty}\frac{1}{t}
\frac{\rm correction}{{\cal O}_W[{\overline V}]}\label{StaticPotential}
\ee
with the correction given by the sum of the exchange, self energy and tadpole
diagrams.

\subsection{Second order corrections}

We will encounter cases where a physical observable (like a 
gauge boson mass for example) due to the choice of the operator that
represents it, is identically vanishing to first order. 
We have to know therefore how to generalize our computations to 
second order in the expansion around the mean-field background. 

To second order we have the expansion
\bea
S_{\rm eff} &=& S_{\rm eff}[{\overline V}, {\overline H}] + 
\frac{1}{2}\left( \frac{\d^2 S_{\rm eff}}{\d H^2}h^2 +
2 \frac{\d^2 S_{\rm eff}}{\d H \d V}hv +
\frac{\d^2 S_{\rm eff}}{\d V^2}v^2 \right)\nonumber\\
&+& \frac{1}{6}\left(  \frac{\d^3 S_{\rm eff}}{\d H^3}h^3  +  
\frac{\d^3 S_{\rm eff}}{\d V^3}v^3 \right)
+  \frac{1}{24}\left(  \frac{\d^4 S_{\rm eff}}{\d H^4}h^4  +  
\frac{\d^4 S_{\rm eff}}{\d V^4}v^4 \right)+\cdots\nonumber\\
\eea
The cross terms in the cubic and quartic terms vanish because of the special
form of $S_{\rm eff}$.
We have a similar expansion for the observable
\bea
{\cal O}[V] &=& {\cal O}[{\overline V}] + 
\frac{\d  {\cal O}}{\d V}v+\frac{1}{2} \frac{\d^2  {\cal O}}{\d V^2}v^2
+\frac{1}{6} \frac{\d^3  {\cal O}}{\d V^3}v^3 + \frac{1}{24} 
\frac{\d^4  {\cal O}}{\d V^4}v^4+\cdots
\eea
Straightforward algebra leads to the result that
the tadpole-free contributions to the observable are
\bea
<{\cal O}> = {\cal O}[\overline V]+\frac{1}{z} 
\int DhDv \left( \frac{1}{2}\frac{\d^2  {\cal O}}{\d V^2} v^2
+ \frac{1}{24}\frac{\d^4  {\cal O}}{\d V^4} v^4 \right) e^{-S^{(2)}}[v,h]
\eea
where the second order correction involves the integral
\bea
&&\frac{1}{z} \int Dv \int Dh\; (v_iv_jv_lv_m) e^{-S^{(2)}}[v,h] = \nonumber
\\
&&(K^{-1})_{ij}(K^{-1})_{lm}+(K^{-1})_{il}(K^{-1})_{jm}+(K^{-1})_{im}(K^{-1})_{jl}
\eea
The contribution $\frac{\d  {\cal O}}{\d V}\frac{\d^3 S_{\rm eff}}{\d V^3}$
leads to a time independent tadpole contribution to the correlator
\eq{2p-correlator} and can be dropped.
We can finally express the physical expectation value as
\bea
<{\cal O}> &=& {\cal O}[\overline V]+  \frac{1}{2}\left(\frac{\d^2  {\cal O}}{\d
  V^2}\right)_{ij}\left(K^{-1} \right)_{ij} \nonumber\\
&+&\frac{1}{24}\sum_{i,j,l,m}\left(\frac{\d^4  {\cal O}}{\d
  V^4}\right)_{ijlm}\Bigl( (K^{-1})_{ij}
(K^{-1})_{lm}+(K^{-1})_{il}(K^{-1})_{jm}+(K^{-1})_{im}(K^{-1})_{jl}\Bigr)\nonumber\\
\label{msecond}
\eea
To extract the masses to second order is straightforward. Again, all time
independent self energies cancel from connected correlators and
at the end the mass is obtained from 
\be
m \simeq \lim_{t\to \infty} \ln \frac{C^{(1)} (t)+C^{(2)} (t)}{C^{(1)} 
(t-1)+C^{(2)} (t-1)},\label{mcor2}
\ee
with
\be
C^{(2)} (t) = 
\frac{1}{24}\sum_{i,j,l,m}\left(\frac{\d^4  {\cal O}(t)}{\d
  V^4}\right)_{ijlm}\Bigl( (K^{-1})_{ij}
(K^{-1})_{lm}+(K^{-1})_{il}(K^{-1})_{jm}+(K^{-1})_{im}(K^{-1})_{jl}\Bigr)
\,.
\label{Wmass2}
\ee

\subsection{Gauge dependence and zero modes}

Gauge dependence of the results turns out to be the most controversial issue in computing 
fluctuations around the mean-field background.
It can be shown \cite{CreutzGF} that for any gauge invariant polynomial 
 $P(U)$ of the links it is true that
 \be 
 \frac{1}{Z}\int \rD U e^{-S_W} P (U) =  \frac{1}{Z}\int \rD U \d (U_l, g)
 e^{-S_W} P (U) \label{gi}
 \ee
with $S_W$ the gauge invariant (for example Wilson plaquette) 
action, and $g$ an $SU(N)$ group element.
The inserted term $\d (U_l, g)$ allows the link $U_l$ to be set equal to $g$.
This implies that one can set any particular link to 
a chosen element which is typically
taken to be the unit element. Iterating this construction 
one can set to unity any number of 
links, as long as there are no closed loops forming in the process.
The argument then is that since the lattice action, the measure and the operator 
"observable" $P(U)$ are explicitly gauge invariant, then 
any correlator is also gauge invariant.
This allows one to compute the correlator in a fixed gauge 
and the previous argument guarantees that this
will not affect the gauge invariant result, as long as one 
can compute exactly the left hand side of eq. (\ref{gi})
analytically.
This is however not possible in general and one either computes the correlator via a 
Monte Carlo simulation or analytically in some approximation.
 
Higher dimensional gauge theories offer an excellent opportunity to combine successfully the
Monte Carlo and mean-field methods because of the expectation that
the expansion around the mean-field background becomes a better approximation
as the number of space time dimensions increases  \cite{Drouffe:1983fv}.
Thus, in a five dimensional gauge theory one can compute the same quantities in
two different ways, one numerical and gauge invariant and 
one analytical but gauge dependent and
if these two agree to a satisfactory degree (as one would expect from general arguments)
one can be confident that the calculations are not out of control.
By "satisfactory degree" we mean that both methods should 
describe the same physics qualitatively
(this is a necessary condition)
and they must be in a good quantitative agreement. 
By good quantitative agreement we mean that
of course the numbers should not be off by order of magnitudes and if the two methods 
agree on the physics, some quantitative disagreement can be perhaps tolerated. 
The next issue is which gauge to choose.

Taking maximal advantage of the gauge fixing possibility amounts to fixing the links to unity
on a spanning (or maximal) tree. The disadvantage of this "maximal" gauge is that it is 
hard to perform explicit calculations with it for non-Abelian groups.
Another simple choice is that of the axial gauge where one fixes to unity links along
only a specific direction.  
An advantage of this gauge is that it is by construction ghost free. 
A great disadvantage is that it leaves part of the local 
gauge invariance unfixed and as a consequence
most observables are plagued by zero modes of the propagator which correspond to 
local gauge transformations that are independent of the gauge fixed coordinate.
In momentum space they appear whenever the momentum $p_j=0$, where $j$ is the gauge fixed direction.
If one restricts the analysis to observables that do not see these zero modes
(i.e. if they couple to sectors of the propagator with 
non-zero eigenvalues only), the axial gauge is actually fine.
If the local zero modes do affect the observables, one 
must fix the residual gauge and then ghosts appear.
A further subtle issue of the axial gauge
concerns the $j$-coordinate dependence of the various quantities in the
axial gauge on a finite lattice. In order that the axial 
gauge is performed, one has to assume 
that the gauge fixed direction is infinite, however one can compute in practice the 
propagator and observables only at a finite extent of the $j$ direction.  
In order to minimize the ambiguities originating from this,
one should take this direction as large as possible. 

Finally, there are more conventional gauges such as the 
Lorentz gauge which is the one we will use here.
It involves ghosts and one must take account of them. 
All gauge invariant correlators 
evaluated in the zero'th order background turn out to be gauge
independent. In some cases, like for the static potential and the scalar 
field's mass, we will be able to
show this analytically and in others, like for the vector field's mass, we
will be able to show it only numerically. 
The only quantity which will have some gauge dependence is the 
first order correction to the free energy.
We do not fix the gauge at zero'th order and
we will make sure that the gauge dependence 
of the first order correction results only in quantitative changes
(i.e. it moves around a bit the minimum)
but does not change the qualitative properties of the system, for 
moderate varying of the gauge fixing parameter.
This of course means that numbers originating from the minimization of the
free energy should not be given any physical meaning.
If this had to be the case, like in a next to leading order calculation, a more
careful treatement would be necessary. Since we will not use the first order
free energy for anything else other than showing stability, we will not need
to carry out any special regularization.

As already stated in  \cite{Drouffe:1983fv}, in any gauge 
with periodic boundary conditions
there are additional zero modes, the global zero modes (torons) which
appear when all lattice momenta vanish simultaneously.  
These zero modes are present and they must be taken into account 
as well \cite{Petersson:1990hq}. 
In our case, they always show up as a zero of the inverse propagator at the
same time with a zero in either the observable or in the
Fadeev-Popov determinant. The result will be always a zero over zero contribution which
can be regularized and a finite part can be extracted. This
finite piece however is volume suppressed and contributes a negligible amount.
Our regularization
of the toron contributions will be thus to simply drop them.
The only other regularization we do is drop an overall constant
contribution from the free energy and which has no physical meaning.

\section{The lattice model and its observables} 

\subsection{The action}

The discretized version of the torus is defined on a five-dimensional
Euclidean lattice with lattice spacing $a$. The points are
labeled by integer coordinates $n\equiv\{n_M\}$.
The dimensionless lengths of the lattice are $L=l/a$ in the 
spatial directions ($M=\m=1,2,3$),
$T$ in the time-like direction ($M=\m=0$) and $N_5 = 2\pi R/a$ in the extra dimension ($M=5$).
Periodic boundary conditions are used in all directions. 
The dimensionless length of a direction on the lattice is 
just the number of points in that direction.
Occasionally, we will use the notation ${\cal N}$ for the 
total number of points in the lattice.
The convention for the fifth coordinate labels is that $n_{5} = 0, \cdots ,{N}_5 -1$ .
The lattice momenta\footnote{
Unless otherwise stated, dimensionless coordinates and momenta are used.}
are correspondingly
\be
p_M = \frac{2\pi}{L_M} k \,,\quad k=0,1,\ldots, {L_M}-1 \,
\ee
and $L_M$ is the dimensionless length of the lattice in direction $M$.
The gauge field is the set of link variables $\{U(n,M)\in SU(N)\}$
and its action is taken to be the Wilson action
\be
S_W[U] = \frac{\b}{2N} \sum_p \; {\tr}\, \{1-U(p)\}, 
\label{wila}
\ee
where the sum runs over all oriented plaquettes $U(p)$.
A plaquette at point $n$ in directions $M$ and $K$ is defined as the product
\be
U(p) = {\rm tr}\left\{ U(n,M) U(n+{\hat M},K) U^\dagger(n+{\hat K},M)
  U^\dagger(n,K)  \right\}
\ee
of links.
A gauge transformation $\Omega$ acts on a link as
\be
U(n,M)\longrightarrow \Omega(n) U(n,M)\Omega^\dagger(n+{\hat M})
\ee
The Wilson action reproduces the 
correct naive continuum gauge action of the fields
$A_M^B$, $B=1,\ldots,N^2-1$ of the five-dimensional gauge potential $A_M$
defined through $U(n,M) = \exp\{a A_M(n)\}$.

We will now apply the general formalism we described to a
specific example: an $SU(2)$ lattice gauge theory in 5 dimensions with
fully periodic boundary conditions.

\subsection{Ghosts}

A general $2\times 2$ complex matrix can be represented by
\be
v_0+i\sum_{A=1}^3v_A\s^A, \hskip .4cm v_0,v_A\in \mathbb{C} \,.
\ee
We choose the Lorentz gauge, which amounts to adding the gauge fixing term
\be 
S_{\rm GF}=\frac{1}{2\xi}\sum_{A=1}^3\sum_{n}\left[f_A(n)\right]^2 \label{gf}
\ee
to the lattice action. 
We will now derive the corresponding Faddeev-Popov term. Notice that we are
fixing the link fluctuations, not just the gauge field fluctuations.
Also, we do not fix the link fluctuations along the Hermitian component
$v_0$. Consistently, it will turn out that the propagator for $v_0$ does not
have poles. A similar statement is implied in \cite{Drouffe:1983fv} for
$SU(N)$ in general.

In order to compute the Faddeev-Popov determinant we have to compute the
variation of the gauge fixing term 
\be
f_A(n) = \sum_{M} [v_A(n,M)-v_A(n-{\hat M},M)]
\ee
under infinitesimal gauge transformations. We parametrize gauge transformation
functions as $\Omega (n) = e^{i \omega ^A(n) \s^A} = 1 + i \omega ^A(n) \s^A $
where the second equality holds for infinitesimal transformations. Then, the
gauge transformation rules of the components of a link are
\bea
\d v_0(n,M) &=& v^A(n,M) \left( \omega^A(n+{\hat M})
-\omega^A(n)\right)\nonumber\\
\d v_C(n,M) &=& - v_0(n,M) \left( \omega^C(n+{\hat M})-\omega^C(n)\right)
\nonumber \\
&-&  \e^{ABC} v^B(n,M)\left( \omega^A(n+{\hat M}) + \omega^A(n) \right)
\eea
and the transformation they induce on the gauge fixing function is
\bea
&& \d f_C(n) = \sum_{M} \left[ \d v_C(n,M) - \d v_C (n-{\hat M},M)\right] =
\sum_{m,A} {\cal M}_{Cn;Am}\omega^A(m)
\eea
with the ghost kernel defined as ${\cal M}_{Cn;Am} = \sum_M {\cal
  M}_{Cn;Am}^{(M)}$ and
\bea
{\cal M}^{(M)}_{Cn;Am} &=& \d_{n,m}\left[\d^{CA} \left(v_0(n,M) + v_0(n-{\hat
    M},M)\right) \right.\nonumber \\
&-&\left.\e^{ABC} \left(v_B(n,M) - v_B(n-{\hat M},M)\right)\right]\nonumber\\
&+& \d_{m,n+{\hat M}} \left[-\d^{CA}v_0(n,M) - \e^{ABC}
  v^B(n,M)\right]\nonumber\\
&+& \d_{m,n-{\hat M}}\left[-\d^{CA}v_0(n-{\hat M},M) + \e^{ABC}
  v^B(n-{\hat M},M)\right].\label{Mgh}
\eea
The ghost action is then
\be
S_{\rm FP} = \sum_{n,m}{\overline c}^A(n) {\cal M}_{An;Bm} c^B(m).
\ee
The quadratic part of this defines the Faddeev-Popov kernel which in a general mean-field
background ${\overline V} = {\overline V}(n,M)\cdot {\bf 1}$ is
\bea
&& {\cal M}_{An;Bm} ({\overline V}) = \nonumber\\
&& \d^{AB} \sum_M 
\left[ \d_{n,m}\;\left({\overline V}(n,M) + {\overline V} (n-{\hat M},M)\right) -
  \d_{m,n+{\hat M}}\;{\overline V}(n,M) - 
\d_{m,n-{\hat M}}\; {\overline V}(n-{\hat M},M)\right]. \nonumber\\
\label{genghost}
\eea
Note that the  gauge boson-gauge boson-ghost vertex 
is proportional to ${\overline V}$. In a constant background \eq{genghost} is
the lattice version of the Laplace operator, while in a non-constant
background its appropriate generalization.
Note also that $\d \ov_0(n)|_{\overline V}=0$ is consistent with not gauge
fixing $v_0$.

\subsection{Observables}

We first define the auxiliary ``lines'' 
\be
l^{(n_5)}({m_0, {\vec m}}) = \prod_{m_5=0}^{n_5-1}U((m_0,{\vec m},m_5); 5) =
\prod_{m_5=0}^{n_5-1}\Bigl[\ov_0 {\bf 1} + v_\a \, ((m_0,{\vec m},m_5); 5) \,\s^\a )\Bigr]
\ee
and
\be
l^{(t)}({t_0,{\vec m},m_5}) = \prod_{m_0=t_0}^{t_0+t -1}U((m_0,{\vec m},m_5); 0) =
\prod_{m_0=t_0}^{t_0+t -1}
\Bigl[\ov_0 {\bf 1} + v_\a \, ((m_0,{\vec m},m_5); 0) \,\s^\a )\Bigr] \,,
\ee
where in the second parts of the equations we inserted the 
mean-field parametrization and we have introduced
the matrices
\be
\s^\a = \{ {\bf 1},\; i\s^A\},\hskip 1cm  {\overline \s}^\a = 
\{ {\bf 1},\; -i\s^A\}, \hskip .5cm \a=0, A
\ee
The line along the extra dimension can start from any spatial point ${\vec m}$
but will be always taken to start from $m_5=0$ while the 
temporal line can start from any $t_0, {\vec m}, m_5$.
The superscript indicates the extent of the line. 
Finally, we will be using the label $m_0$ or $t, t', t''$ etc. for
the temporal components according to convenience.
 
\begin{figure}[!t]
\centerline{\epsfig{file=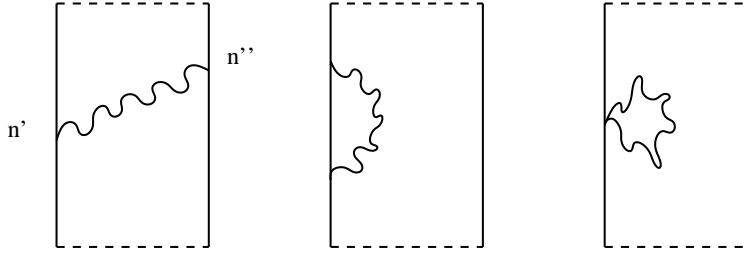,width=10cm}}
\caption{
Contributions to the static potential on the torus: from left 
to right, gauge boson exchange, self energy and tadpole.
\label{f_potpt}}
\end{figure}

The static potential will be computed using the averaged version over 
the 4d starting position $(t_0,\vec{m})$ 
\be
{\cal O}^{(t-n_5)}_{W} =  \frac{1}{T L^3}\sum_{t_0,\vec{m}}
{\cal O}_W^{(t-n_5)}(t_0, {\vec m})	\label{Os1}
\ee
of the Wilson loop observable
\be
{\cal O}_{W}^{(t-n_5)}(t_0, {\vec m}) = 
{\rm tr} \Bigl\{ l^{(t)}{(t_0,{\vec m},0)}\, l^{(n_5)}{(t+t_0, {\vec m})}\,
l^{(t)\dagger}{(t_0,{\vec m},n_5)}\, 
l^{(n_5)\dagger}{(t_0, {\vec m})}\Bigr\}\,. \label{Os2}
\ee
Similarly to $l^{(n_5)}$ one can define spatial lines 
$l^{(n_k)}$ along the $k=1,2,3$ directions as well.
In fig. \ref{f_potpt} the diagrams contributing to the static potential can be seen. 
%
%

For the masses we first define the Polyakov loop
\be
P^{(0)}{(t, {\vec m})} = l^{(N_5)}{(t, {\vec m})}  \label{P5loop}
\ee
in terms of which we define the operator
\be
\Phi^{(0)}{(t, {\vec m})} = P^{(0)}{(t, {\vec m})} -P^{(0)\dagger}{(t, {\vec
    m})}. \label{A5torus}
\ee 
The scalar gauge invariant operator
\be
{\cal O}_0 (t, {\vec m}) = {\rm tr} \{ P^{(0)}(t, {\vec m})\}\label{torusH}
\ee
is then used to determine the Scalar ("Higgs")
\footnote{We will call the scalar Higgs even though we will not 
be interested here in the phenomenon of
spontaneous symmetry breaking. We have in mind the analogous 
observable on an orbifold geometry which
in the presence of a vev for the scalar can trigger symmetry 
breaking \cite{OrbMean}.} observable, determined 
by the averaged over space and time location of the operator
\be
{\cal O}_{H} (t) = \frac{1}{T} \sum_{t_0}\frac{1}{L^6} \sum_{{\vec m}',{\vec
    m}''} {\cal O}_0(t_0, {\vec m}'){\cal O}_0(t_0+t, {\vec m}'')\label{con1}
\ee
Next, we construct the displaced Polyakov loop
\be
W^{(0),A}_k(t, {\vec m}) = {\overline \s}^A \, U((t,{\vec
  m},0);k)\, \Phi^{(0)\dagger}{(t, {\vec m}+{\hat k})}\, U((t,{\vec m},0);k)^\dagger
\, \Phi^{(0)}{(t, {\vec m})}\label{gaugeboson}
\ee
which is an operator with a vector and a gauge index.
The gauge covariant quantity 
\be
{\cal O}^A_{k}(t, {\vec m}) = {\rm tr} \Bigl\{W^{(0),A}_k(t, {\vec m})\Bigr\}
\label{vector}
\ee
will be used to define our Vector (``$W$'') mass observable, via the 2-point
function
\be
{\cal O}_{V} (t) = \frac{1}{T} \sum_{t_0}\frac{1}{L^6} \sum_{{\vec m}',{\vec
    m}''} \sum_A \sum_{k,l}{\cal O}^A_{k}(t_0, {\vec m}')\,
\d_{kl}\,{\cal O}^A_{l}(t_0+t, {\vec m}'')\label{con2}
\ee
where $A=1,2,3$ is a sum over the gauge index and $k,l=1,2,3$ a sum over the
spatial Euclidean index.
We have set the lattice spacing $a=1$ for simplicity. 
It will be reintroduced when we discuss the results, where it has a physical meaning.

\begin{figure}[!t]
\centerline{\epsfig{file=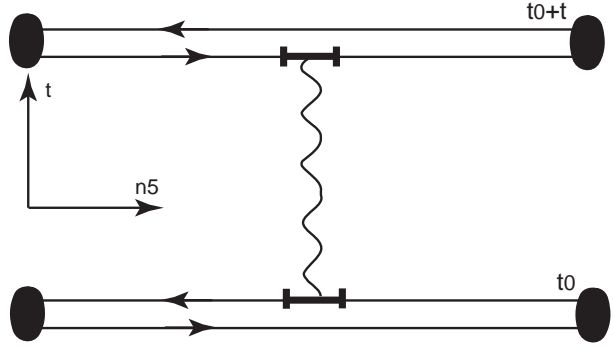,width=8cm}}
\caption{
Contributions to the mass of the scalar at first order in the mean-field.
The blobs are arbitrary lattice points, identified on the torus.
\label{Higgs1order}}
\end{figure}

\begin{figure}[!t]
\centerline{\epsfig{file=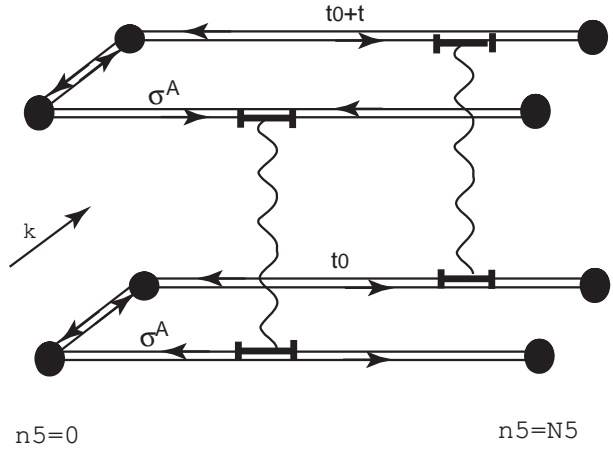,width=8cm}}
\caption{
Contribution to the gauge boson mass at second order in the mean-field. 
Points at a fixed $n_\m$, at $n_5=0$ and $n_5=N_5$ are identified.
\label{Displaced}}
\end{figure}

The choice eq. (\ref{torusH}) gives a non-trivial result for the Higgs mass already at
first order. A consequence of this choice is that since 
it is essentially a single gauge boson exchange
diagram, see Fig. \ref{Higgs1order}, there is no $L$-dependence built in it by
construction. Therefore, by measuring the mass using 
this observable, we conveniently obtain its
infinite $L$ limit value. If desired, the second order
correction could be added which will bring in an intrinsic $L$-dependence.  
Here, we will restrict ourselves to the first order expression.
%
%
In Fig. \ref{Displaced} the second order contribution to the gauge boson mass  is shown. 
%
%

These choices are of course not unique but not arbitrary either. Their general form
is fixed by their transformation properties under charge conjugation ($C$),
three dimensional coordinate inversion ($P$)
and by their Lorentz (spin $J$) and gauge (isospin $I$) indices \cite{Montvay:1984wy}.
The basic quantity is the field $\Phi^{(0)}$
which represents the extra dimensional component of the gauge
field (which we have called Higgs for short). 
It is easy to see that its continuum limit is \cite{Irges:2006hg}
\be
\Phi^{(0)} = 4N_5 a A_5 + O(a^3).
\ee
We will be interested in taking the continuum limit at finite
$R=\frac{N_5a}{2\pi}$. In this limit, $\Phi^{(0)}$ is fixed. 
The continuum limit of ${\cal O}_k^A$ is on the other hand \cite{Irges:2006hg}
\be
{\cal O}_k^A = a\; {\rm tr}\{{\overline \s}^A  (D_k
\Phi^{(0)\dagger})\Phi^{(0)}\} \,,
\ee
where $D_k\Phi^{(0)} = \partial _k + [A_k,\Phi^{(0)}]$ is the covariant derivative.

The charge conjugation $C$ leaves the coordinates invariant and acts on links as
\be
U^* = \s^2\, U\, \s^2 \,,
\ee
while parity $P$ acts on the coordinates as $Pn=n_P=(n_0,-{\vec n},n_5)$ and
on the links as
\bea
P\, U(n,k) &=& U^\dagger (n_P-{\hat k},k) \,, \nonumber\\
P\, U(n,0) &=& U(n_P,0) \,, \nonumber\\
P\, U(n,5) &=& U(n_P,5) \,.
\eea
By construction, ${\cal O}_0$ of \eq{torusH} has $C=P=1$ (scalar) and ${\cal
  O}_{k}^A$ of \eq{vector} has $C=1, P=-1$ (vector). 
The isospin is defined by the transformation of the operators under global
gauge transformations. For the definition of the spin of lattice operators
we refer to \cite{MMbook}. It is easy to see that ${\cal O}_0$ has $I=J=0$ and 
${\cal O}_{k}^A$ has $I=J=1$.

\section{Free energy, static potential and mass formulae in the SU(2) theory}

In this section we continue our discussion on the $SU(2)$ theory.
In expressions valid though for general $SU(N)$ we keep an explicit $N$. 

\subsection{At zero'th order}

The $SU(2)$ link matrices have the representation
\bea
 U(n,M) & = & u_0(n,M){\bf 1} + i \sum_{A=1}^3 u_A(n,M)\sigma^A \,,\quad
 u_\a \in\mathbb{R}\,,\quad u_\a u_\a=1 \,,
\eea
where $\a=0,1,2,3$.
The auxiliary matrices $V$ and $H$ are arbitrary complex $2\times 2$ matrices
and can be represented by 4 complex numbers as
\bea
 V(n,M) & = & v_0(n,M){\bf 1} + i \sum_A v_A(n,M)\sigma^A \,,\quad
 v_\a \in\mathbb{C} \,,\\
 H(n,M) & = & h_0(n,M){\bf 1} - i \sum_A h_A(n,M)\sigma^A \,,\quad
 h_\a \in\mathbb{C} \,.
\eea
A standard computation gives for the effective action $u$ the result
\be
u(H(n,M)) = -\ln\left( \frac{2}{\rho} I_1(\rho) \right) \,,\quad
\rho(n,M) = \sqrt{\sum_\a (\Real h_\a(n,M))^2} 
\label{fe}
\ee
with $I_1$ the Bessel function of the $I$ type
and evidently it does not depend on the imaginary parts of the $h_\a$'s.
The $\delta$ function that we insert in the partition function is  
\bea
1 & = & \prod_n\prod_M\prod_\a \int \rd\Real v_\a\rd\Imag v_\a 
\, \delta[(\Real v_\a) -u_\a]\delta(\Imag v_\a) 
\nonumber \\
& = & \prod_n\prod_M\prod_\a \int_{-i\infty}^{+i\infty}
\frac{\rd\Real h_\a\rd\Imag h_\a}{(2\pi i)^2}
\rd\Real v_\a\rd\Imag v_\a\, \nonumber\\
& & 
\exp\{-(\Real h_\a) [(\Real v_\a) -u_\a] - (\Imag h_\a)(\Imag v_\a)\} \,.
\eea
Since $u(H)$ does not depend on $\Imag h_\a$ the integration over
$\Imag h_\a$ will simply give $\delta(\Imag v_\a)$ and eliminate these
degrees of freedom from the computation so finally we can take
\bea
 v_\a \in\mathbb{R} \; & \mbox{and}\; &
 h_\a \in\mathbb{R} \,.
\eea 
The effective action is therefore
\bea
\Seff & = & -\frac{\beta}{2}\sum_{n}\sum_{M<K}\Real\tr{V(n;M,K)} 
\nonumber \\
& & + \sum_{n}\sum_M\left( u(\rho(n,M)) 
+ \sum_\a h_\a(n,M) v_\a(n,M) \right) \,, \label{seff}
\eea
where
$V(n;M,K)$ is the plaquette product at position $n$ in directions $M$ and $K$.
When we calculate corrections 
\begin{itemize}
\item
we set
\bea
h_0(n,M)& \longrightarrow& \bar{h}_0 + h_0(n,M) \,, \\
v_0(n,M)& \longrightarrow& \bar{v}_0 + v_0(n,M) \,,
\eea
where $\bar{h}_0$ and $\bar{v}_0$ are the solution of the mean-field
saddle-point equations
\bea
&& \ov_0 = \frac{I_2(h_0)}{I_1(h_0)} \,, \nonumber\\
&& \overline {h}_0 = 2(d-1)\b \ov_0^3 \,,\label{zerothMF}
\eea
where $d=5$ in our case.
If we insert the mean-field solution in the zero'th order effective action
\eq{seff} we obtain the zero'th order expression for the free energy per
lattice site
\be
{ F}^{(0)} =
-\frac{\beta d(d-1)}{2}\ov_0^4 + d u(\oh_0) + d \oh_0\ov_0 \,.
\ee
\item
we have to fix the gauge; we take the Lorentz gauge which in the continuum
amounts to adding the gauge fixing term 
$1/(2\xi)\partial_M A^B_M\partial_N A^B_N $ to the Lagrangian. As mentioned, the
lattice version of this is \eq{gf}.
\item
the derivatives of the effective action are
derivatives with respect to the components $h_\a(n,M)$ and $v_\a(n,M)$ .
\item
the derivatives of the observables are
derivatives with respect to the components $v_\a(n,M)$ .
\end{itemize}

\subsection{The Faddeev-Popov determinant}

On the torus where the mean-field value of the links is universal, 
the ghost propagator is a simple application of \eq{genghost}:
\be
{\cal M}_{An;Bm} ({\overline V}) = i\, \d^{AB}{\overline V}  \sum_M 
\left( 2\d_{n,m} - \d_{m,n+{\hat M}} -\d_{m,n-{\hat M}}\right) \equiv -i\; \d^{AB}
 {\overline V}\left(\partial^*_M \partial_M\right)_{n,m}
\ee
We then have (for $SU(N)$) the Faddeev-Popov term
\be
(-i)^{N^2-1}\left({\rm det}\left(\ov_0 \partial^*_M \partial_M
\right)\right)^{N^2-1}
\ee
In momentum space we will use the form
\be
\D_{\rm FP} = \left[\prod_{p} \left({\ov_0} \sum_M  {\hat p}_M^2\right)\right]^{N^2-1}, 
\hskip 1cm {\hat p}_M= 2\sin(p_M/2)\, .\label{FPtorus}
\ee
In the above, apart from a Fourier transformation we dropped a power of $-i$ 
from $\D_{\rm FP}$ since it merely contribute an overall constant to the free
energy. Instead, we choose to keep the factor of $\ov_0$ with the analogous
consequences, as discussed in sect. \ref{PhTr}.

 \subsection{The propagator}

Next we compute the matrix \eq{Kmatrix} in
Fourier space. The kernels which define the propagators 
in coordinate space are matrices
\be
K(n',M',\a';n'',M'',\a'') \,.
\ee
In Fourier space they are
\bea
\tilde{K}(p',M',\a';p'',M'',\a'') = \frac{1}{\cal N}\sum_{n',n''}{\rm e}^{ip'n'+ip'_{M'}/2}
{\rm e}^{-ip''n''-ip''_{M''}/2} K(n',M',\a';n'',M'',\a'') \,,\nonumber\\ \label{Fourier}
\eea
where the factor $1/{\cal N}$ has been inserted to guarantee that the Fourier
transform of the identity $\d_{n'n''}$ is the identity $\delta_{p'p''}$ and
$\tr\{\tilde{K}\}=\tr\{K\}$.

The matrix $K^{(vh)}$ is equal to the unit matrix.
A straightforward calculation yields the result
\be
\tilde{K}(p',M',\a';p'',M'',\a'') = \delta_{p'p''}\delta_{\a'\a''}
C_{M'M''}(p',\a') \,, \label{Ktilde}
\ee
with the notation
\be
C_{M'M''}(p',\a')=\left[A\d_{M'M''}+B_{M'M''}(1-\d_{M'M''})\right] \,,
\label{Cmatrix}
\ee
and ($b_1$ and $b_2$ are defined in \eq{b1b2} below)
\be
A  =  -\left[\frac{1}{b_2}(1-\d_{\a'0})+\frac{1}{b_1}\d_{\a'0}\right]
-2\b \ov_0^2 \left[\sum_{N\ne M'}\cos{(p'_N)} 
+ (1-\d_{\a'0})\frac{1}{\xi} \sin^2(p'_{M'}/2)\right] \,,
\label{Aprop_t}
\ee
and
\bea
B_{M'M''} & = &  -4\b {\overline v_0}^2
\Biggl[\delta_{\a'0}
\cos\left(\frac{p'_{M'}}{2}\right)\cos\left(\frac{p'_{M''}}{2}\right)
+ y/2(1-\delta_{\a'0})
\sin\left(\frac{p'_{M'}}{2}\right)\sin\left(\frac{p'_{M''}}{2}\right)\Biggr]
\nonumber \\
&& \label{Bprop_t}
\eea
where
\bea
&& b_1 = -\frac{1}{\oh_0 I_1({{\overline h}_0})}\left(I_2({{\overline h}_0})
-\oh_0\left(\frac{I_2({{\overline h}_0})^2}{I_1({{\overline h}_0})}-
I_3({{\overline h}_0})\right)\right) \,, \nonumber\\
&& b_2 = -\frac{\ov_0}{{\overline h}_0} \,
\label{b1b2}
\eea
and $y=2-1/\xi$.
For convenience we rescaled the gauge fixing parameter $\xi$.
This propagator was presented in \cite{Drouffe:1983fv} in the axial
gauge. 

\subsection{The Free Energy}

If we write \eq{seffquadratic} in momentum space,
the free energy \eq{fe1def} can be evaluated as
\be
F^{(1)}=F^{(0)}+\frac{1}{2{\cal N}} \ln \left[\prod_{p\neq0}
{\rm det} \Bigl(-{\bf 1} + {\tilde K}_{\a'=0}^{(hh)}{\tilde K}_{\a'=0}^{(vv)}\Bigr)
{\rm det} \Bigl(-{\bf 1} + {\tilde K}_{\a'\ne 0}^{(hh)}
{\tilde K}_{\a'\ne 0}^{(vv)}\Bigr)^3(\hat{p}^2)^{-6}\right]
\label{FreeEnergy1} \,.
\ee
The toron contribution $p=0$ is dropped according to our regularization
scheme.

The determinants in \eq{FreeEnergy1} are invariant under the flip of the sign
of any of the components of the momentum $p$. We compute their contributions
to the free energy by summing the logarithms of the absolute value of their
eigenvalues and we keep track of their signs. The flip symmetry then implies that
the overall sign of the determinant is decided by the momenta whose components
are 0 or $\pi$.

\subsection{The Static Potential}

The observable we will use to compute the static potential is
the Wilson loop extending $n_5$ points along the extra dimension
and $t$ points in time. We have defined the observable in eq. (\ref{Os1}).
The objective here is to obtain the correction as it appears in \eq{StaticPotential}.
The calculation of Fig. \ref{f_potpt} yields the first 
order correction to the tree level result
\be
{\cal O} [\overline V]=2({\overline v_0})^{2(t+n_5)} \, ,
\ee
which can be brought in the form
\bea
&&
\frac{1}{2}\frac{t}{L^3 N_5}
2(\ov_0)^{2(t+n_5)-2}\sum_{p_1',\,p_2',\,p_3',\,p_5';\,p_0'=0}
\nonumber \\
&&
\left\{[2\cos(p_5'n_5)+2]C^{-1}_{00}(p',0)
+ 3[2\cos(p_5'n_5)-2]\frac{1}{C_{00}(p',1)}\right\} \,.
\eea
Here we have used the fact that for $\a'\neq0$ and $p_0'=0$ the matrix
${\tilde K}$ is block diagonal, in particular $C^{-1}_{00}=1/C_{00}$.
This can be seen by inspection of the matrix ${\tilde K}_{\a'\ne 0}=$
\be
- 2\b \ov_0^2
\left(\begin{array} {cc} -4+\sum^\prime c_{M'}-\frac{1}{\xi} \sin^2 (\frac{p'_{0}}{2}) \,\,\;\; ys_{0/2}s_{1/2}
 \,\,\;\; ys_{0/2}s_{2/2} \,\,\;\; ys_{0/2}s_{3/2}  \,\,\;\; ys_{0/2}s_{5/2} \\ 
ys_{1/2}s_{0/2}  \,\,\;\;  -4+\sum^\prime c_{M'}- \frac{1}{\xi} \sin^2 (\frac{p'_{1}}{2})\,\,\;\; 
ys_{1/2}s_{2/2} \,\,\;\; ys_{1/2}s_{3/2} \,\,\;\; ys_{1/2}s_{5/2}\\ 
ys_{2/2}s_{0/2} \,\,\;\; ys_{2/2}s_{1/2} \,\,\;\; 
-4+\sum^\prime c_{M'}- \frac{1}{\xi} \sin^2 (\frac{p'_{2}}{2})\,\,\;\; ys_{2/2}s_{3/2}  \,\,\;\; ys_{2/2}s_{5/2}\\ 
ys_{3/2}s_{0/2}  \,\,\;\; ys_{3/2}s_{1/2} \,\,\;\; 
ys_{3/2}s_{2/2} \,\,\;\; -4+\sum^\prime c_{M'}- \frac{1}{\xi} \sin^2 (\frac{p'_{3}}{2}) \,\,\;\; ys_{3/2}s_{5/2}\\ 
ys_{5/2}s_{0/2}  \,\,\;\; ys_{5/2}s_{1/2} \,\,\;\; 
ys_{5/2}s_{2/2} \,\,\;\; ys_{5/2}s_{3/2}\,\,\;\;-4+\sum^\prime c_{M'}- \frac{1}{\xi} \sin^2 (\frac{p'_{5}}{2})\\ 
\end{array} \right)\d_{p'p''},\nonumber\\
\ee
$\sum^\prime $ means sum over all values of $M'$ except the one that
corresponds to its column/row index 
(the index ordering is $M=0,1,2,3,5$) and $s_{M/2}\equiv \sin (p_{M'}'/2)$
(similarly, $c_{M'} \equiv \cos{p_M'}$ and $c_{M/2}\equiv \cos (p_{M'}'/2)$).

A crucial cancellation happens for $\a'\neq0$.
As can be seen directly from \eq{Aprop_t} in this case,
if $p'=0$ then $A=0$ and the
matrix $C$ has a zero mode. This "toron" is precisely cancelled by
the factor $[2\cos(p_5'n_5)-2]$, where the $-2$ comes from adding
the self-energy and tadpole contributions. 
Another important observation is that the static potential which
selects the value $p_0'=0$ is $\xi$ independent. 
Also, it is easy to check that in the limit
\be
\b\longrightarrow \infty , \hskip 1cm \ov_0\longrightarrow 1 \label{PT}
\ee
one recovers the proper form of the perturbative static potential. 
This is because the $\a'=0$ part of the propagator is of order $1/\b$
with respect to its $\a'\ne 0$ part (which itself is of order $1/\b$), 
for large $\b$. 

For the static potential we arrive at the final expression
\bea
V(r)  &=& -2\log(\ov_0)\nonumber\\
&-&\frac{1}{2\ov_0^2}\frac{1}{L^3N_5}
\times\left\{\sum_{p_{M\ne 0}',p_0'=0}\left[\frac{1}{4}\sum_{N\ne 0}(2\cos(p_N'r)+2)\right] 
C^{-1}_{00}(p',0) \right. \nonumber\\
&&\left. + 3\sum_{p_{M \ne 0}',p_0'=0}\left[\frac{1}{4}\sum_{N\ne 0}(2\cos(p_N'r)-2)\right]
\frac{1}{C_{00}(p',1)}\right\} \,.\label{StaticTorus}
\eea
In the above we have used translational invariance to average over Wilson loops of spatial
size $r$ along the $M=1,2,3,5$ directions.

\subsection{The Scalar and Gauge Boson masses}

In this section we compute the scalar and vector masses to the order
illustrated in \fig{Higgs1order} and \fig{Displaced} respectively.
At this order there are no ghost contributions.
The computation of the mass of the scalar observable on the torus starts from
\eq{Higgsmass1}, the contraction of the second derivative of the
observable with the propagator. The observable is computed using \eq{con1} and
\eq{torusH} Fourier transformed and the propagator given by
\eq{Ktilde}.

We define the very useful quantity\footnote{
The argument of $\ov_0(m)$ refers to a more general situation where the
background can depend on the position along the extra dimension, like on
the orbifold \cite{Knechtli:2005dw}. Integer values for $m$ label four
dimensional links and half-integer values the links along the extra dimension.
On the torus $\ov_0({\hat r})=\ov_{05}$.}
\be
\D^{(m_5)}(n_5) = \sum_{r=0}^{m_5-1}\frac{\d_{n_5r}}{\ov_0({\hat r})}\,, \hskip
1cm {\hat r} = r+1/2 \,, \label{Deltasymb}
\ee
and its Fourier transform
\be
{\tilde \D}^{(m_5)}(p) = \sum_{r=0}^{m_5-1}
\frac{e^{ip{\hat r}}}{\ov_0({\hat r})} \,. \label{Deltasymb_fou}
\ee
The final expression for the scalar mass observable 
in Fig. \ref{Higgs1order} turns out to be
\be
C_H^{(1)}(t) = \frac{4}{\cal N}(P_0^{(0)})^2\sum_{p_0'}\cos{(p_0't)}
\sum_{p_5'}|{\tilde \D}^{(N_5)}(p_5')|^2{\tilde K}^{-1}
\Bigl((p_0',{\vec 0},p_5'),5,0; (p_0',{\vec 0},p_5'), 5,0\Bigr)
\,, \label{HT1mass}
\ee
where $P_0^{(0)}$ is the Polyakov loop evaluated on the zero'th order background.
It is interesting to note that the component of the propagator that
contributes to \eq{HT1mass} in gauge space does not have a pole at the toron
$p'=0$. The toron contribution is in fact the dominant one.
Also, because the only component of the propagator that gives a contribution
to the mass is along the 0-direction and the latter is not gauge fixed, it is
exactly gauge independent.

For the $W$ gauge boson mass we define the contraction 
\bea
&& {\overline K}^{-1}((p_0',{\vec p}'),5,\a)=
\sum_{p_5',p_5''}{\tilde \D}^{(N_5)}(p_5'){\tilde \D}^{(N_5)}(-p_5'') K^{-1}(p'',5,\a; p',5,\a)\nonumber\\
\eea
with the property
\be
{\overline K}^{-1}((p_0',{\vec p}'),5,\a)={\overline K}^{-1}((p_0',-{\vec p}'),5,\a).
\ee
In addition, we define a second useful contraction as
\bea
&& {\overline {\overline K}}^{-1}(t,{\vec p}',\a) = \sum_{p_0'}e^{ip_0't}
{\overline K}^{-1}((p_0',{\vec p}'),5,\a) .
\eea
A calculation similar to the one for the scalar, gives for Fig. \ref{Displaced}
the result
 \be
C_{V}^{(2)}(t)=
\frac{768}{{\cal N}^2} (P_0^{(0)})^4 (\ov_0(0))^4 
\sum_{{\vec p}'} \sum_k \sin^2(p_k')
\left({\overline {\overline K}}^{-1}(t,{\vec p}',1)\right)^2 \,.\label{WT2mass}
\ee
The toron here is cancelled due to the $\sin^2{(p_k')}$ factor.
Here gauge independence is not obvious because it is gauge fixed components of
the propagator that contribute. A careful numerical investigation shows that
the vector mass is actually gauge independent to an accuracy of $10^{-11}$.

In Appendix \ref{appa} we give more details about the calculations leading to
\eq{HT1mass} and \eq{WT2mass}.

\subsection{Anisotropy}

The only regime of parameters where dimensional reduction is guaranteed is when the physical size 
of the extra dimension becomes much smaller than the spatial dimensions, $R<<l$. 
On the lattice, such a scenario
can be realized when an anisotropy parameter is introduced along the extra dimension.
To this effect, following \cite{Ej}, we define the anisotropy
factor
 \be
 \g = \sqrt{\frac{\b_5}{\b_4}}
 \ee
whose zero'th order value is $\g=a_4/a_5$,
the ratio of the lattice spacings along four dimensional hyperplanes 
and along the extra dimension and
\be
 \b_4 = \frac{2Na_5}{g_5^2}\,, \hskip 1cm \b_5 = \frac{2Na_4^2}{g_5^2a_5} \,.
\ee
Then, 
\be
\b_5 = \b \g \,, \hskip .5 cm \b_4 = \frac{\b}{\g} \,.\label{betas}
\ee
A quantity that is worth keeping in mind is 
the four dimensional effective bare coupling which rescales to
\be
g_4^2 = \frac{g_5^2}{2\pi R} = \frac{2N\g}{N_5\b}\label{g4}
\ee
and it can in principle become non-perturbative for large enough anisotropy factor if 
the other parameters are kept fixed. 
We are also interested in the ratio
 \be
 \frac{N_5}{L} = \frac{2\pi R}{a_5}\frac{a_4}{l} = \frac{2\pi R}{l} \g
 \ee
which controls dimensional reduction via compactification, provided that it is kept fixed while 
$\g$ is increased.
We distinguish three regimes of the anisotropy parameter when  $N_5/L=1$:
\begin{itemize}
\item $\g=1$. This defines the isotropic lattice which,
in the limit of an infinite lattice, should represent the "non-compact phase " of the 
continuum gauge theory. 
\item $\g >> 1$. This is a regime where the size of the extra dimension is small with respect to the 
spatial length and in the limit $L\to \infty$ it represents the "compact
phase" of the continuum gauge theory.
\item $\g<1$. This is a situation where in the $L\to \infty$ limit the continuum gauge theory
has a large extra dimension.  If dimensional reduction is realized in this phase, it must be due to 
a localization effect.
\end{itemize}
The $\g\neq1$ cases are interesting in case there is dimensional reduction. Then
one expects the
four dimensional effective coupling to behave as a confining/asymptotically
free coupling. If this is indeed the case, such a property should be reflected
by the static potential.

The parameters we choose to parametrize anisotropy are the coupling $\b$ and 
the anisotropy factor $\g$.
In terms of these, we rewrite the Wilson action as
\be
S_W =
\frac{\b}{2N} \Bigl[ \frac{1}{\g}\sum_{\rm 4d-p} w(p) \Bigl(1-{\rm tr}\{U_p \} \Bigr)
+ {\g}  \sum_{\rm 5d-p} \Bigl(1-{\rm tr}\{U_p \}\Bigr) \Bigr],\label{anisaction}
\ee
where the first term contains the effect of all plaquettes 
along the four dimensional slices of the five
dimensional space and the second term contains the effect 
of plaquettes having two of their sides along the extra dimension.
  
The partition function in the presence of anisotropy gets modified. All
derivatives of plaquettes containing a link pointing in the extra dimension
will be now different. There will be two mean values for the links,
$\ov_0$ and $\ov_{05}$ determined by the extrimization of
\be
\frac{S_{\rm eff}[{\overline V}{\overline H}]}{{\cal N}}=
-\b_4 \frac{(d-1)(d-2)}{2} \ov_0^4 - \b_5 (d-1) \ov_0^2 \ov_{05}^2 + (d-1)
u({\overline h}_0) + u({\overline h}_{05}) + (d-1) {\overline h}_0 \ov_0 +{\overline h}_{05} \ov_{05}
\ee
which yields the conditions
\bea
&& \ov_0 = - u({\overline h}_0)'\,, \hskip 1cm 
{\overline h}_0=\frac{6\b}{\g}\ov_0^3+2\b\g \ov_0 \ov_{05}^2 \,,\nonumber\\
&& \ov_{05} = - u({\overline h}_{05})'\,, \hskip 1cm 
{\overline h}_{05}=8\b\g\ov_0^2 \ov_{05} \,,\label{MF_a}
\eea
for $d=5$.

The anisotropy factor cancels out in scalar products when written in terms
of dimensionless momenta and coordinates
\be
pn = p_5n_5+\sum_{\mu}p_\mu n_\mu 
   = (p_5a_5)(x_5/a_5) + \sum_\mu (p_\mu a_4)(x_\mu/a_4) \,,
\ee
where in the second equality we explicitely put the lattice spacings.

The modifications to the propagators are slightly more involved.
The propagators in the anisotropic vacuum become
\bea
&& {\tilde K}^{(hh)} =  
-\delta_{p'p''}\delta_{\a'\a''}\frac{I_2({\overline h}_{0})}
{{\overline h}_{0} I_1({\overline h}_{0})} 
\left[1-\e\cdot\frac{{\overline h}_{0}}{I_2({\overline h}_{0})}
\left(\frac{I_2^2({\overline h}_{0})}
{I_1({\overline h}_{0})}-I_3({\overline h}_{0})\right)\right]\cdot 
\left(\begin{array} {cc} 1 \hskip 2cm 
\\ \hskip .5cm 1 \hskip 1.5cm \\ \hskip 1cm 1\hskip 1cm  \\ \hskip 1.5cm
1 \hskip .5cm \\ 
\hskip 2cm 0\end{array} \right)
\nonumber\\
&&  -\delta_{p'p''}\delta_{\a'\a''}
\frac{I_2({\overline h}_{05})}{{\overline h}_{05} I_1({\overline h}_{05})}
\left[1-\e\cdot\frac{{\overline h}_{05}}{I_2({\overline h}_{05})}\left(\frac{I_2^2({\overline h}_{05})}
{I_1({\overline h}_{05})}-I_3({\overline h}_{05})\right)\right]\cdot 
\left(\begin{array} {cc} 0 \hskip 2cm 
\\ \hskip .5cm 0 \hskip 1.5cm \\ \hskip 1cm 0\hskip 1cm  \\ \hskip 1.5cm
0 \hskip .5cm \\ 
\hskip 2cm 1\end{array} \right)
\eea
where $\e=1$ for $\a'= 0$ and $\e=0$ for $\a'\ne 0$. 
For later reference, we note that the limit of the factor in 
the second line of the above expression for ${\overline h}_{05}\to 0$ is $-1/4$.
Also,
\bea
&&{\tilde K}^{(vv)}_{\a'\ne 0}=
\delta_{p'p''}\delta_{\a'\a''}(-2\frac{\b}{\g}{\ov}_0^2)\cdot \nonumber\\
&&\left(\begin{array} {cc} \sum^\prime c_{M'}-\frac{1}{\xi} s^2_{0/2} \hskip 1cm  ys_{0/2}s_{1/2}
 \hskip 1cm ys_{0/2}s_{2/2} \hskip 1cm ys_{0/2}s_{3/2}  \hskip 1cm y_5s_{0/2}s_{5/2}\\ 
ys_{1/2}s_{0/2}  \hskip 1cm \sum^\prime c_{M'}-\frac{1}{\xi}  s^2_{1/2}\hskip 1cm
ys_{1/2}s_{2/2} \hskip 1cm ys_{1/2}s_{3/2} \hskip 1cm y_5s_{1/2}s_{5/2}\\ 
ys_{2/2}s_{0/2} \hskip 1cm ys_{2/2}s_{1/2} \hskip 1cm 
\sum^\prime c_{M'}-\frac{1}{\xi}  s^2_{2/2}\hskip 1cm
ys_{2/2}s_{3/2}   \hskip 1cm y_5s_{2/2}s_{5/2}\\ 
ys_{3/2}s_{0/2}  \hskip 1cm ys_{3/2}s_{1/2} \hskip 1cm
ys_{3/2}s_{2/2} \hskip 1cm \sum^\prime c_{M'}- \frac{1}{\xi}  s^2_{3/2}\hskip 1cm
y_5s_{3/2}s_{5/2}\\ 
y_5s_{5/2}s_{0/2} \hskip .5cm  y_5s_{5/2}s_{1/2} \hskip .5cm
y_5s_{5/2}s_{2/2} \hskip .5cm  y_5s_{5/2}s_{3/2}\hskip .5cm
\g^2 \left(\sum^\prime c_{M'}- \frac{1}{\xi}  s^2_{5/2}\right)\\ 
\end{array} \right)\nonumber\\ 
&&\hskip 1cm {\rm where}\;\;\; y = 2-\frac{1}{\xi}, \;\;\; 
y_5 = 2\g^2 \frac{\ov_{05}}{{\ov}_0}-\frac{\g}{\xi}\label{Kvva1}
\eea
and
\bea
&&{\tilde K}^{(vv)}_{\a' = 0}=
\delta_{p'p''}\delta_{\a'\a''}(-2\frac{\b}{\g}{\ov}_0^2)\cdot \nonumber\\
&&\left(\begin{array} {cc} \sum^\prime c_{M'} \hskip 1cm  2c_{0/2}c_{1/2}
 \hskip 1cm 2c_{0/2}c_{2/2} \hskip 1cm 2c_{0/2}c_{3/2}  \hskip 1cm 2c_{0/2}c_{5/2}
\g^2 \frac{\ov_{05}}{{\ov}_0} \\ 
2c_{1/2}c_{0/2}  \hskip 1cm \sum^\prime c_{M'} \hskip 1cm
2c_{1/2}c_{2/2} \hskip 1cm 2c_{1/2}c_{3/2} \hskip 1cm 2c_{1/2}c_{5/2}\g^2
\frac{\ov_{05}}{{\ov}_0} \\ 
2c_{2/2}c_{0/2} \hskip 1cm 2c_{2/2}c_{1/2} \hskip 1cm 
\sum^\prime c_{M'} \hskip 1cm
2c_{2/2}c_{3/2}   \hskip 1cm 2c_{2/2}c_{5/2}\g^2 \frac{\ov_{05}}{{\ov}_0}\\ 
2c_{3/2}c_{0/2}  \hskip 1cm 2c_{3/2}c_{1/2} \hskip 1cm
2c_{3/2}c_{2/2} \hskip 1cm \sum^\prime c_{M'} \hskip 1cm
2c_{3/2}c_{5/2}\g^2 \frac{\ov_{05}}{{\ov}_0} \\ 
2c_{5/2}c_{0/2} \g^2 \frac{\ov_{05}}{{\ov}_0} \hskip .5cm  2c_{5/2}c_{1/2} \g^2 \frac{\ov_{05}}{{\ov}_0}\hskip .5cm
2c_{5/2}c_{2/2} \g^2 \frac{\ov_{05}}{{\ov}_0}\hskip .5cm  2c_{5/2}c_{3/2}
\g^2 \frac{\ov_{05}}{{\ov}_0} \hskip .5cm
\g^2 \sum^\prime c_{M'}\\ 
\end{array} \right).\nonumber\\
\eea
The only point not explicitly shown in these formulas is that 
in the diagonal elements, 
$c_5 = \g^2 \frac{{\ov}_{05}^2}{{\ov}_{0}^2} \cos{(p_5')} $.

In \eq{Kvva1} we have already implemented the anisotropic version of the gauge
fixing term 
\bea
S_{\rm GF}=\frac{1}{2\xi}\sum_{A=1}^3
\sum_n\Biggl[\sum_{\m}\Bigl(v_A(n,\m)-v_A(n-{\hat
  \m},\m)\Bigr) + \g \Bigl(v_A(n,5)-v_A(n-{\hat 5},5)\Bigr)\Biggr]^2
\eea
(up to a redefinition of $\xi$).
The resulting Faddeev-Popov determinant splits accordingly to two parts:
\be
\D_{\rm FP} =
\left[\prod_{p} 4\left( \ov_0 \sum_{\m} \sin^2(\frac{a_4p_{\m}}{2})+
\g \ov_{05} \sin^2(\frac{a_5p_{5}}{2})\right)\right]^{N^2-1} \,.
\ee

As far as the observables are concerned, the modifications to the free energy
and the mass formulae are simple. For the free energy \eq{FreeEnergy1} all the
information about the anisotropy is contained in the propagator and the
Faddeev-Popov determinant. The correlators \eq{HT1mass} and \eq{WT2mass} for
the masses do not change. On an anisotropic lattice there are two inequivalent
Wilson loops. One along the four dimensional hyperplanes for which the
static potential is given by
\bea
V(r)  &=& -2\log(\ov_0)\nonumber\\
&-&\frac{1}{2\ov_0^2}\frac{1}{L^3N_5}
\times\left\{\sum_{p_{M\ne 0}',p_0'=0}\left[\frac{1}{3}\sum_k(2\cos(p_k'r)+2)\right] 
C^{-1}_{00}(p',0) \right. \nonumber\\
&&\left. + 3\sum_{p_{M \ne 0}',p_0'=0}\left[\frac{1}{3}\sum_k(2\cos(p_k'r)-2)\right]
\frac{1}{C_{00}(p',1)}\right\} \,.\label{StaticTorusa4d}
\eea
The other Wilson loop is along the extra dimension and gives the potential
\bea
&&V(r)  = -2\log(\ov_0)\nonumber\\
&&-\frac{1}{2\ov_0^2}\frac{1}{L^3N_5}
\sum_{p_{M\ne 0}',p_0'=0}\left\{\left[2\cos(p_5'r)+2\right] 
C^{-1}_{00}(p',0) + 3\left[2\cos(p_5'r)-2\right]
\frac{1}{C_{00}(p',1)}\right\} \,. \nonumber
\label{StaticTorusaed}
\eea
The matrix $C$ is defined as in \eq{Ktilde}.

For completeness and as a check, the perturbative limit $\b\to\infty$,
$\ov_0\to1$, $\ov_{05}\to1$
of the static potential at distance $r$ in direction $M$
is easily computed to be
\be
a_4V_{\rm pert}(r) \sim \frac{g_5^2}{a_4} \sum_{p'_M\ne 0, p_0'=0} 
\frac{\cos(p'_Mr)-1}{\sum_k(1-\cos(p'_ka_4))+\g^2(1-\cos(p'_5a_5))}
\ee
and for $\g\to \infty$ one is left with a pure Coulomb potential at $p_5'=0$
(for clarity we have explicitly put in the lattice spacings).

\section{The phase diagram}

In the anisotropic background the phase diagram can be split in three regimes 
according to the solution to which
the numerical iteration \eq{MF_a} converges. Where it converges to 
$\ov_0=\ov_{05}=0$ we define the confined phase.
Where it gives a solution of the form $\ov_0, \ov_{05} \ne 0$ we define the
deconfined phase and where it gives a solution of the form $\ov_0\ne 0$, $\ov_{05}=0$
we define the "layered phase" \cite{Fu:1983ei}. 
The latter is a regime where the four dimensional coupling is 
deconfined and the five dimensional coupling confined. 
%
\begin{figure}[!t]
\begin{minipage}{8cm}
\centerline{\epsfig{file=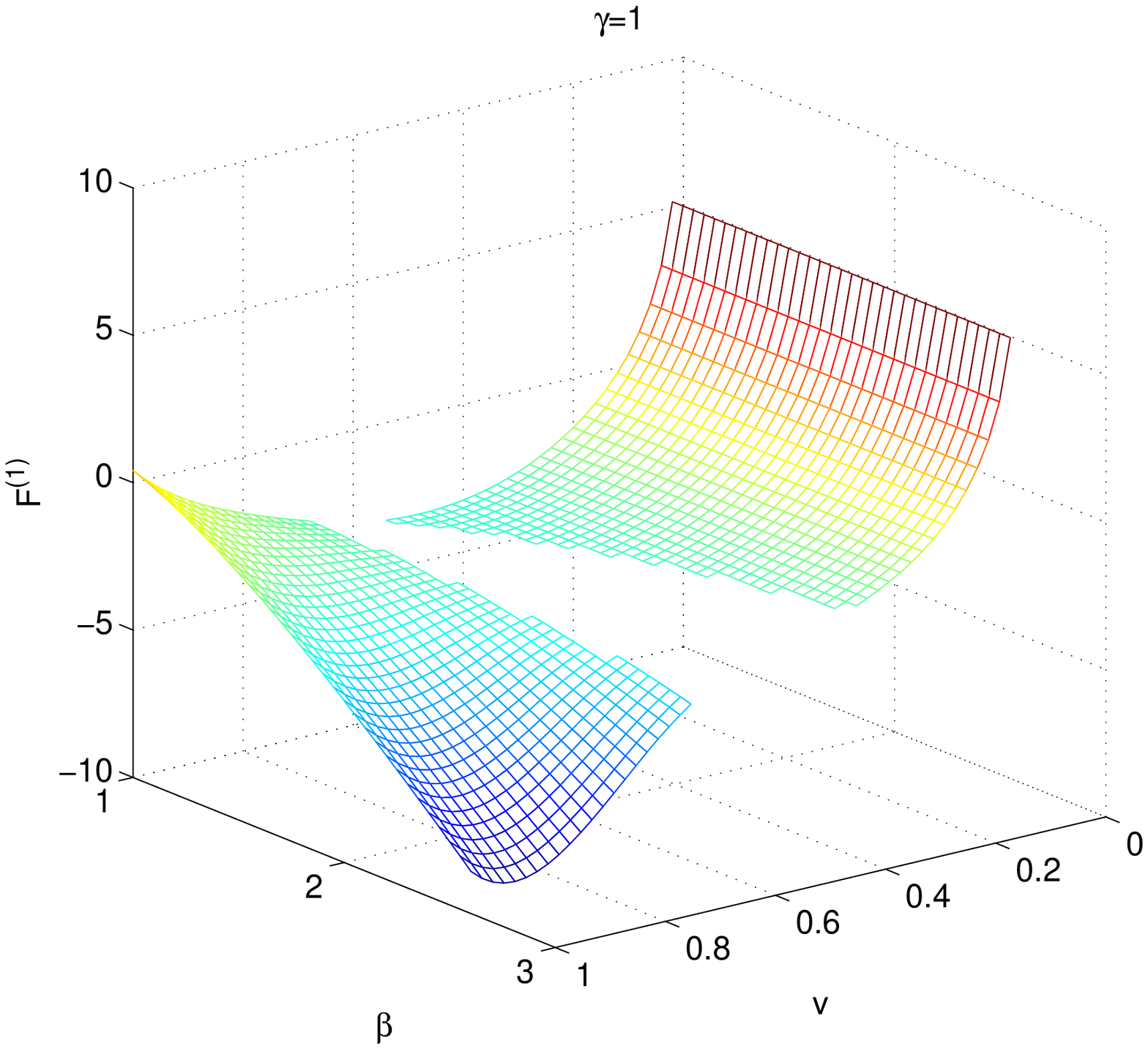,width=8cm}}
\end{minipage}
\begin{minipage}{8cm}
\centerline{\epsfig{file=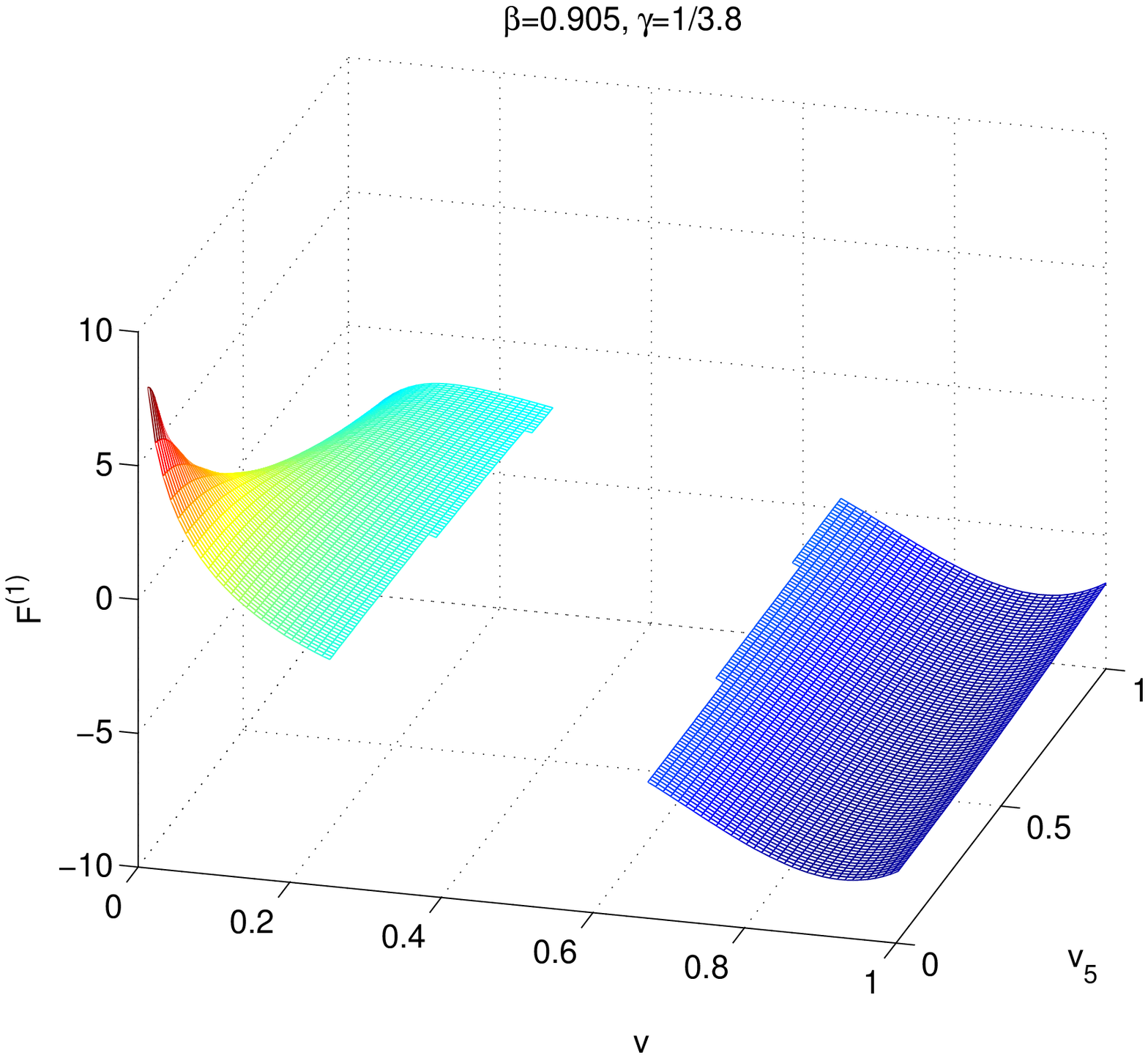,width=8cm}}
\end{minipage}
\caption{\small
Left: The free energy at first order \eq{FreeEnergy1} computed on isotropic
$T=L=N_5=10$ lattices
as a function of $\beta$ and the background value $v$. 
For $\beta\ge1.55$ the minimum is at $v\ge0.80$ (at 0th order we found a
non-trivial minimum for $\beta>2.12$). For $\beta<1.55$ the minimum jumps to 
values $0.42\le v\le0.48$.
Right: The free energy at first order in the layered
phase for $\beta=0.905$ and $\gamma=1/3.8$ computed on a $T=L=N_5=10$ lattice. 
The minimum found at 0th order ($\ov_0=0.9030$, $\ov_{05}=0$) is 
stable.
Where the free energy is complex, it is not plotted.
\label{ffeisolay}}
\end{figure}
%
%
\begin{figure}[!t]
\begin{minipage}{8cm}
\centerline{\epsfig{file=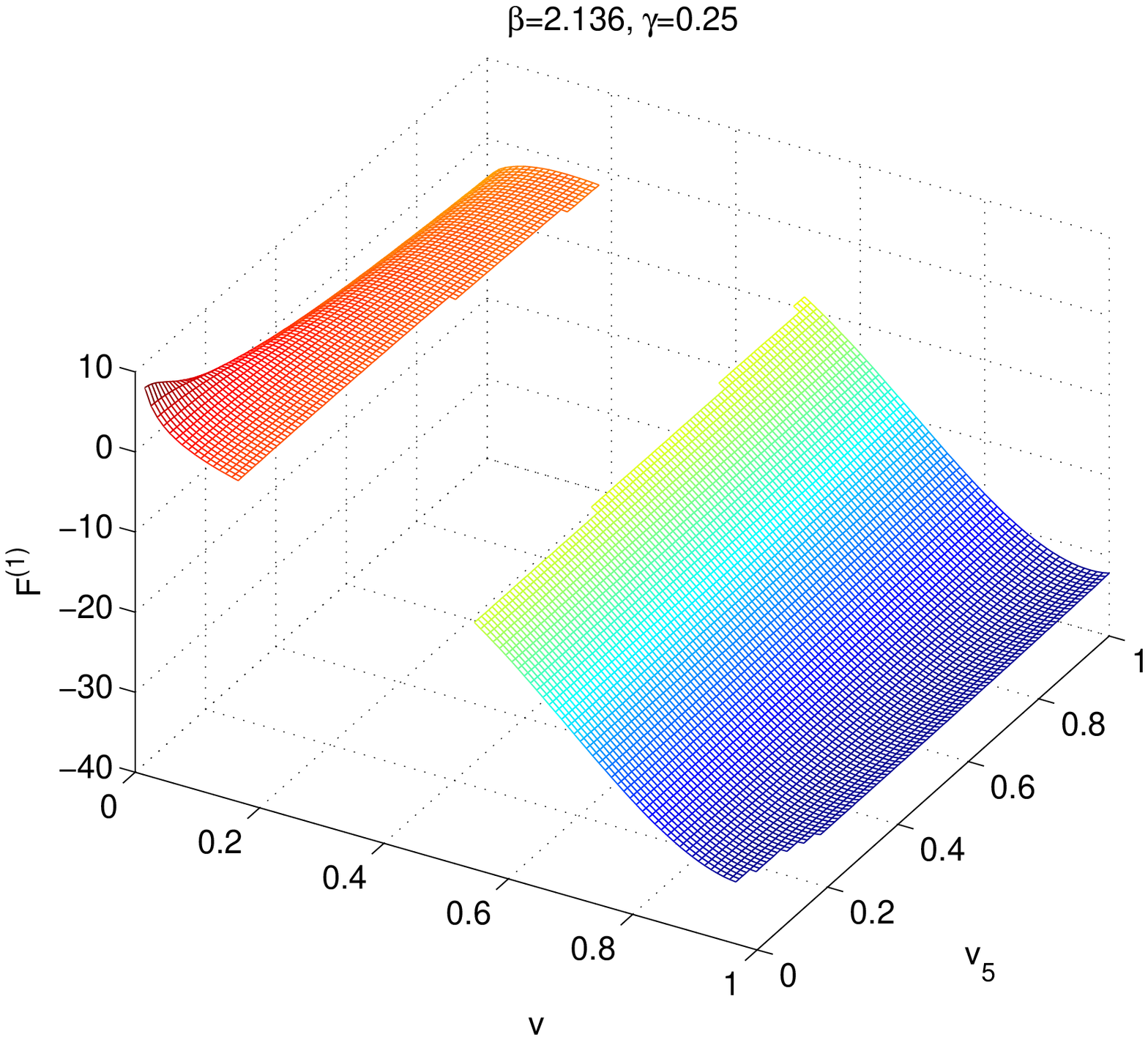,width=8cm}}
\end{minipage}
\begin{minipage}{8cm}
\centerline{\epsfig{file=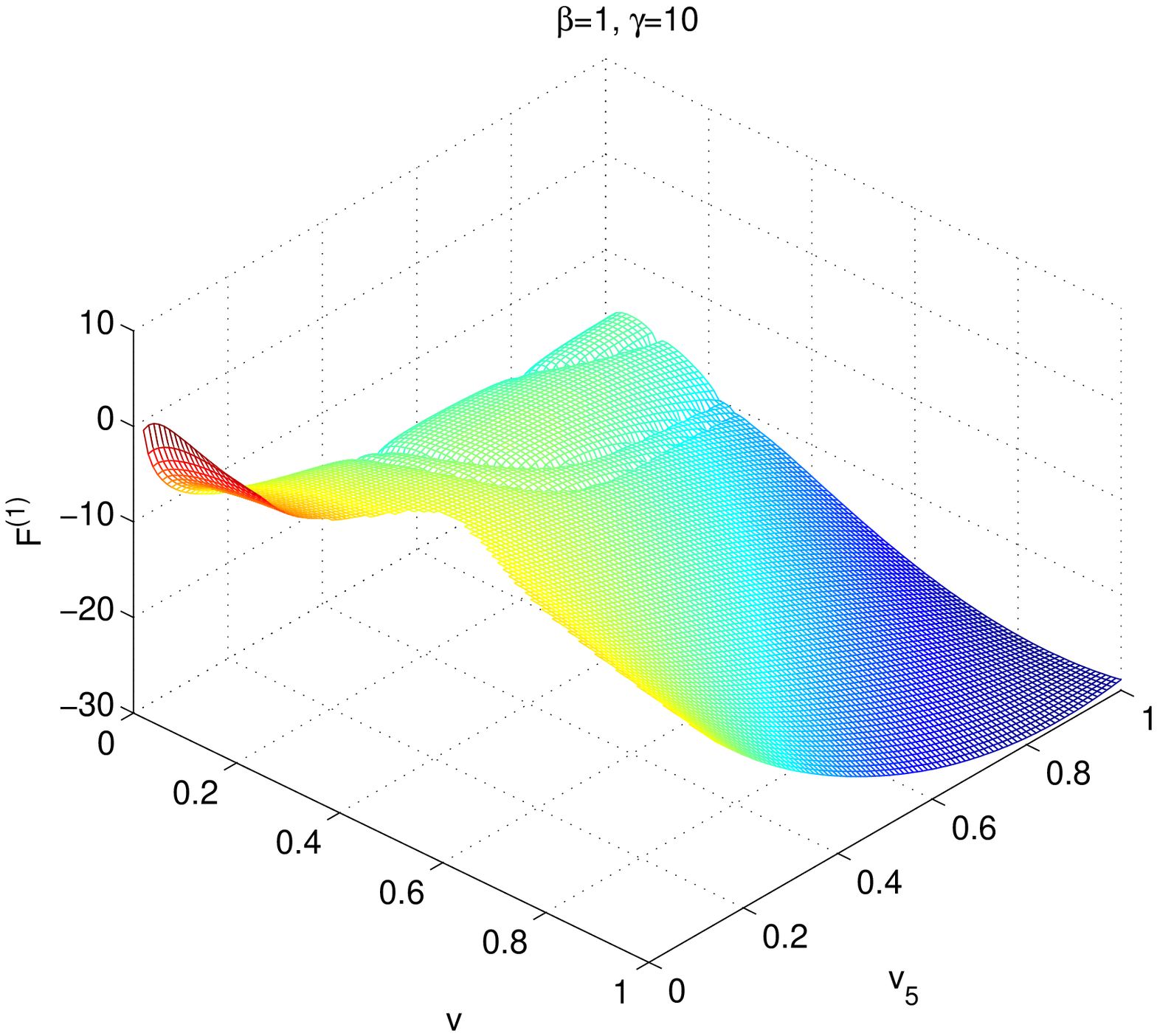,width=8cm}}
\end{minipage}
\caption{\small
Left: The free energy at first order \eq{FreeEnergy1} in the d-compact phase
for $\beta=2.136$ and $\gamma=0.25$ computed on a $T=L=N_5=10$ lattice
as a function of the background values $v$ and $v_5$. 
The minimum found at 0th order ($\ov_0=0.9679$ and $\ov_{05}=0.0288$) is
stable.
Right: The free energy at first order in the compact
phase for $\beta=1$ and $\gamma=10$ computed on a $T=L=N_5=10$ lattice. 
The minimum found at 0th order ($\ov_0=0.9179$, $\ov_{05}=0.9773$) is not
stable, it moves to a ``phase'' with $v_5=1$ (and $v=0.86$).
Where the free energy is complex, it is not plotted.
\label{ffedcc}}
\end{figure}
%
This is an approach that gives a first approximation to the phase diagram. 
One then has to check whether the solution
of \eq{MF_a} indeed corresponds to the global minimum of the free energy
\eq{FreeEnergy1}.

The mean-field method is blind to the confined phase of the five dimensional
theory so we do not have to say anything about it.
According to the first order free energy the layered phase is stable.
The 0th order background (or saddle-point) solution 
$\ov_0=0.9030, \ov_{05}=0$ remains as a minimum when
1st order corrections are included,  see right of \fig{ffeisolay}. 

For now, the deconfined phase will be the regime of our interest.
The isotropic lattice is stable as the left of \fig{ffeisolay} shows.
Clearly this is a case not interesting from the point of view of dimensional
reduction so we will not have to say much about it.
The right plot of \fig{ffedcc} shows the free energy
for $\beta=1$ and $\gamma=10$ (compact phase)
computed on a $T=L=N_5=10$ lattice as a function of the background values
$v$ and $v_{5}$.
The component of the 0th order background along the four dimensional
hyperplanes $\ov_0=0.9179$ moves to $v=0.86$ (precision 0.01).
The background along the extra dimension however, which at zero'th order is
$\ov_{05}=0.9773$ is unstable since the minimum 
moves to $v_5=1.00$ (with an uncertainty less than
$0.01$) when the 1st order correction is included.
In fact, the same instability is observed everywhere inside the compact phase
at small $\b$:
the $\ov_{05}$ component of the background runs to 1,
the value that is the solution to the background equations at
$\b=\infty$. This instability could be an inherent property of the system or
just a property of the first order approximation, we can not tell for sure 
until a higher order correction to the free energy is computed.

The left plot of \fig{ffedcc} shows the free energy for 
$\beta=2.136$ and $\gamma=0.25$
computed on a $T=L=N_5=10$ lattice.
As it will turn out this is in an especially interesting regime of the deconfined phase which
we call the 'd-compact' phase.
The 0th order background $\ov_0=0.9679$, $\ov_{05}=0.0288$ is stable,
including the first order correction we get a minimum at
essentially the same values. Furthermore,
this result is quantitatively unchanged on a $T=L=N_5=6$ lattice.
In general, the d-compact phase, for $\g$ not too small, seems to
have a deeper from the  zero'th order minimum at $v=0$ and $v_5=1$.
This global minimum however gradually disappears as $\g$ is lowered, as 
it is the case on the left of  Fig. \ref{ffedcc}.
The local, first order minimum on the other hand
is consistent with the zero'th order result and is
separated from the global minimum by a regime where the free energy is complex. 
We conclude this part of the discussion by saying that
all the caveats regarding the interpretation of the 
$\xi$-dependent free energy mentioned before, remain. 

%
\begin{figure}[!t]
\centerline{\epsfig{file=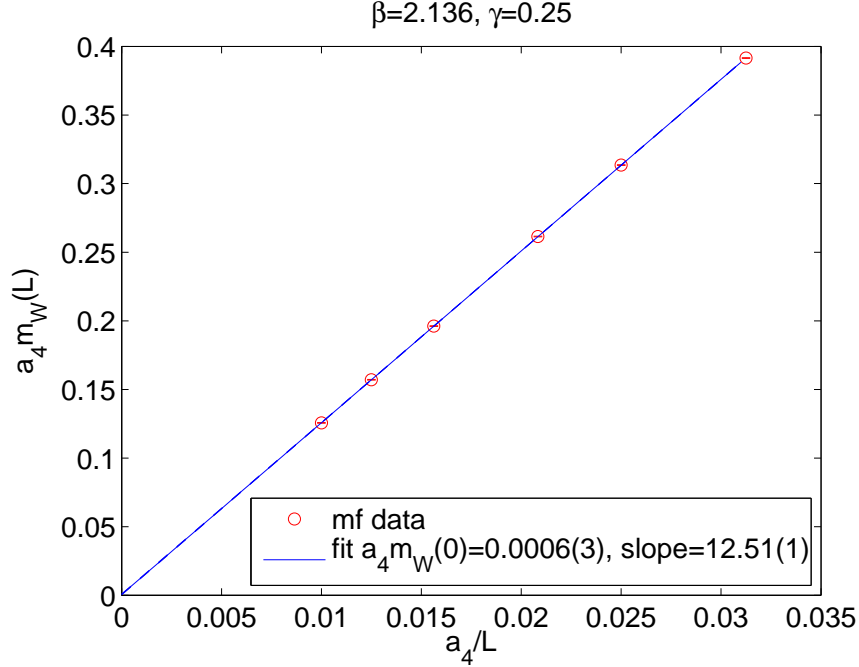,width=12cm}}
\caption{\small 
The gauge boson mass as a function of $1/L$. A linear extrapolation in the
infinite $L$ limit gives zero within errors.
\label{f_mWL}}
\end{figure}
%

Next, using our local (Wilson loop) and global
(Higgs and $W$ masses) quantities we will characterize more precisely the
various regimes of the phase diagram. We keep always $N_5=L$.
The first observable we use is the $W$ gauge boson mass from \eq{WT2mass}. 
In \fig{f_mWL}
we present a scaling study of the behavior of the mass in lattice units as a
function of $1/L$ for the point $\b=2.136$, $\g=0.25$. 
The errors represent the uncertainty in the plateau value of \eq{mcor2}.
A linear fit in $1/L$
gives an extrapolation $a_4m_{\rm W}=0.0006(3)$ for $L\to\infty$ with
$\chi^2/\mbox{d.o.f.}=0.32$. The slope is $12.51(1)$.
Adding a quadratic term to the fit confirms the conclusion
that the mass seems to be consistent with zero within errors.
We can not exclude of course a tiny (possibly exponentially suppressed) mass.
We will come back to this issue when we analyze the static potential.
A scan of the phase diagram reveals that the $W$ mass is actually nearly
independent of any parameter other than $L$. 
There is a very mild $\b$ and $\g$ dependence but it does not seem easy at the
moment to attribute any quantitative physical significance to it. 
Therefore, we conclude that the
deconfined phase, at least in regimes where the system is five dimensional
(i.e. not dimensionally reduced), is in a Coulomb phase.
Note that since the background is along the Hermitian component of the links
it can not possibly play the role of a symmetry breaking vev. That is, 
in the absence of an explicit vev we can not probe the Higgs phase, unlike in
a Monte Carlo simulation where choosing the appropriate input parameters they
system can be in a Higgs phase or not.
%
\begin{figure}[!t]
\centerline{\epsfig{file=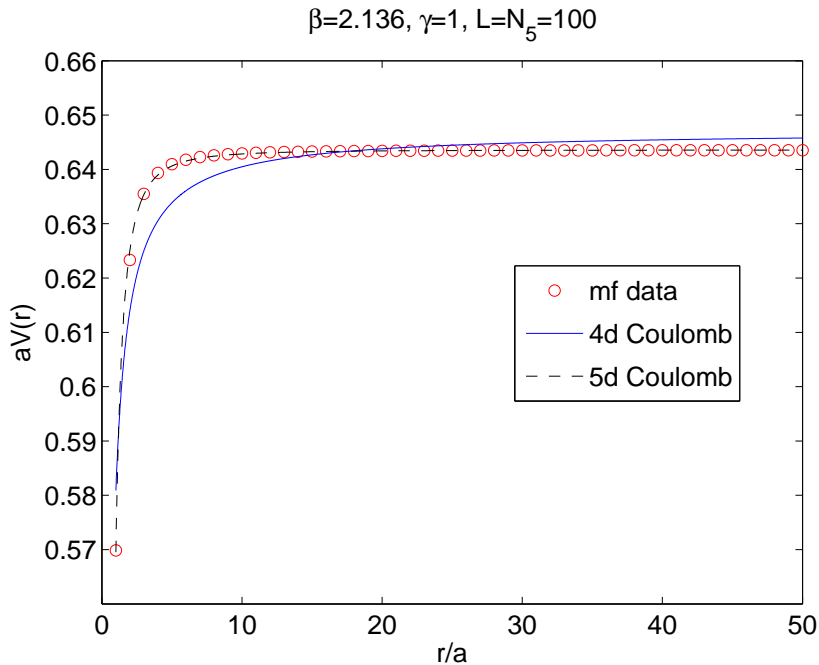,width=12cm}}
\caption{\small 
The static potential on the isotropic lattice. It is fitted by a five dimensional Coulomb
form, as expected from the corresponding continuum theory.
\label{f_potgamma1}}
\end{figure}
%

The second of our observables that we will use is the static potential 
derived from Wilson loops oriented along the four dimensional slices of the
lattice which are orthogonal to the extra dimension\footnote{
Note that on the anisotropic lattice there are two types of inequivalent
Wilson loops.
One along the four dimensional slices and one along the extra dimension.
We discuss here the former.}.
The points obtained will be fitted to both four dimensional
($c_0+c/r$, where $c_0$ and $c$ are the fitted parameters)
and five dimensional 
($d_0+d/r^2$, where $d_0$ and $d$ are the fitted parameters)
Coulomb forms.
This will decide wether the theory is in a dimensionally reduced phase or not,
as already mentioned.
Let us fix $\b=2.136$.
In order to obtain a first, qualitative picture of dimensional reduction
we perform global fits to four dimensional or
five dimensional Coulomb potentials and assign to the potential points
fictitious relative errors\footnote{
This is justified if we want to compare fits at different values of $\g$.}
and perform a comparative study of the $\chi^2$ of the fits.
A much more careful study of the local curvature of the potential will be
performed later. A scan in $\g$ of the global fits 
shows that for $0.35 < \g < 4$ there is no dimensional reduction.
In particular on a lattice with $L=N_5=100$ for $\g=1$ one sees a perfect
global five dimensional Coulomb fit, see \fig{f_potgamma1}.
For $\g > 4$, as expected, we start seeing a four dimensional Coulomb law which becomes
better as $\g$ increases
when only the points corresponding to small $r$ are fitted. For large $r$ we
see deviations. We call this phase the compact phase.
According to the free energy analysis the compact phase for small $\beta$ is unstable to this order
in the mean-field expansion, therefore we do not analyze further its properties.
The somewhat surprising fact is that for $\g < 0.35$ the potential turns again
four dimensional Coulomb, until the layered phase is hit, at around $\g=0.25$. A similar picture can be
obtained for other values of $\b$ in the deconfined phase. 
The narrow band for $\gamma<1$ which, at least according to our mean-field
approach, apparently describes a phase where the extra dimension is large but
the system is dimensionally reduced, we call the "d-compact phase". 
%
\begin{figure}[!t]
\centerline{\epsfig{file=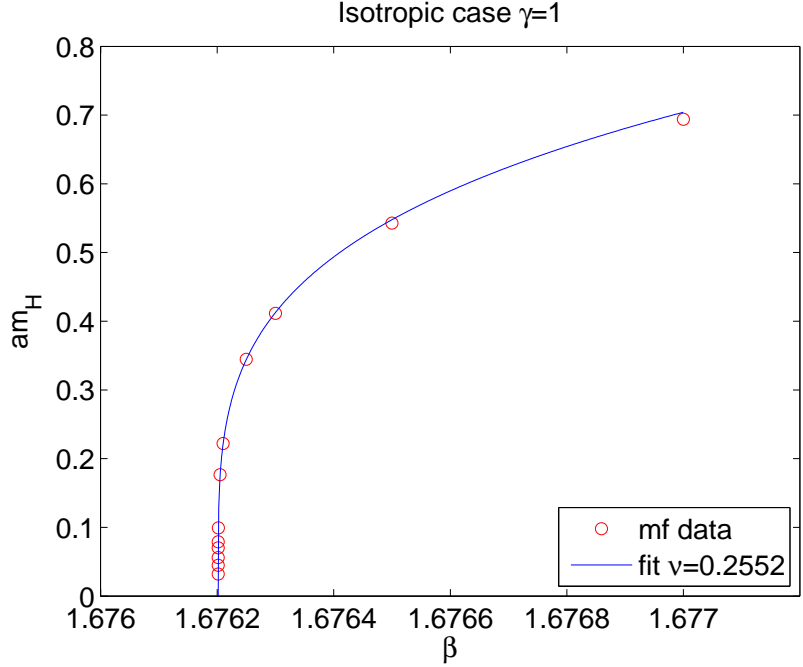,width=12cm}}
\caption{\small 
The scalar mass in lattice units on the isotropic lattice
shows a ferromagnet type of behavior with exponent $\n=1/4$.
\label{f_mHg1}}
\end{figure}
%
%
\begin{figure}[!t]
\centerline{\epsfig{file=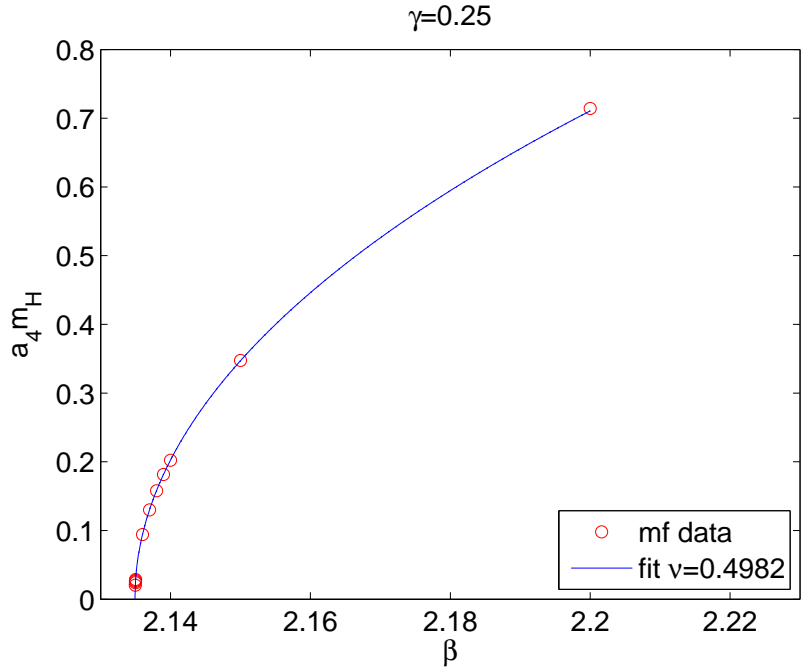,width=12cm}}
\caption{\small 
The scalar mass in lattice units $a_4m_H$ in the d-compact phase
shows a ferromagnet type of behavior with critical exponent $\n=1/2$.
\label{f_mHg1o4}}
\end{figure}
%

We now turn to the Higgs observable. By construction it is $L$-independent so
we do not have to scale it with $L$. It turns out to be also $N_5$ independent
and to have a strong $\b$ and $\g$ dependence. At a generic point in the
interior of the phase diagram the scalar is heavy in lattice units. It starts
loosing weight as boundaries between phases are approached. In \fig{f_mHg1}
we show its behavior as the isotropic phase transition is approached.
The way the scalar mass approaches zero
reminds us of the magnetization of a ferromagnet as a function of the
temperature as the Curie temperature is approached where a second order phase
transition occurs. Even though in our case the phase transition is definitely
known to be of first order, the resemblance of the mass dependence is close
enough so that we attempt to fit the data points to a critical law of the form
\be
am_H\sim (1-\beta_c/\beta)^\n
\ee
which yields $\n=0.2470$ when we take $\beta_c=1.6762017$. When we take
$\beta_c=1.6762016$ we obtain $\n=0.2546$ instead. The average value 
from the two fits $(0.2546+0.2470)/2=0.251\pm 0.004$ allows us to safely 
approximate the exponent to be
\be
\n =\frac{1}{4}
\ee
That this can not be a second order phase transition is also supported by the
fact that the mass does never really go to zero. It reaches a small value and
it stops there. 
The same behavior can be verified along the whole line that separates the
confined and deconfined phases.
%
\begin{figure}[!t]
\centerline{\epsfig{file=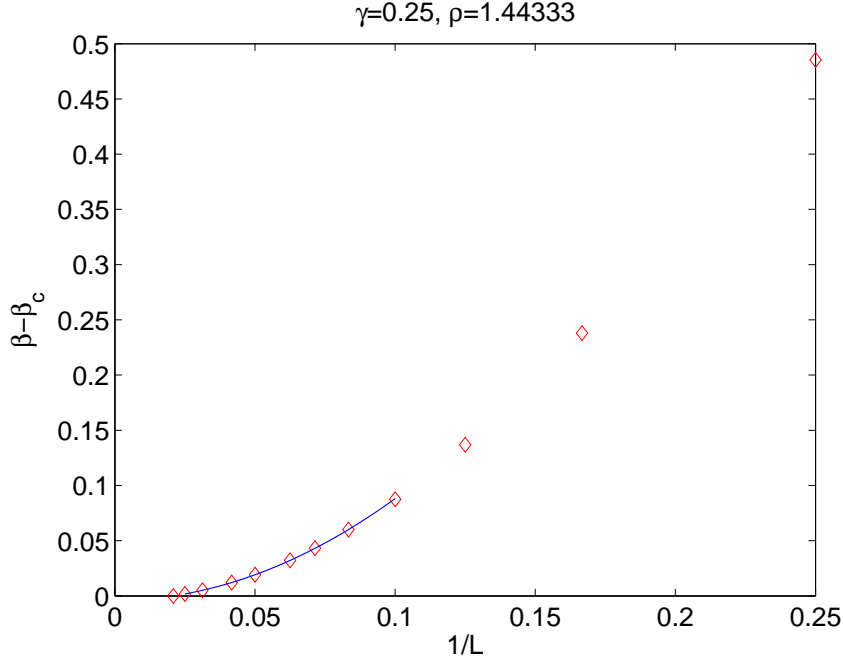,width=12cm}}
\caption{\small
Tuning of the bare coupling $\b$ as a function of $L$ when the phase 
transition is approached at constant $\rho=m_W/m_H=1.44333$ and $\g=0.25$, 
$N_5=L$. The solid line is a fit according to \eq{RGeq}.
\label{f_RG}}
\end{figure}
%

A qualitative difference appears on the boundary between the d-compact and
layered phases. To begin, the scalar mass in lattice units goes all the way
down to zero as the layered phase is approached. Moreover, in \fig{f_mHg1o4}
we measure a critical exponent 
\be
\n = \frac{1}{2}
\ee
when we set $\beta_c=2.1349$,
indicating that the d-compact phase describes different physics from 
the rest of the deconfined phase.  
We note that this is the critical exponent of the four dimensional Ising model and 
it is believed that if the $SU(N)$ theory should have a second order
phase transition in $d$+1 dimensions, its order parameter is the $Z_N$ values
of the Polyakov loop and its critical exponent should be $\n=1/N$
\cite{Svetitsky:1982gs}. Furthermore, if there is a tricritical point then
above it the phase transition is expected to turn into a first order one with
"exponent" 1/4. We have also checked that the free energy, as the phase
transition is approached looses its local maximum and it becomes flatter
around its global minimum, both facts consistent with a second order phase
transition \cite{ZJbook}.
In our case the second derivative of the free energy at
the minimum approaches zero as the phase transition is reached. 
Whether this is the real physical picture or it is just an
artificial effect that the mean-field method produces, is not clear. Until a
Monte Carlo simulation study is performed we can not be sure. Finally, we have
examined the passing from the layered phase into the confined phase. As the
scalar mass\footnote{
The Polyakov loop \eq{P5loop} is zero in the layered phase.
In order to measure the scalar mass there, we rotate the
Polyakov loop in the $k=3$ direction.}
gets stuck at a large value, this can not be a second order phase
transition. In this case the indication is that it is a first order phase
transition, like in the $U(1)$ theory \cite{Dimopoulos:2006qz}.

For now we assume that what we observe is a second order phase transition
and we study the scaling behavior as we approach it.
First, we tune the values of $\b$ as the phase transition is approached
keeping $\rho=m_W/m_H$ fixed (we keep $N_5=L$ and $\g$ fixed as usual) as a
function of the lattice size, see \fig{f_RG}.
\footnote{It is important to notice that the ratio $\rho$ is independent on
  $N_5$ since both $m_W$ and $m_H$ turn out to be independent on $N_5$. This
  means that as the lattice spacing goes to zero as the phase transition is
  approached one can keep $R$ fixed and therefore, when $L=N_5$, also $l$
  fixed. 
}
The behavior fits well to the form
\be
\b = c_0 + \frac{c_1}{L^2} = c_0 + c_1\frac{a_4^2}{l^2} \,,\label{RGeq}
\ee
which implies that the bare coupling squared $g_0^2=4/\b$ approaches its
critical value as a quadratic function of the lattice spacing.
%
\begin{figure}[!t]
\centerline{\epsfig{file=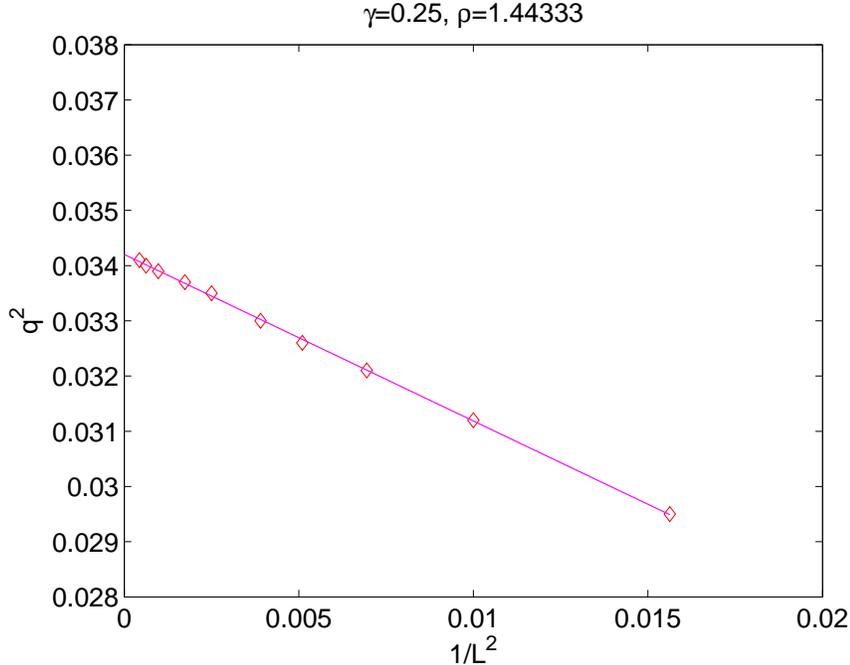,width=12cm}}
\caption{\small
Scaling of the charge $q^2$ defined in \eq{qcharge} as the phase transition is
approached along the values $\b(L)$ of \fig{f_RG} at constant $\rho=1.44333$.
\label{f_RGcharge}}
\end{figure}
%

Second, for the values of $\b(L)$ at constant $\rho=1.44333$ we calculate the
static potential $V(r)$ in the four-dimensional slices orthogonal to the extra
dimension and fit it to the four dimensional Coulomb form
\be
V(r) = -\frac{q^2}{r} + \mbox{const} \,. \label{qcharge}
\ee
The fits become better as $L$ increases. The fitted parameter $q^2$ can be
interpreted as a charge. Since the static potential is renormalized up to an
additive divergent constant \cite{willreno}, $q^2$ is a physical quantity and
\fig{f_RGcharge} shows its scaling behavior. The points $L=16\ldots48$ closest
to the phase transition can be very well fitted by
\be
q^2 = -\frac{0.30}{L^2}           +0.0342
\ee
with $\chi^2/\mbox{d.o.f}=3.7/4$. The fit is shown in \fig{f_RGcharge}.
Adding a $1/L$ term to the fits give
$q^2 = -0.37/L^2 +0.0054/L +0.0341$ with
$\chi^2/\mbox{d.o.f}=3.0/3$,\footnote{
We assign artificial relative errors of 0.001 to the potential in order to
compare the fits.}
so we
conclude that the continuum limit of $q^2$ is approached up to quadratic
cut-off effects as expected \cite{Necco:2001xg}.

\subsection{The static potential \label{subs_stat}}

We now turn to a closer examination of the static potential for $\gamma\neq1$.
We concentrate on the d-compact phase and we choose to approach the 2nd order
phase transition keeping
\be
\gamma = 0.25 \nonumber
\ee 
fixed. We choose the values
$\beta=2.13495$, $2.136$ and $2.3$, which correspond to Higgs masses of
$a_4m_H=0.020$, $0.094$ and $1.12$ respectively.
In the d-compact phase, the relations (for $N_5=L$)
\be
a_5 = a_4/\gamma \quad \mbox{and} \quad R/l = 1/(2\pi\gamma)
\ee
indicate that the extra dimension is large compared to the size of the
four-dimensional space. We take the continuum limit at fixed anisotropy, as
indicated by the arrow in \fig{f_5Dphasediagram}. It is
interesting that only if $\gamma<1$ a continuum limit exists, that means that
the anisotropy survives in the limit and has physical consequences.
We measure the static potentials along four-dimensional hyperplanes orthogonal
to the extra dimension and along the extra dimension.
%
\begin{figure}[!t]
\centerline{\epsfig{file=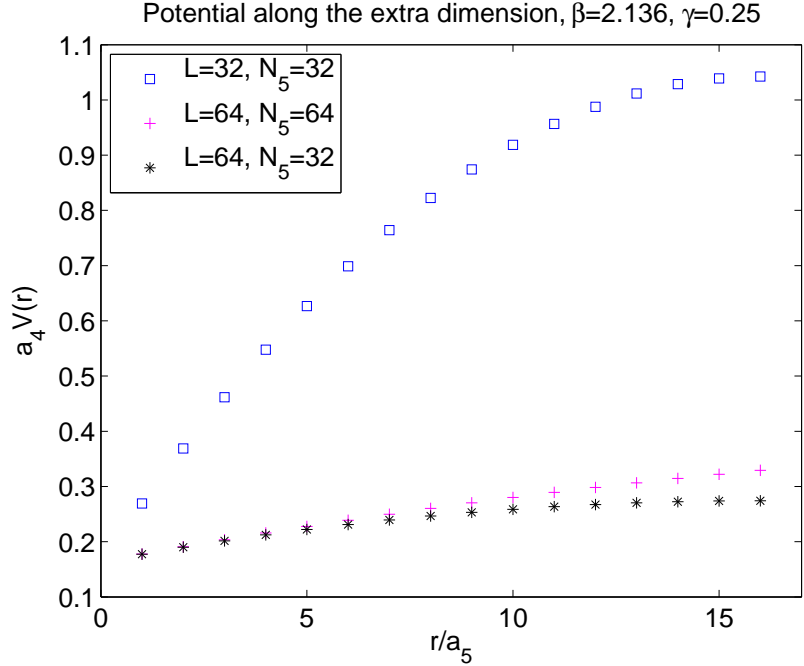,width=12cm}}
\caption{\small
The static potential derived from the Wilson loop along the extra dimension is
independent on $N_5$ and has a strong $L$ dependence. As $L$ increases, its
slope decreases.
\label{f_pot_L_N5_dep1}}
\end{figure}
%
%
\begin{figure}[!t]
\begin{minipage}{8cm}
\centerline{\epsfig{file=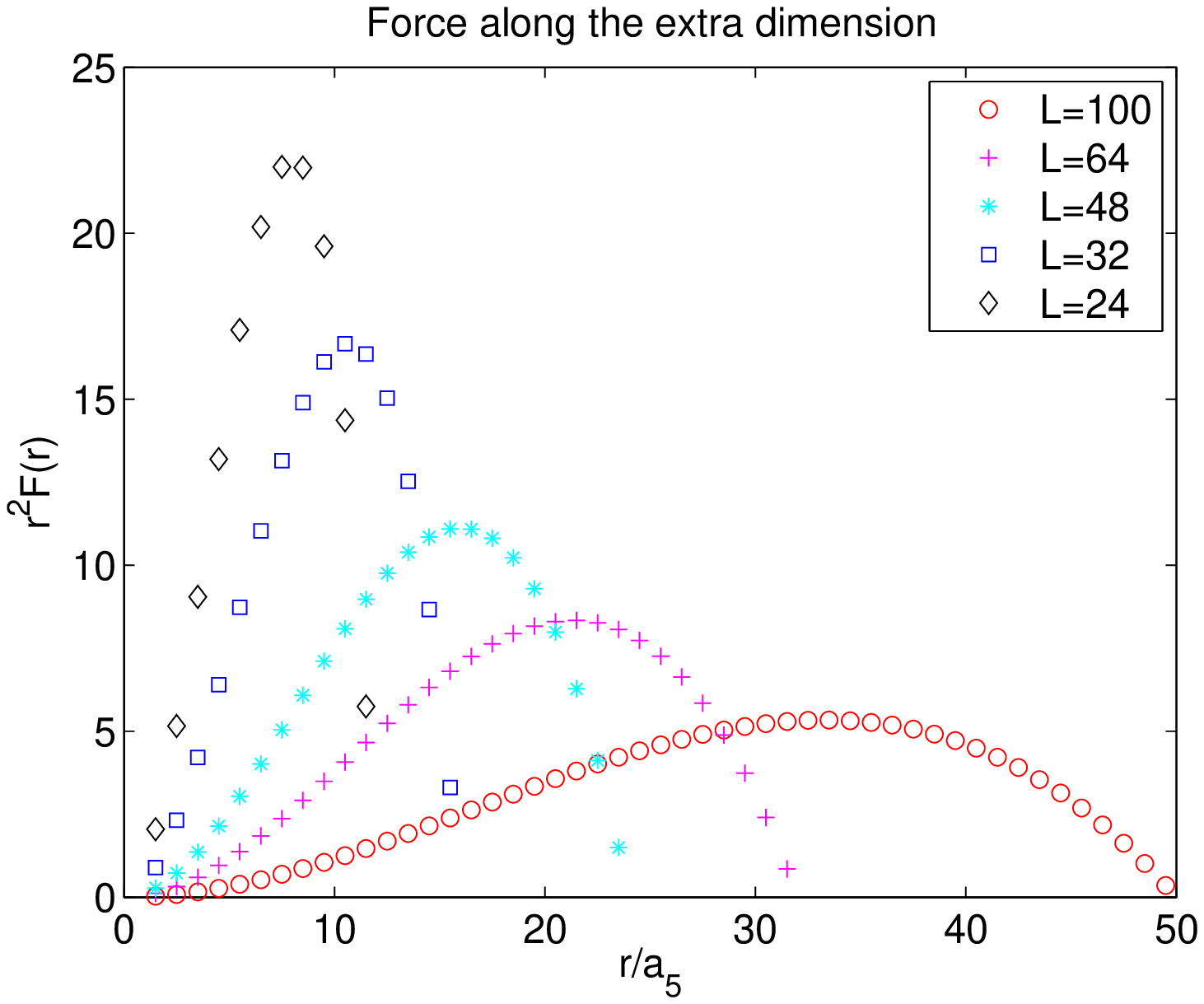,width=8cm}}
\end{minipage}
\begin{minipage}{8cm}
\centerline{\epsfig{file=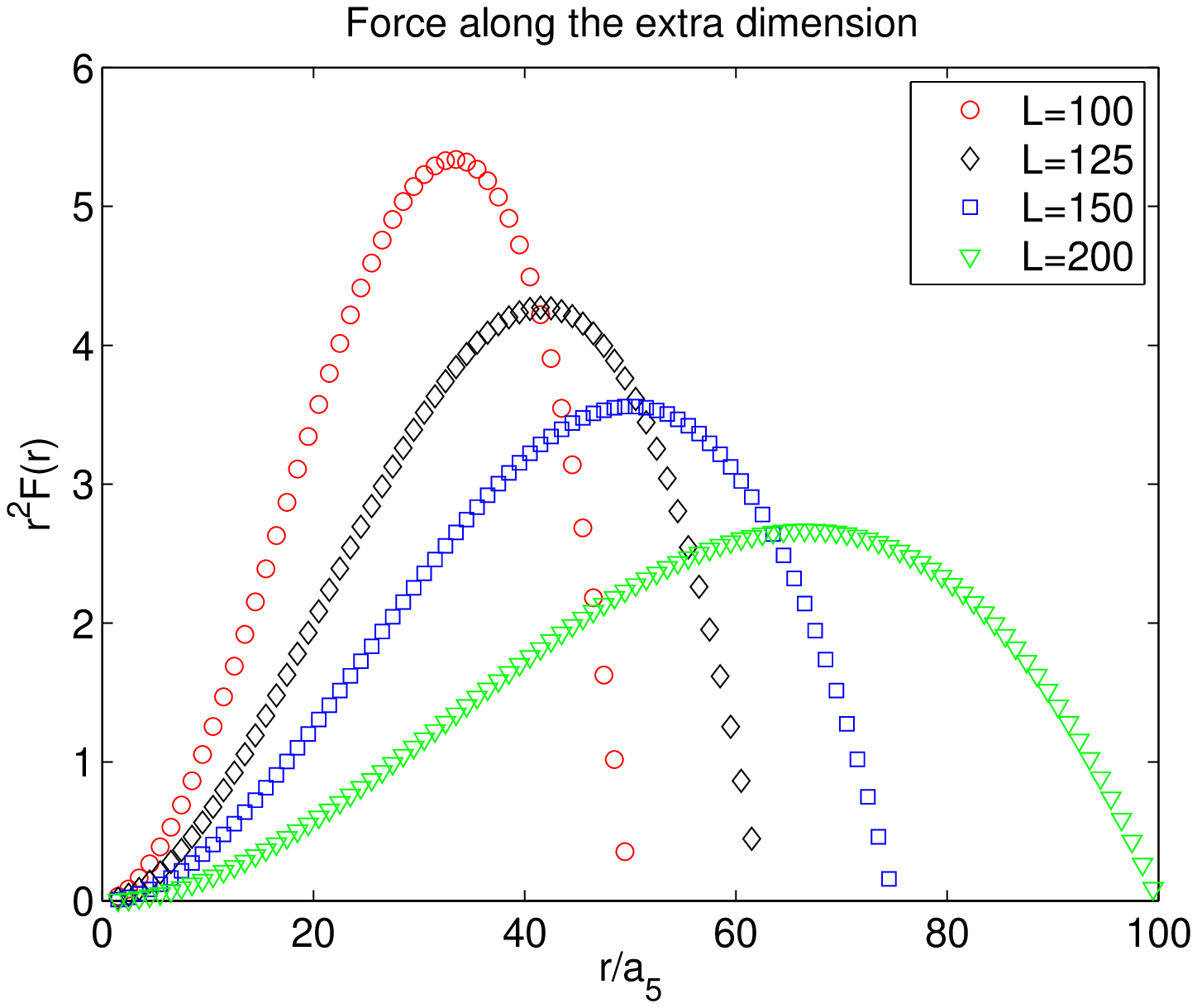,width=8cm}}
\end{minipage}
\caption{\small
The dimensionless force between two static charges separated in the extra
dimension. Apparently it is a pure finite size effect that approaches zero in
the infinite volume limit.
\label{f_forceextra}}
\end{figure}
%

We start by looking at the potential along the extra dimension.
In \fig{f_pot_L_N5_dep1} we plot the static potential at $\beta=2.136$
for various values of $N_5$ and $L$ equal to $32$ or $64$. 
Due to the periodicity, we measure the static potential up to
distances $L/2$ and we can therefore
compare distances up to $16$ in lattice units.
One can see that it is essentially independent\footnote{Barring
small effects due to the periodicity of the lattice.} on $N_5$ and
that its slope strongly decreases as $L$ increases. These two facts give rise
to the suspicion that it may not represent a physical interaction.
To check this, we plot in \fig{f_forceextra} the physical dimensionless force
\be
{\overline r}^2 F({\overline r}) = {\overline r}^2 
\left[V(r) - V(r-a_5)\right]/a_5, \hskip .5cm 
{\overline r} = r-\frac{1}{2}a_5 + O(a_5^2)
\ee
derived from the potential, keeping $N_5=L$ and increasing $L=24, 32, 48, 64,
100$ (left) and $L=100, 125, 150, 200$ (right). 
The plots show that the force goes to zero as the $L\to \infty$ is
approached suggesting that the four dimensional hyperplanes decouple in
this limit and the system reduces to an array of non-interacting ``branes''
where gauge interactions are localized. 

\begin{figure}[!t]
\begin{minipage}{8cm}
\centerline{\epsfig{file=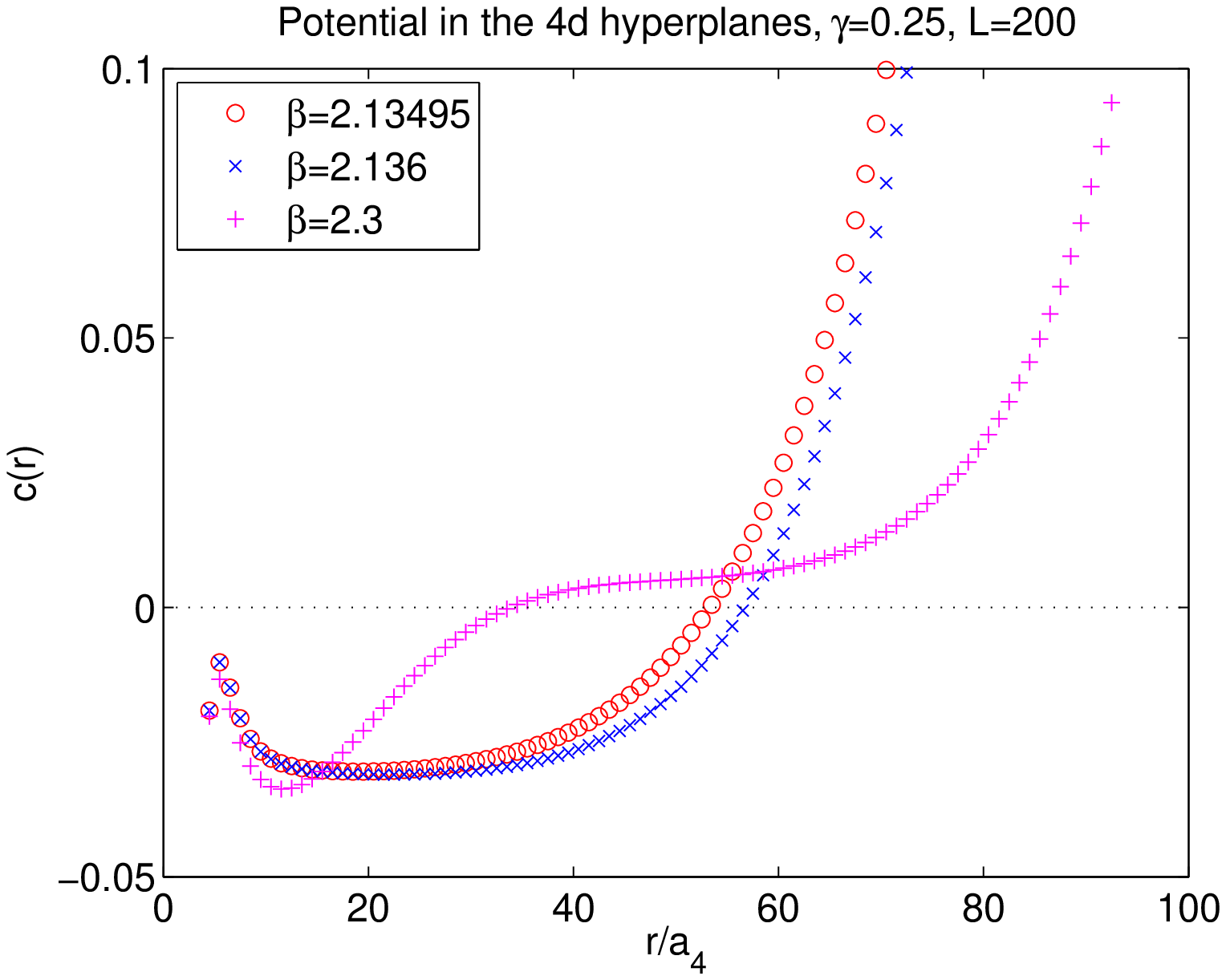,width=8cm}}
\end{minipage}
\begin{minipage}{8cm}
\centerline{\epsfig{file=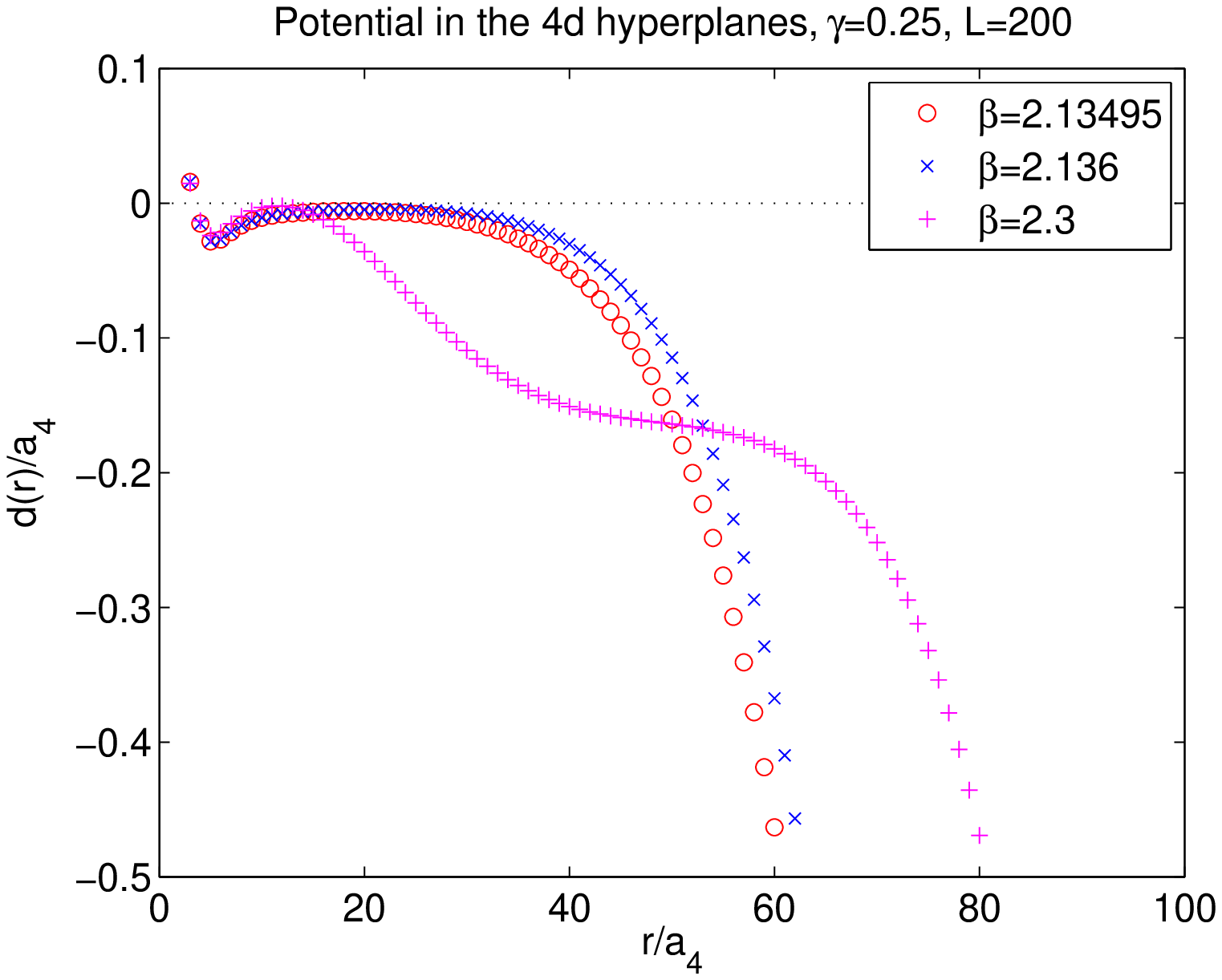,width=8cm}}
\end{minipage}
\caption{\small
Analysis of the static potential along four-dimensional hyperplanes in the
d-compact phase on lattices of size $L=200$ at fixed anisotropy $\gamma=0.25$.
Left: The coefficient $c(r)$ of a four dimensional Coulomb term.
Right: The coefficient $d(r)/a_4$ of a five dimensional Coulomb term.
These coefficients are determined from local fits to the form in \eq{fitV}. 
\label{f_cdcoeff}}
\end{figure}

We continue by looking at the potentials along four-dimensional hyperplanes.
%
\begin{figure}[!t]
\begin{minipage}{14cm}
\centerline{\epsfig{file=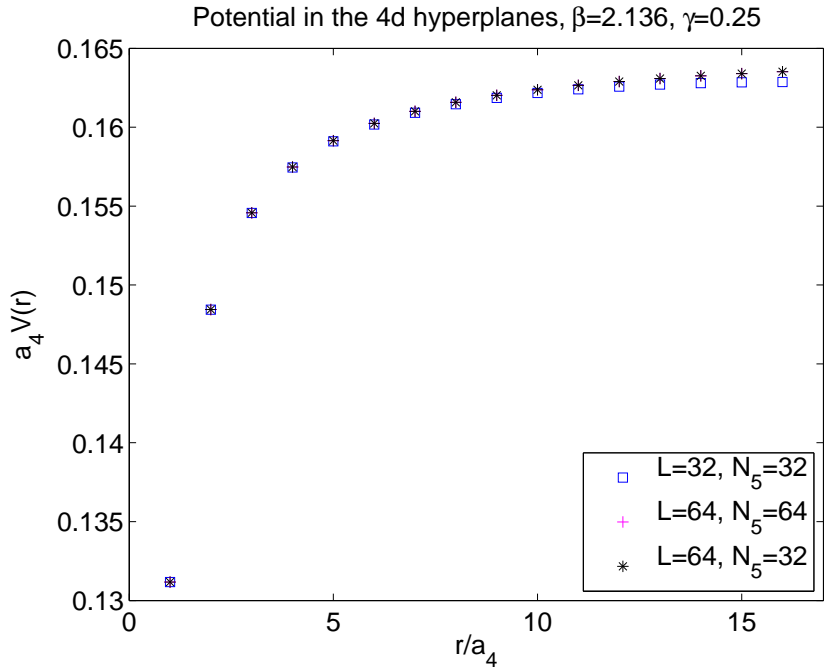,width=12cm}}
\end{minipage}
\caption{\small
The static potential derived from four dimensional Wilson loops is $N_5$ and
$L$ independent.
\label{f_pot_L_N5_dep2}}
\end{figure}
%
First, in \fig{f_pot_L_N5_dep2} we check $N_5$ and $L$ dependence.
As the figure shows, these potentials are essentially independent 
of $N_5$ and $L$. Thus, it is likely that they reflect physical interactions.
Recall that the global fits of the potential along the four dimensional
hyperplanes and the
analysis of the extra dimensional force suggested that in the large $L$
limit the potential along the four dimensional planes
should describe four dimensional physics to a good approximation. 
%
\begin{figure}[!t]
\begin{minipage}{14cm}
\centerline{\epsfig{file=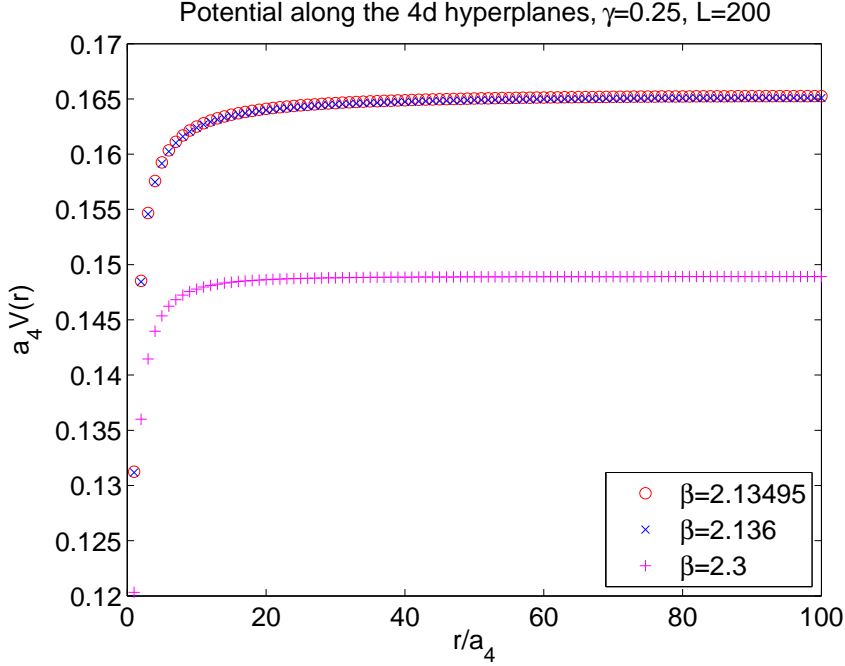,width=12cm}}
\end{minipage}
\caption{\small
The static potential along the four dimensional planes computed
on $L=200$ lattices approaching the phase
transition in the d-compact phase and keeping $\gamma=0.25$ fixed. 
\label{f_pot_200}}
\end{figure}
%
In order to show this we analyze the potentials on $L=200$ lattices for
$\beta=2.13495$, $\beta=2.136$ and $\beta=2.3$, shown in \fig{f_pot_200}.
We perform local fits to the form
\be
V(r) = c_0 + \sigma r + \frac{c}{r} + \frac{d}{r^2} + l \ln(r) \,.
\label{fitV}
\ee
The inclusion of a logarithmic correction is motivated for example in
\cite{logterm}.
From the dimensionless potential $a_4V(x)$, $x=r/a_4$, we derive
the auxiliary quantities
\bea
Z_1(x) & = & x^3 [V(x+1) - V(x-1)]/2 \,, \\
Z_2(x) & = & Z_1(x+1) + Z_1(x-1) - 2Z_1(x) \,, \\
Y_i(x) & = & x^i [V(x+1) + V(x-1) - 2V(x)] \,,\; i=2\,,3\,,4\,, \\
Y_1(x) & = & x^3 [Y_2(x+1) - Y_2(x-1)]/2 \,,
\eea
in terms of which we can estimate the coefficients in \eq{fitV} to be
\bea
\overline{\sigma}(x+0.5) = 
\sigma(x+0.5)a_4^2 & = & [Z_2(x+1) - Z_2(x)]/6 \,, \\
c(x+0.5)           & = & -[Y_1(x+1) - Y_1(x)]/2 \,,\\
\overline{d}(x) =
d(x)/a_4           & = & x^3[Y_3(x+1) + Y_3(x-1) - 2Y_3(x)]/12 \,,\\
\overline{l}(x) =
l(x)a_4            & = & -[Y_4(x+1) + Y_4(x-1) - 2Y_4(x)]/2 \,.
\eea
In \fig{f_cdcoeff} we plot the coefficients $c$ and $d$ of the four dimensional
and five dimensional Coulomb term respectively.
%
%
%
\begin{figure}[!t]
\begin{minipage}{8cm}
\centerline{\epsfig{file=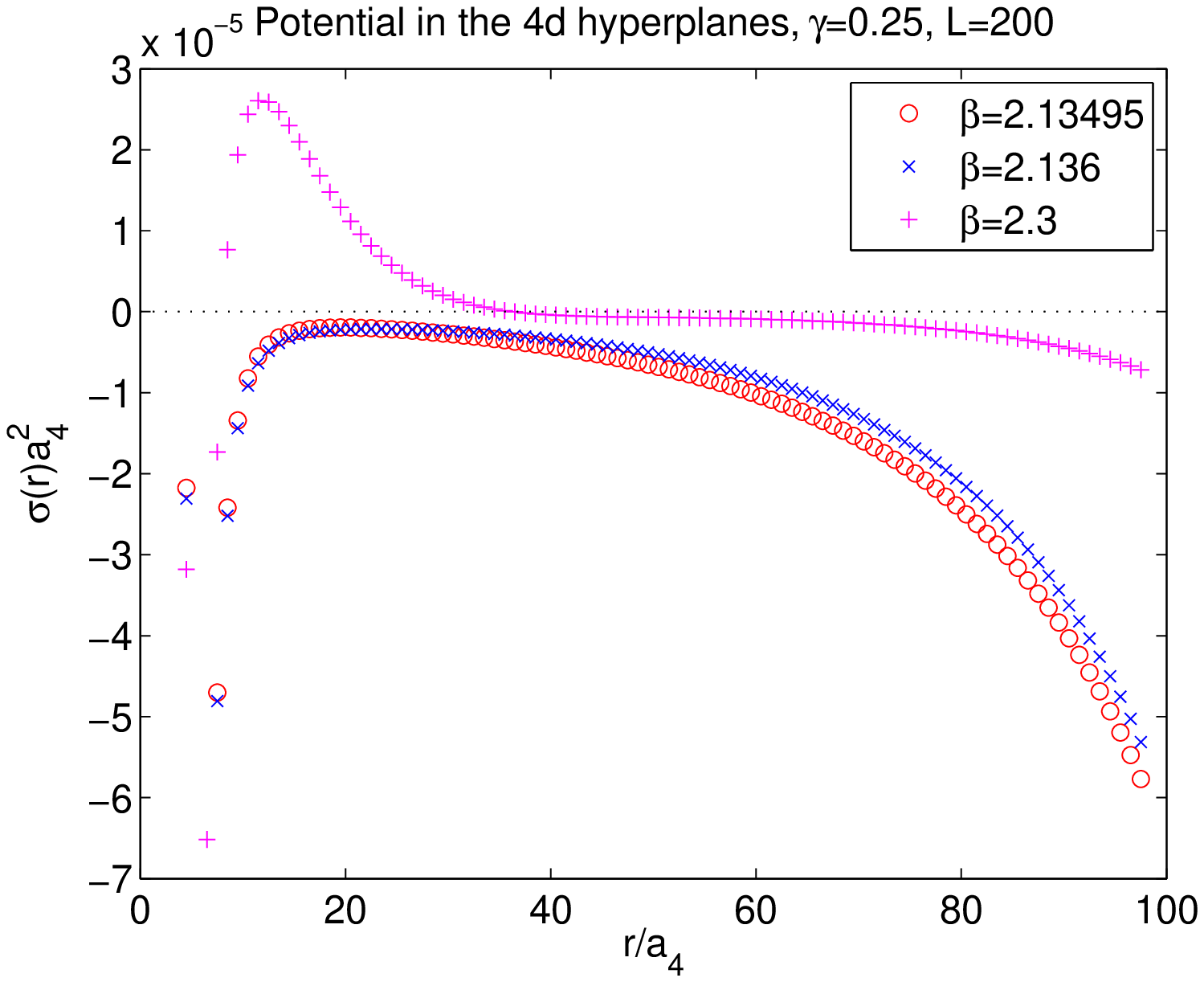,width=8cm}}
\end{minipage}
\begin{minipage}{8cm}
\centerline{\epsfig{file=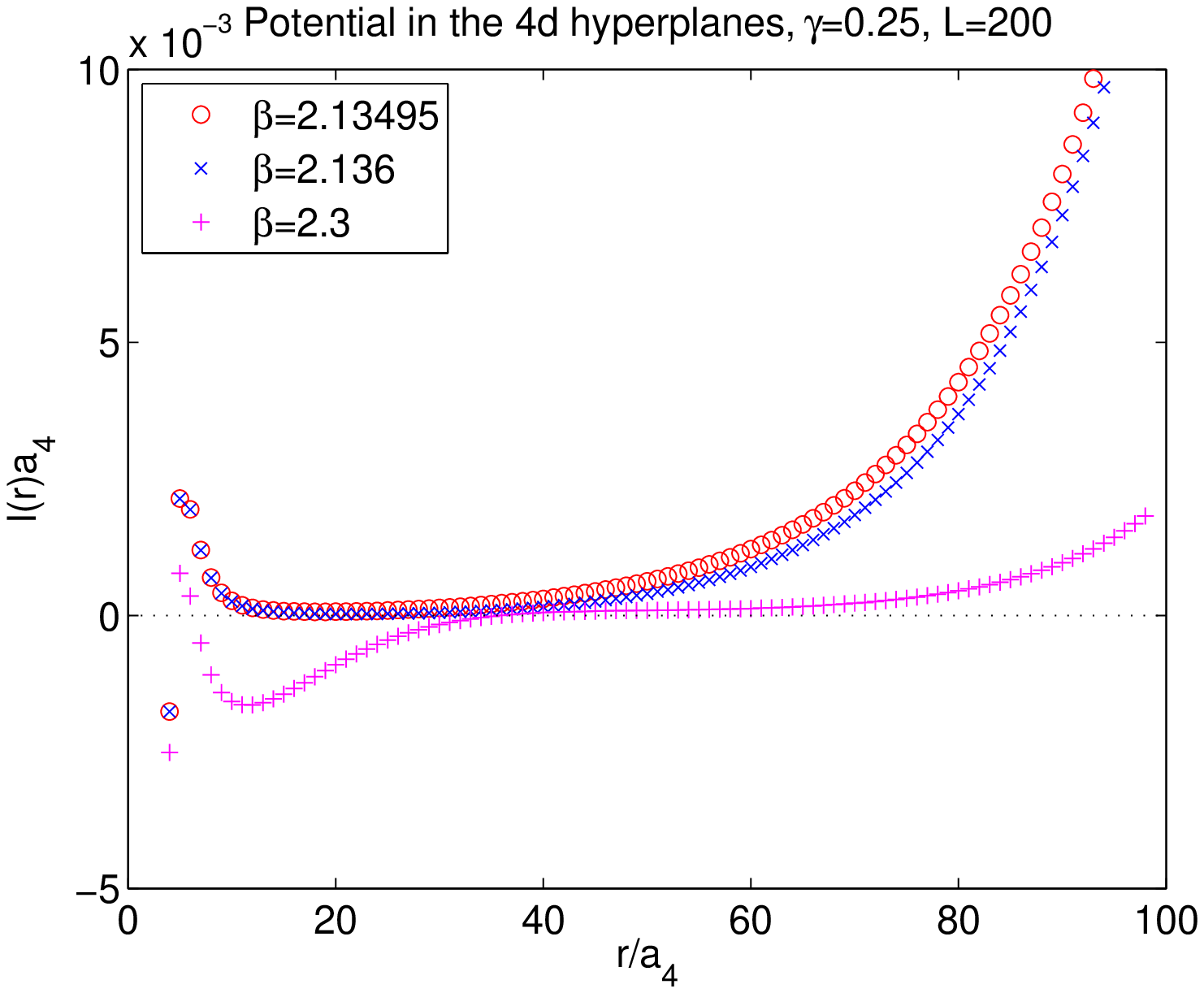,width=8cm}}
\end{minipage}
\caption{\small
Analysis of the static potential along four-dimensional hyperplanes in the
d-compact phase on lattices of size $L=200$ at fixed anisotropy $\gamma=0.25$.
Left: The string tension $\sigma(r)a_4^2$.
Right: The coefficient $l(r)a_4$ of a logarithm term.
These coefficients are determined from local fits to the form in \eq{fitV}.
\label{f_sigmalogcoeff}}
\end{figure}
%
Away from the phase transition,
at $\beta=2.3$, we observe almost plateaus for the coefficients 
$\overline{d}\in[0.15,0.18]$ and $c\in[0.004,0.007]$ in the range of
distances $x=40\ldots 60$. As can be seen in \fig{f_sigmalogcoeff}, the
coefficients $\overline{\sigma}$ and $\overline{l}$ are basically zero in this
same range of distances. Therefore, the potential is a balance between
a four-dimensional and a five-dimensional Coulomb term and we interpret this
as the onset of dimensional reduction from five to four dimensions.

In fact, as we go closer to the phase transition at $\beta=2.136$ and
$\beta=2.13495$ the potential becomes four dimensional.
In the range of distances $x=15\ldots 35$ we see in \fig{f_cdcoeff}
a plateau for the
four dimensional Coulomb coefficient $c\in[-0.031,-0.029]$, 
where the coefficient of
the five-dimensional Coulomb term is much smaller $d\in[-0.01,-0.005]$.
In the same range, see \fig{f_sigmalogcoeff}, the coefficient of the
logarithmic term $\overline{l}$ is basically
zero and the string tension has a plateau at a very small but negative value
$\overline{\sigma}\in[-3\times10^{-6},-2\times10^{-6}]$.
%
\begin{figure}[!t]
\begin{minipage}{14cm}
\centerline{\epsfig{file=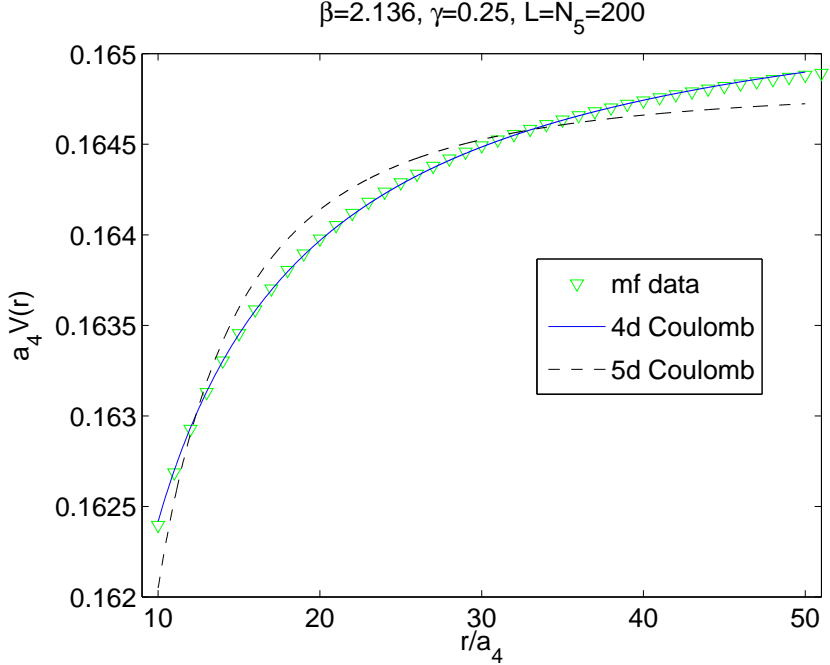,width=12cm}}
\end{minipage}
\caption{\small
Coulomb fits of the static potential along four-dimensional hyperplanes
close to the phase transition in the d-compact phase. The range
$r/a_4=10\ldots50$ is fitted to a four of five dimensional Coulomb form. 
\label{Coulombplot}}
\end{figure}
%

In \fig{Coulombplot}
we fit the potential at $\beta=2.136$ in the range $x=10\ldots50$ globally 
to purely four or five dimensional Coulomb forms (i.e. we set $\sigma=d=l=0$
in \eq{fitV}).
Clearly, the four dimensional Coulomb is an excellent 
fit while the five dimensional Coulomb law is exlcuded. 
Also, these plots exclude a possible large contribution from the presence of a
light excited vector state which would contribute an extra Yukawa term in the 
short (and therefore also in the long) distance part of the potential.
As already briefly mentioned in the beginning of this section, 
this was our first indication for dimensional reduction in the d-compact phase.
Together with the vanishing of the extra dimensional force, it can be taken as
a strong evidence for dimensional reduction via localization.

\subsection{Discussion \label{subs_stat_dis}}

We have now to compare these results with the behavior that we expect based on
the underlying physical picture.
If conventional Kaluza-Klein 
dimensional reduction works and the low-energy theory contains a massless
gauge boson and a Higgs particle (the assumption here is that higher
excitations are separated by a large mass gap), then this low-energy theory
corresponds to the four-dimensional Georgi-Glashow ($SU(2)$ gauge theory with
adjoint Higgs) model.
In \cite{Shrock:1987bx} the phase diagram of this model is discussed, it has
a confined phase and a Higgs phase where the gauge symmetry is broken to
$U(1)$ \cite{Lee:1985yi}. Note that in this model string breaking
cannot happen when the static charges are in the fundamental representation and
the matter in the adjoint.
In order to attempt an interpretation of the long distance part of the static
potential that is consistent with such a physical picture we will assume
that the data in this regime is not unphysical. Since we have no guarantee
that this is the case, the following discussion must be read with great caution.
Also, in the following we will take seriously the short distance analysis of this section
which lead to the conclusion that the system in the d-compact phase is truly dimensionally reduced.

On one hand, the fact that 
we do not see a non-vanishing mass for the gauge bosons (no Higgs mechanism)
implies that our four dimensional effective theory should land in a confined
phase. On the other hand one would expect a mass gap to appear in the confined
phase which means that the mass of a gauge invariant vector operator like 
\eq{con2}, representing a vector glueball state, should be non-zero. 
To reconcile this apparent contradiction we recall the argument of 
the first reference of \cite{NPdimred} where the dimensional reduction of a
five dimensional gauge theory\footnote{
In the context of D-theory the scalar field is decoupled, this can be
achieved by choosing orbifold boundary conditions.}
in a Coulomb phase is discussed. There, 
in the dimensionally reduced state, the non-perturbatively
generated mass gap behaves like
\be
m_G a \sim e^{-k \frac{1}{{\hat g}_4^2}}
\ee 
($k$ is a constant) when ${\hat g}_4\to 0$ and $a\to 0$.
We have denoted the asymptotically free coupling by ${\hat g}_4$ to
distinguish it from our effective coupling $g_4$ defined in \eq{g4} and which 
approaches a non-zero critical value as $a\to 0$. The mass gap in lattice
units is exponentially suppressed near the continuum limit.  Thus, in order to
interpret our results in the context of dimensional reduction we must assume
that also in our case such a mechanism is at work.
This is an assumption since our data is not able to distinguish a zero from an
exponentially small mass. 

Based on these arguments we expect to see in the large distance behavior of
the potential a string tension contribution, which rises linearly with the
distance. Instead the local fits based on the assumed form of the potential
give a very small but negative string tension, see \fig{f_sigmalogcoeff}. 
This could signal a breakdown of the meanfield at large distances, 
absence of dimensional reduction 
(these go against our working assumptions though) or that there are 
important terms in the confining string potential which we have neglected.  
In order to circumvent this ambiguity we have assumed the presence of the
universal L\"uscher term $-\pi/(8 r)$ in $d=5$ dimensions
\cite{Luscherterm}, subtracted it from the data 
and performed global fits of
only the long distance part of the rest, varying $L$.
In this case we obtained a positive string tension. The same analysis applied
to the long distance part of the isotropic potential shown in
\fig{f_potgamma1} gave a similar result though, 
with almost the same value for the string tension. This means that the
ambiguity persists.
%
\begin{figure}[!t]
\begin{minipage}{8cm}
\centerline{\epsfig{file=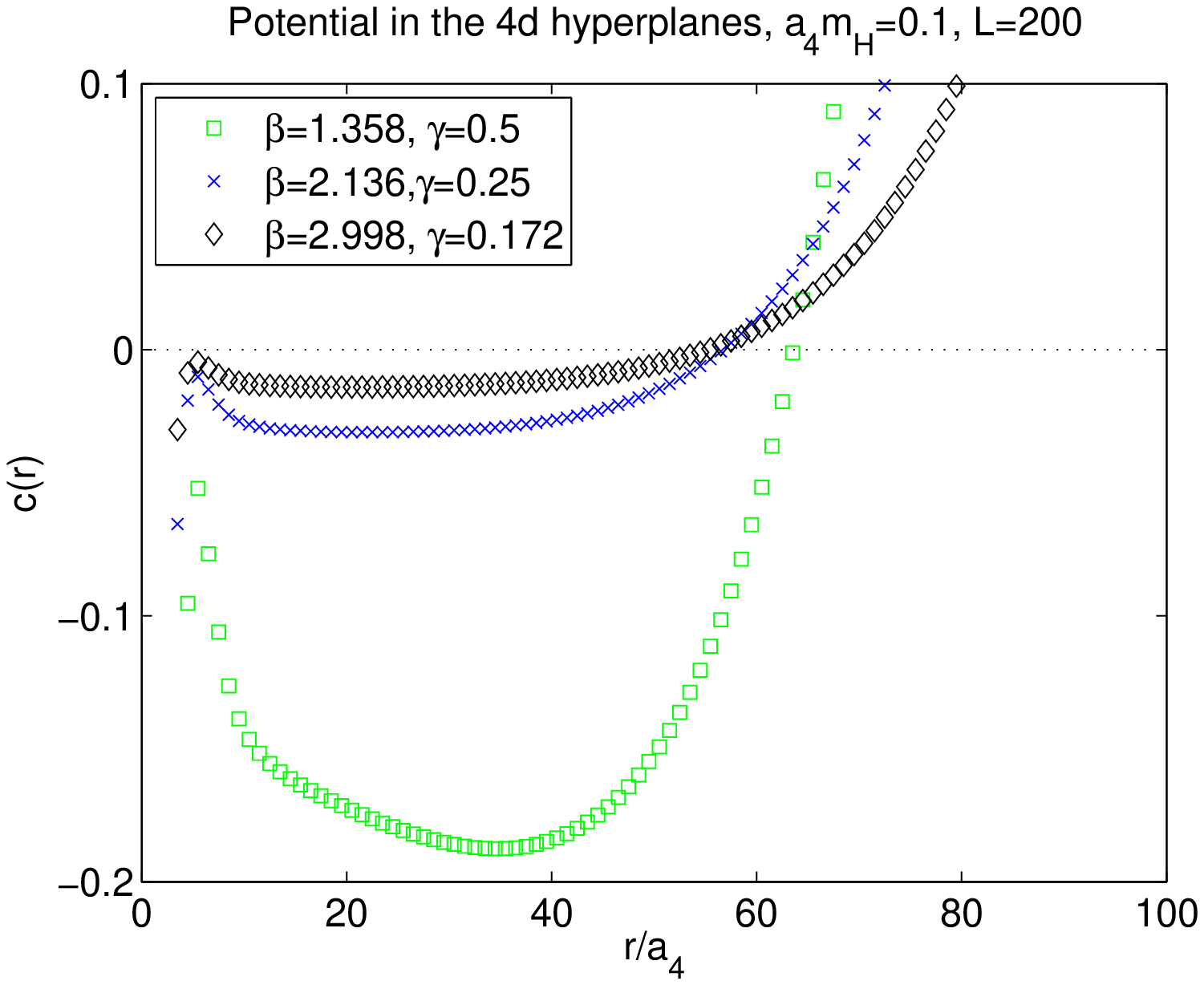,width=8cm}}
\end{minipage}
\begin{minipage}{8cm}
\centerline{\epsfig{file=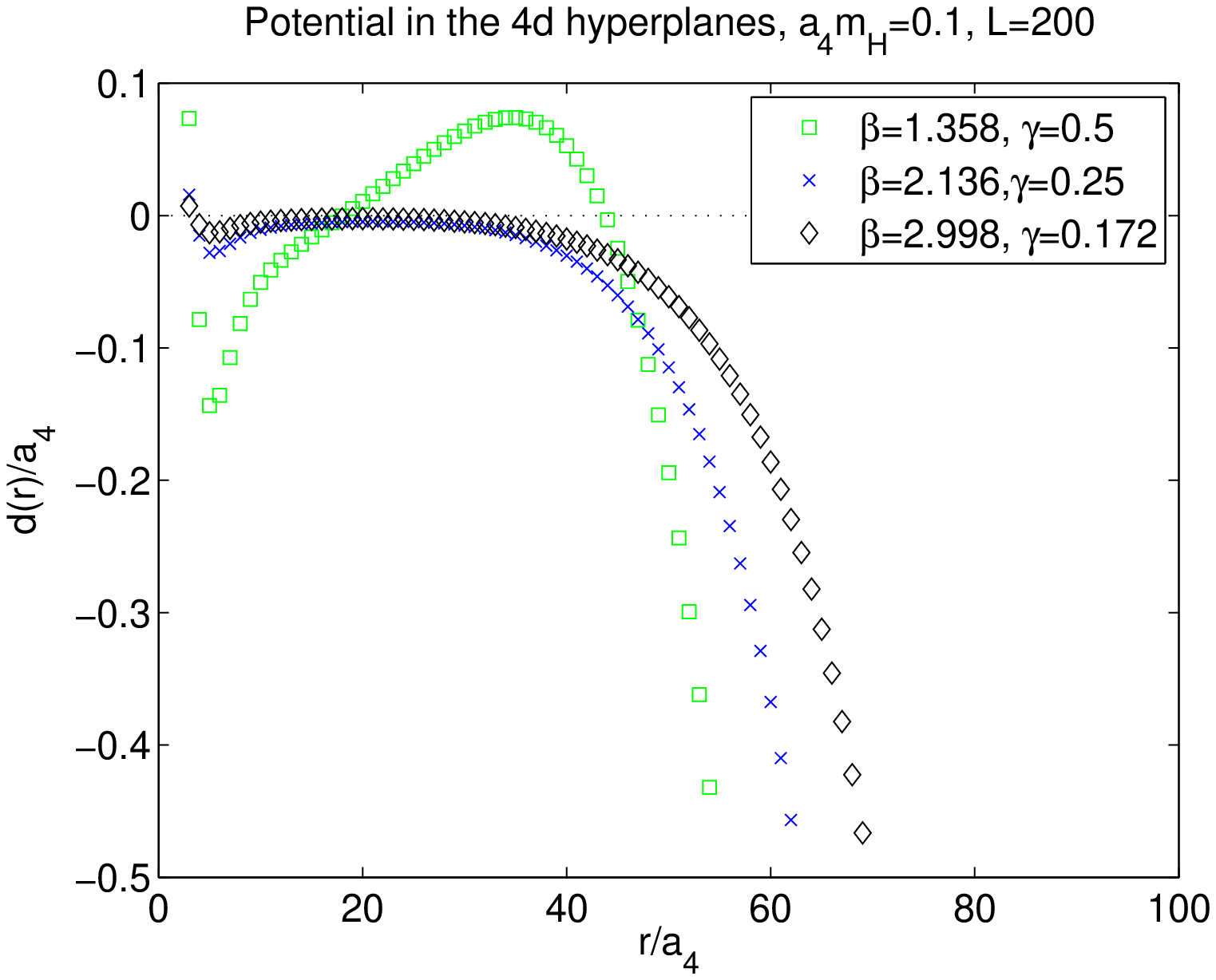,width=8cm}}
\end{minipage}
\caption{\small
Analysis of the static potential along four-dimensional hyperplanes in the
d-compact phase on lattices of size $L=200$ at fixed inverse correlation length
$a_4m_H=0.1$.
Left: The coefficient $c(r)$ of a four dimensional Coulomb term.
Right: The coefficient $d(r)/a_4$ of a five dimensional Coulomb term.
These coefficients are determined from local fits to the form in \eq{fitV}. 
\label{f_cdcoeff_lcp}}
\end{figure}
%
%
\begin{figure}[!t]
\begin{minipage}{8cm}
\centerline{\epsfig{file=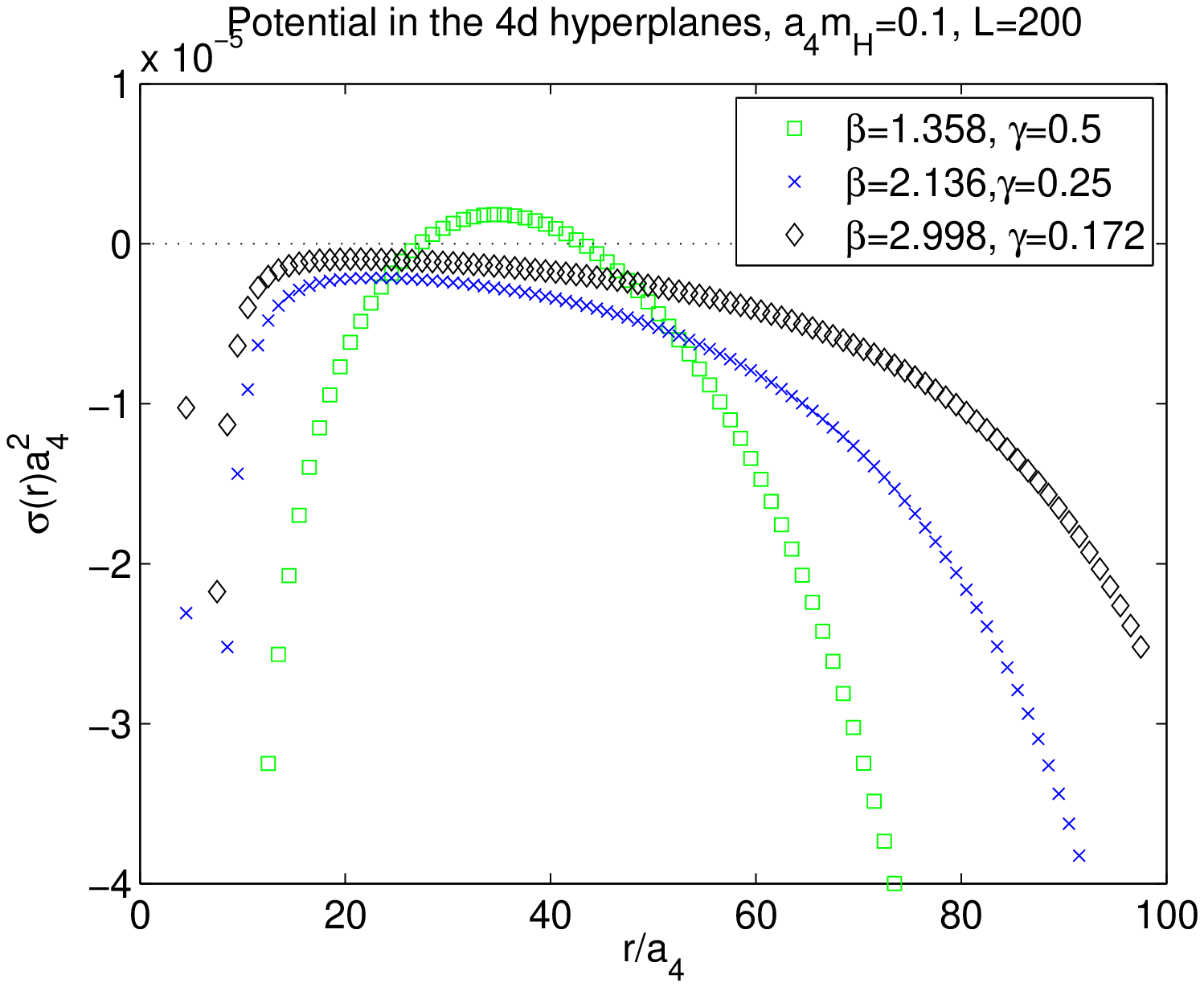,width=8cm}}
\end{minipage}
\begin{minipage}{8cm}
\centerline{\epsfig{file=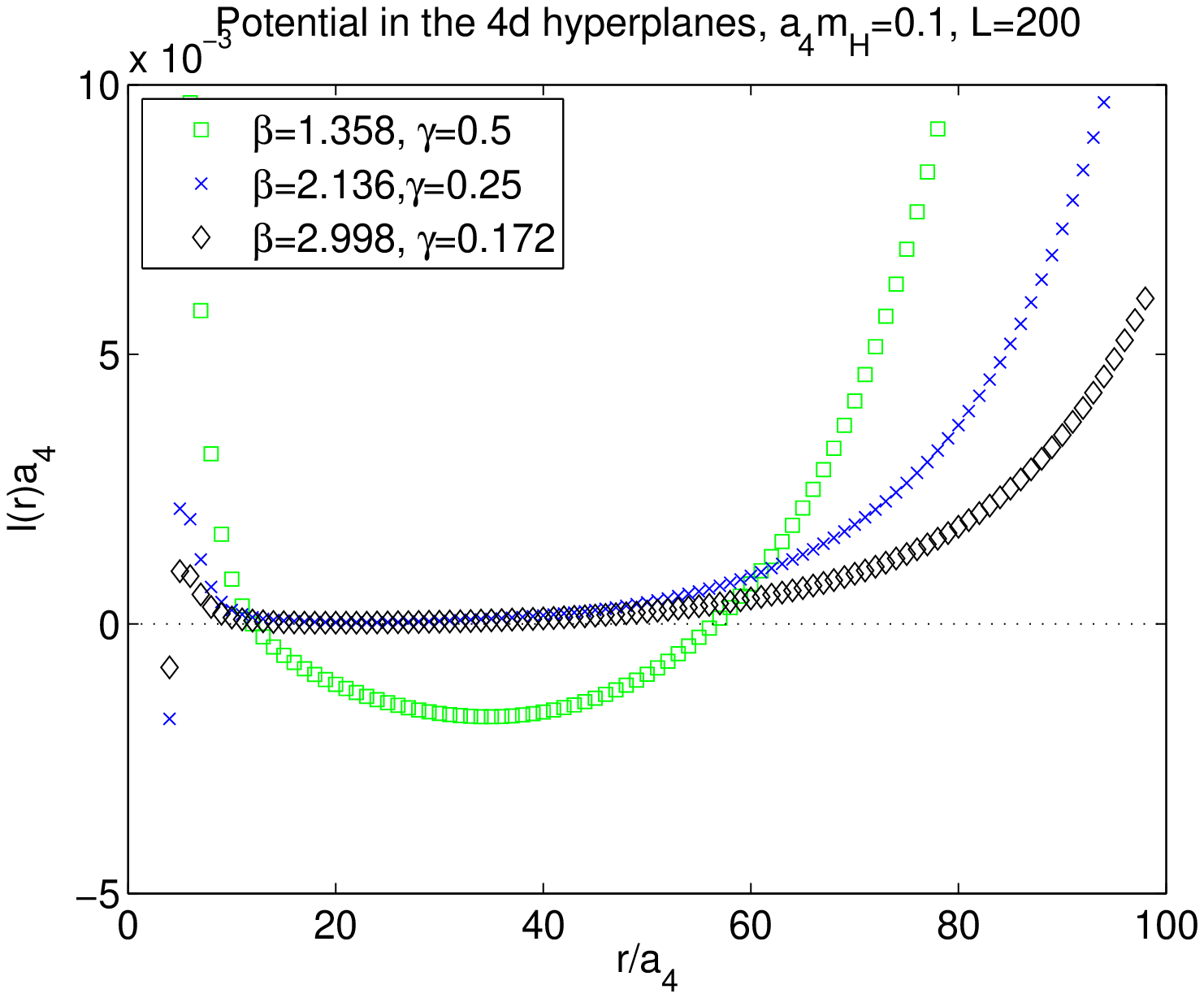,width=8cm}}
\end{minipage}
\caption{\small
Analysis of the static potential along four-dimensional hyperplanes in the
d-compact phase on lattices of size $L=200$ at fixed inverse correlation length
$a_4m_H=0.1$.
Left: The string tension $\sigma(r)a_4^2$.
Right: The coefficient $l(r)a_4$ of a logarithm term.
These coefficients are determined from local fits to the form in \eq{fitV}.
\label{f_sigmalogcoeff_lcp}}
\end{figure}
%

\begin{figure}[!t]
\centerline{\hskip -7cm\epsfig{file=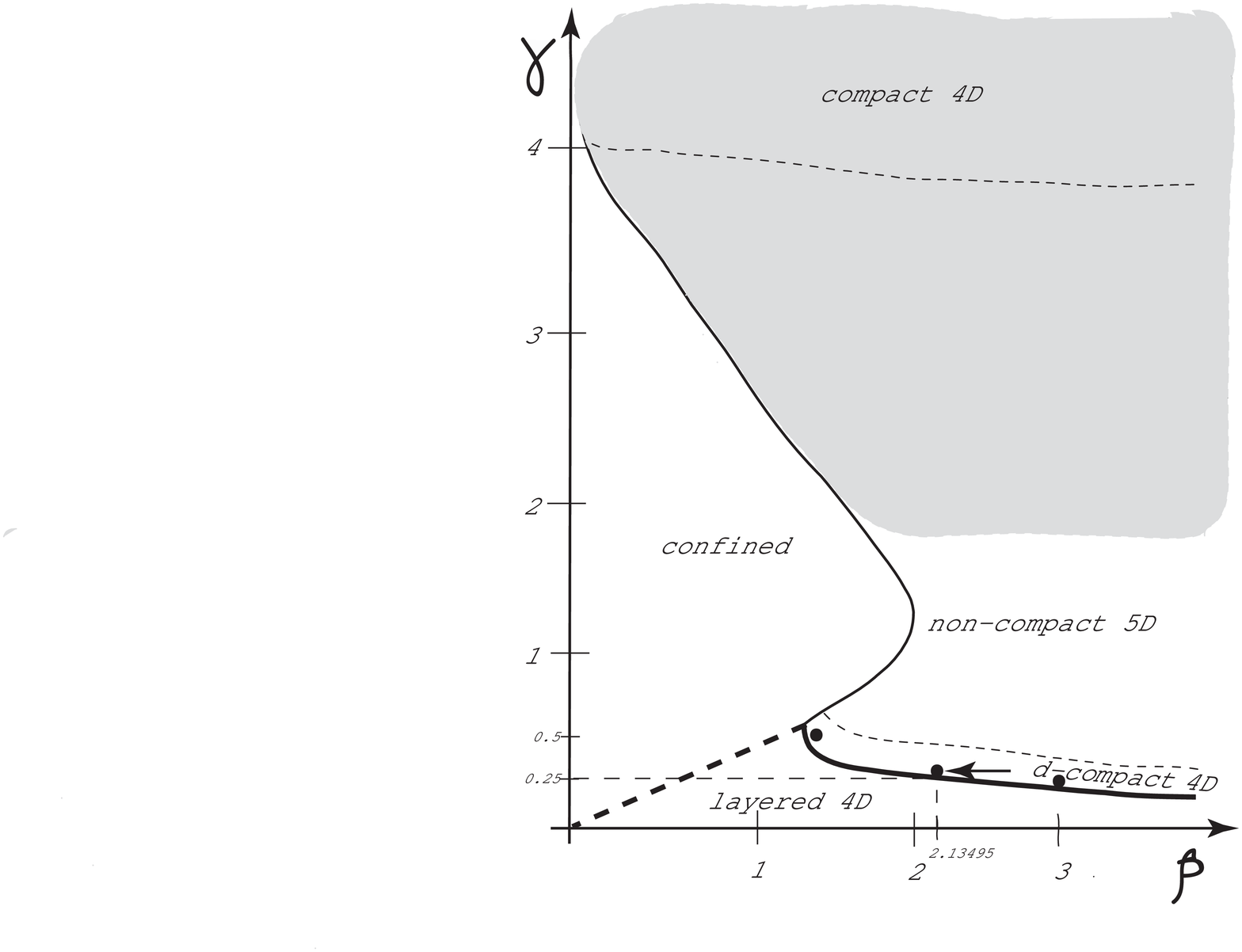,width=14cm}}
\caption{\small 
The phase diagram for lattice $SU(2)$ gauge theory on an anisotropic 
torus, according to the mean-field method. In bold, the line of second order
phase transition of critical exponent 1/2 is plotted. The bold dashed line 
that separates the layered phase from the confined phase 
is a line of first order phase transition. 
The normal line indicates also first order phase transitions.
The dashed lines indicate cross-over regions. 
The gray shaded area is unstable according to the first order free energy.
The arrow shows the approach to the continuum limit for the discussion of
the static potential at fixed anisotropy in Sect. \ref{subs_stat}. The blobs
represents three points at fixed inverse correlation length $a_4m_H$
discussed in Sect. \ref{subs_stat_dis}.
}
\label{f_5Dphasediagram}
\end{figure}

In order to check whether it is possible to find a unambiguous signal for a
positive string tension we computed the potential along the line of phase
transitions separating the d-compact from the layered phase. We choose three
points in the $(\beta,\gamma)$ parameter space,
represented by blobs in \fig{f_5Dphasediagram}: $(1.358,0.5)$, $(2.136,0.25)$
and $(2.998,0.172)$. These points are at the ``same'' distance from the phase
transition if we measure it by the inverse correlation length $a_4m_H$ 
(the Higgs mass in lattice units), which for the chosen points is
$0.101$, $0.094$ and $0.103$ respectively. We present the results for the
coefficients $c$, $\overline{d}$ in \fig{f_cdcoeff_lcp} and for the
coefficients $\overline{\sigma}$, $\overline{l}$ in \fig{f_sigmalogcoeff_lcp}.
While not much difference is seen between the points $(2.136,0.25)$
and $(2.998,0.172)$, for the point $(1.358,0.5)$ closest to the {\em
  tricritical point} of the phase diagram (the point where the compact,
layered and d-compact phases meet) we get a positive string tension in the
range of distances $r/a_4=27\ldots43$ with average value
$1.2\times10^{-6}$. The value of the string tension increases as $L$
increases.
We also note that
the coefficient $c$ of the $1/r$ term increases by almost an order of
magnitude as we get closer to the tricritical point.
These observations deserve to be studied further. 
%
%
%

We conclude this section by summarizing our results for the phase diagram
of the anisotropic five dimensional $SU(2)$ gauge theory in
\fig{f_5Dphasediagram} and by making a couple of remarks regarding
the validity of our mean-field results. 

Firstly, the results we presented for our physical
quantities (static potential, masses) are independent of the gauge fixing
parameter $\xi$. This is an immediate consequence of the fact that these
quantities are gauge invariant and provides a check of our programs.
The free energy depends\footnote{
The gauge dependence of the free energy is inherent in the meanfield
approximation. Already at 0th order its value depends whether we fix or not
the gauge.}
on $\xi$ but not the conclusions we derive from it. We checked
that the instability of the compact phase and the stability of the d-compact
phase are independent on the choice of $\xi$. The precise location of the
minimum of $F^{(1)}$ depends on $\xi$: for example at $\beta=2.136$ and
$\gamma=0.25$ on a $T=L=N_5=10$ lattice, the minimum is at $v=\ov_0$ and
$v_5=0.48$ ($\xi=1$) or $v=\ov_0$ and $v_5=0.42$ ($\xi=10$).
Thus we would not base a serious criticism of our results on that.

Secondly, it is difficult to predict the effects
of higher order
corrections and/or the general domain of validity of the mean-field
approximation. The mean-field is known to produce fake physics in certain
cases, so it may generate 
consistent but unphysical phase transitions. This
criticism concerns mainly the line of second order phase transitions
separating the d-compact from the layered phase, where we have carried out
most of our analysis. 
This is indeed possible but our working assumption has been that whenever the
mean-field is non-trivial, it describes a physical property.
The existence of such an ultraviolet fixed point is suggested by the epsilon
expansion \cite{UV} but has been elusive so far in Monte Carlo
simulations. This also deserves further study.

\section{Conclusions}

Using the mean-field approach we explored the phase diagram of a five
dimensional $SU(2)$ gauge theory with periodic boundary conditions. 
On the anisotropic lattice, we found two regimes where dimensional
reduction seems to be at work. 

One, the compact phase at large values of the anisotropy parameter is found
to be unstable for moderately small values of $\beta$ down to the phase
transition into the confined phase when the free energy is computed at first
order. In the limit $\beta\to\infty$ the mean-field reduces to standard
perturbation theory where the compact phase does not have an instability but
the theory becomes trivial.

The other regime, the d-compact phase, is at small values of the anisotropy
parameter as approaching the phase where the coupling along the fifth dimension 
becomes confined -- the layered phase. The latter phase is metastable to first
order in the free energy.
The transition between the d-compact and the layered phase is found to be
a second order phase transition. This allows us to take the continuum limit 
of physical quantities. We explicitly computed the continuum limit of a
physical charge.
In a dimensionally reduced state we are entitled to view the spectrum in a
basis of four dimensional quantum numbers. In this basis, ours is an $SU(2)$
gauge-Higgs system with the latter in the adjoint representation. 
Approaching the
critical line, the Higgs mass remains finite while we are removing the
cut-off. This is the first example of a 
four dimensional theory with this property as far as we
know and is an independent check of results from Monte Carlo 
simulations \cite{Irges:2006zf} and effective field theory
approaches \cite{DelDebbio:2008hb}. 
The gauge boson mass vanishes within errors in the limit of infinite volume
at any fixed lattice spacing. We cannot exclude an exponentially small mass
expected if the system reduces to an effective four dimensional theory in a
confined phase. 

We investigated dimensional reduction in the d-compact phase by computing the
static potential. The static potential
along the extra dimension seems to vanish in the infinite 
volume limit, a fact which would support a
scenario of dimensional reduction via localization in the d-compact phase.
Moreover we found that as we get closer to the phase transition, the potential
along the four dimensional hyperplanes turns into a four dimensional Coulomb
form at short to intermediate distances.
The unambiguous observation of a positive string tension contribution to the
mean-field potential along the four dimensional planes at large distances
remains elusive. But as the tricitical point is
approached, local fits indicate a small but clear plateau were the string
tension is positive.

It remains to see what a fully non-perturbative Monte Carlo computation
will tell, in particular concerning the existence of a second order phase
transition at small values of the anisotropy parameter. 
Our next analytical step following this work is to change the boundary
conditions along the fifth dimension from periodic to orbifold \cite{OrbMean}.

{\bf Acknowledgments.} We wish to thank Burkhard Bunk for sending his notes on
$U(1)$ mean-field gauge theory, Peter Hasenfratz for comments,
Martin L\"uscher for an important clarification,
Antonio Rago for discussions, 
Rainer Sommer for correspondence on the perturbative static potential and 
Peter Weisz for discussions and for his comments on a draft of our paper
prior to publication.
N. I. acknowledges the support of the Alexander von Humboldt
Foundation via a Fellowship for Experienced Researcher.


\bigskip

\begin{appendix}
\section{Meanfield corrections to Scalar and Gauge Boson masses
\label{appa}}

In this Appendix we describe in detail the computations of the diagram
in \fig{Higgs1order} and \fig{Displaced} leading to the formulae given
by \eq{HT1mass} and \eq{WT2mass} respectively.

\subsection{The Scalar mass to first order}

The first order correction to the Higgs mass is given by \eq{Higgsmass1} with
the operator ${\cal O}$ defined by \eq{torusH}.
The second derivative to be computed is
\bea
&& \frac{\d {\rm tr}\{P^{(0)}(t_0,{\vec m}')\}}{\d v_{\a_2}(n_2,M_2)}
\frac{\d {\rm tr}\{P^{(0)}(t_0+t,{\vec m}'')\}}{\d v_{\a_1}(n_1,M_1)} =
\D^{(N_5)}((n_5)_1)\D^{(N_5)}((n_5)_2)(P_0^{(0)})^2 \nonumber\\
&&4\d_{(n_0)_1,t_0+t}\d_{(n_0)_2,t_0}\d_{{\vec n}_1,{\vec m}''}\d_{{\vec
    n}_2,{\vec m}'}\d_{M_15}\d_{M_25}\d_{\a_10}\d_{\a_20} \,,
\eea
where $P_0^{(0)}$ is the background value of the Polyakov loop
and the symbol $\D^{(N_5)}(n_5)$ is defined in \eq{Deltasymb}.
After averaging over starting time $t_0$ and space positions ${\vec m}'$,
${\vec m}''$, the Fourier transform is
\bea
&& \frac{4}{\cal
  N}\sum_{n_1,n_2}e^{ip'n_1}e^{-ip''n_2}\frac{1}{T}\sum_{t_0}\frac{1}{L^6}
\sum_{{\vec m}',{\vec m}''}\D^{(N_5)}((n_5)_1)\D^{(N_5)}((n_5)_2) \nonumber\\
&&\Bigl[ \d_{(n_0)_1,t_0+t}\d_{(n_0)_2,t_0}\d_{{\vec n}_1,{\vec m}''}\d_{{\vec
    n}_2,{\vec m}'}+ \d_{(n_0)_1,t_0}\d_{(n_0)_2,t_0+t}\d_{{\vec n}_1,{\vec
    m}'}\d_{{\vec n}_2,{\vec m}''}\Bigr]\nonumber\\
&=& \frac{4}{\cal N}{\tilde \D}^{(N_5)}(p_5')
{\tilde \D}^{(N_5)}(-p_5'')\Bigl[ \frac{1}{T}
  \sum_{t_0}e^{ip_0't_0}e^{-p_0''(t_0+t)}\frac{1}{L^6}
\sum_{{\vec m}',{\vec m}''}e^{i{\vec p}'{\vec m}'}e^{-i{\vec p}''{\vec m}''}
\nonumber \\
&& +\frac{1}{T}
  \sum_{t_0}e^{ip_0'(t_0+t)}e^{-p_0''t_0}\frac{1}{L^6}
\sum_{{\vec m}',{\vec m}''}e^{i{\vec p}'{\vec m}''}e^{-i{\vec p}''{\vec m}'}
\Bigr]\nonumber\\
&=& \frac{8}{\cal N} \d_{p_0'p_0''}\cos{(p_0't)}\d_{{\vec p}',{\vec
    0}}\d_{{\vec p}'',{\vec 0}}{\tilde \D}^{(N_5)}(p_5')
{\tilde \D}^{(N_5)}(-p_5'') \,.
\eea
We have also added the term that corresponds to interchanging the index
$1\leftrightarrow 2$. Putting everything together and contracting with the
propagator, we have the final expression for the first order scalar mass
observable \eq{HT1mass}.

\subsection{The Gauge Boson  mass to second order}

Next we turn to the $W$ bosons. As discussed, the contribution to the gauge
boson's masses comes only at second order and is given in \eq{Wmass2} with
the operator ${\cal O}$ defined by \eq{con2}.
What we will need is the derivative of the displaced Polyakov loop
$W^{(0),A}_k$ defined in \eq{gaugeboson},
which is essentially determined by the single derivative of $\Phi^{(0)}$,
\eq{A5torus}.
All terms with derivatives acting on the links $U$ and any terms with
more than one derivative acting on $\Phi^{(0)}$ vanish identically because
of $\Phi^{(0)}_0=0$ when evaluated on the background.
The single derivative is simply
\be
\frac{\d \Phi^{(0)}(t, {\vec m})}{\d v_\a (n,M)}=
2 \S^\a P_0^{(0)}\d_{n_0 t} 
\d_{{\vec n},{\vec m}}\d_{M5}\D^{(N_5)}(n_5) \,. \label{onederphi}
\ee
On the torus, we have to take $\S^0=0$ and $\S^A=i\s^A$.\footnote{
By introducing the matrices $\S^\a$, \eq{onederphi}
is applicable to the orbifold case as well.
}
The non-vanishing contributions to the fourth derivative in \eq{Wmass2} are
of the form
\bea
&& \frac{\d\, W_k^{(0),A} (t_0, {\vec m}')}{\d v_{\a_1}(n_1,M_1)\d
  v_{\a_2}(n_2,M_2)}\frac{\d\, W_l^{(0),A} (t_0+t, {\vec m}'')}{\d
  v_{\a_3}(n_3,M_3)\d v_{\a_4}(n_4,M_4)} =\nonumber\\
&& 16 (P^{(0)}_0)^4\ov_0(0)^4 \d_{(n_0)_1, t_0} \d_{(n_0)_2, t_0}\d_{(n_0)_3,
  t_0+t} \d_{(n_0)_4, t_0+t}\d_{M_15}\d_{M_25}\d_{M_35}\d_{M_45}\nonumber\\
&& \D^{(N_5)}((n_5)_1)\D^{(N_5)}((n_5)_2)\D^{(N_5)}((n_5)_3)\D^{(N_5)}((n_5)_4)
\nonumber\\
&&
\Biggl( {\rm tr} \Bigl\{ {\overline \s}^A
\S^{\a_1\dagger}\S^{\a_2}\Bigr\}{\rm tr} \Bigl\{ {\overline \s}^A
\S^{\a_3\dagger}\S^{\a_4}\Bigr\}
\d_{{\vec n}_1,{\vec m}'+{\hat k}}\d_{{\vec n}_2,{\vec m}'}
\d_{{\vec n}_3,{\vec m}''+{\hat l}}\d_{{\vec n}_4,{\vec m}''}\nonumber\\
&& + (1\leftrightarrow2) + (3\leftrightarrow4) + 
(1\leftrightarrow2,3\leftrightarrow4) \Biggr) \,,\label{2der}
\eea
where $(1\leftrightarrow2)$ means that $(\alpha_1,n_1)$ in the second last
line of \eq{2der} has to be exchanged with $(\alpha_2,n_2)$ and similarly
for the other exchanges of indices.
According to \eq{Wmass2} we have to compute the schematic expression
\bea
&&\frac{1}{24}\sum_{(n_1,M_1,\a_1)\cdots (n_4,M_4,\a_4)}\frac{1}{T}\sum_{t_0}
\frac{1}{L^6}\sum_{{\vec m}', {\vec m}''}\, 6 \,\, \frac{\d^2W}{\d v_1\d
  v_2}\frac{\d^2W}{\d v_3\d v_4}\nonumber\\
&& \Bigl[ K^{-1}(1;2) K^{-1}(3;4)+K^{-1}(1;3) K^{-1}(2;4)+K^{-1}(1;4)
K^{-1}(2;3)\Bigr] \label{4der}
\eea
in momentum space. The factor 6 is a symmetry factor that accounts for the
different choices of pairs of indices out of $v_1$, $v_2$, $v_3$ and $v_4$
when taking the double derivatives. 
The first term in the bracket corresponds to time independent (self energy)
contributions so it can be dropped, since it cancels out from the connected
correlator. We will rewrite each of the two remaining terms as 
\bea
&&\sum_{m,n}O_1(m;n)K^{-1}(n;m) \times \sum_{r,s}O_2(r;s)K^{-1}(s;r)\nonumber\\
&=& {\rm tr}(O_1K^{-1})\times{\rm tr}(O_2K^{-1})=
    {\rm tr}({\tilde O}_1{\tilde K}^{-1})\times
    {\rm tr}({\tilde O}_2{\tilde K}^{-1}) \,. \label{traceprod}
\eea
For easier reference, we call the two relevant contractions 1 and
2 (the second and third term in the bracket of \eq{4der} respectively).
Also, we label the four terms in the last part of eq. (\ref{2der}) with
labels from $I$ to $IV$. 

In order to keep our formulae as general as possible we will not assume
momentum conservation in the fifth direction (like on the orbifold) and
we will consider a propagator that can have off-diagonal terms in gauge
space (like it can happen when the mean-field background is not simply
proportional to the unit matrix).

A term of the form
\be
O(m;n)=O(m_0,n_0)O(m_5,n_5)O({\vec m},{\vec n})
\ee
will transform in momentum space into
\bea
{\tilde O}(p';p'') = {\tilde O}(p_5',p_5'')\frac{1}{T}
\sum_{m_0,n_0}e^{ip_0'm_0}e^{-ip_0''n_0}O(m_0,n_0)\frac{1}{L^3}
\sum_{{\vec m},{\vec n}}e^{i{\vec p}'\cdot {\vec m}}e^{-i{\vec p}''\cdot {\vec
n}}O({\vec m},{\vec n}) \,,
\eea
(and analogously using $q';q''$ for the operator in the other factor of
\eq{traceprod}) where the part along the extra dimension is the universal term
\be
{\tilde O}(p_5',p_5'') = \frac{1}{N_5^2} 
{\tilde \D}^{(N_5)}(p_5'){\tilde \D}^{(N_5)}(-p_5'') \,.
\ee
The temporal Fourier transform gives for the first contraction
\be
O(3;1): \;\;\; \frac{1}{T} e^{ip_0'(t_0+t)}e^{-ip_0''t_0}\,,\hskip 1cm 
O(4;2): \;\;\; \frac{1}{T} e^{iq_0'(t_0+t)}e^{-iq_0''t_0}\,,
\ee
and for the second contraction
\be
O(4;1): \;\;\; \frac{1}{T} e^{ip_0'(t_0+t)}e^{-ip_0''t_0}\,,\hskip 1cm 
O(3;2): \;\;\; \frac{1}{T} e^{iq_0'(t_0+t)}e^{-iq_0''t_0}\,.\nonumber\\
\ee
Both products result into
\bea
\frac{1}{T}\sum_{t_0}\frac{1}{T^2}e^{it_0(p_0'-p_0''+q_0'-q_0'')}
e^{it(p_0'+q_0')}\d_{p_0'',p_0'}\d_{q_0'',q_0'}=
\frac{1}{T^2}e^{it(p_0'+q_0')}\d_{p_0'',p_0'}\d_{q_0'',q_0'} \,,\label{timeFT}
\eea
where we made conservation of temporal momentum,
enforced upon taking the trace with the propagator,
explicit.
Next we turn to the spatial Fourier transforms.
Each of the two propagator contractions contains four terms, which we list
here inserting conservation of spatial momentum.

\noindent
Contraction $1I$:
\be
\frac{1}{L^6}\sum_{{\vec m}',{\vec m}''}\frac{1}{L^6}\d_{{\vec p}',{\vec
    p}''}\d_{{\vec q}',{\vec q}''}e^{i{\vec p}'({\vec m}''+{\hat l}-({\vec
      m}'+{\hat k}))}e^{i{\vec q}'({\vec m}''-{\vec m}')}
= \frac{1}{L^6}\d_{{\vec p}',{\vec p}''}\d_{{\vec q}',{\vec q}''}\d_{{\vec
    p}',-{\vec q}'}e^{i(p_l'-p_k')} \,. \nonumber
\ee
Contraction $1II$:
\be
\frac{1}{L^6}\sum_{{\vec m}',{\vec m}''}\frac{1}{L^6}\d_{{\vec p}',{\vec
    p}''}\d_{{\vec q}',{\vec q}''}e^{i{\vec p}'({\vec m}''+{\hat l}-{\vec
    m}')}e^{i{\vec q}'({\vec m}''-({\vec m}'+{\hat k}))}
= \frac{1}{L^6}\d_{{\vec p}',{\vec p}''}\d_{{\vec q}',{\vec q}''}\d_{{\vec
    p}',-{\vec q}'}e^{i(p_l'+p_k')} \,.\nonumber
\ee
Contraction $1III$:
\be
\frac{1}{L^6}\sum_{{\vec m}',{\vec m}''}\frac{1}{L^6}\d_{{\vec p}',{\vec
    p}''}\d_{{\vec q}',{\vec q}''}e^{i{\vec p}'({\vec m}''-({\vec m}'+{\hat
    k}))}e^{i{\vec q}'(({\vec m}''+{\hat l})-{\vec m}')}
= \frac{1}{L^6}\d_{{\vec p}',{\vec p}''}\d_{{\vec q}',{\vec q}''}\d_{{\vec
    p}',-{\vec q}'}e^{-i(p_l'+p_k')} \,.\nonumber
\ee
Contraction $1IV$:
\be
\frac{1}{L^6}\sum_{{\vec m}',{\vec m}''}\frac{1}{L^6}\d_{{\vec p}',{\vec
    p}''}\d_{{\vec q}',{\vec q}''}e^{i{\vec p}'({\vec m}''-{\vec
    m}')}e^{i{\vec q}'(({\vec m}''+{\hat l})-({\vec m}'+{\hat k}))}
= \frac{1}{L^6}\d_{{\vec p}',{\vec p}''}\d_{{\vec q}',{\vec q}''}\d_{{\vec
    p}',-{\vec q}'}e^{-i(p_l'-p_k')} \,.\nonumber
\ee
Contraction $2I$:
\be
\frac{1}{L^6}\sum_{{\vec m}',{\vec m}''}\frac{1}{L^6}\d_{{\vec p}',{\vec
    p}''}\d_{{\vec q}',{\vec q}''}e^{i{\vec p}'({\vec m}''-({\vec
      m}'+{\hat k}))}e^{i{\vec q}'(({\vec m}''+{\hat l})-{\vec m}')}
= \frac{1}{L^6}\d_{{\vec p}',{\vec p}''}\d_{{\vec q}',{\vec q}''}\d_{{\vec
    p}',-{\vec q}'}e^{-i(p_l'+p_k')} \,.\nonumber
\ee
Contraction $2II$:
\be
\frac{1}{L^6}\sum_{{\vec m}',{\vec m}''}\frac{1}{L^6}\d_{{\vec p}',{\vec
    p}''}\d_{{\vec q}',{\vec q}''}e^{i{\vec p}'({\vec m}''-{\vec
    m}')}e^{i{\vec q}'(({\vec m}''+{\hat l})-({\vec m}'+{\hat k}))}
= \frac{1}{L^6}\d_{{\vec p}',{\vec p}''}\d_{{\vec q}',{\vec q}''}\d_{{\vec
    p}',-{\vec q}'}e^{-i(p_l'-p_k')} \,.\nonumber
\ee
Contraction $2III$:
\be
\frac{1}{L^6}\sum_{{\vec m}',{\vec m}''}\frac{1}{L^6}\d_{{\vec p}',{\vec
    p}''}\d_{{\vec q}',{\vec q}''}e^{i{\vec p}'(({\vec m}''+{\hat l})-({\vec
    m}'+{\hat k}))}e^{i{\vec q}'({\vec m}''-{\vec m}')}
= \frac{1}{L^6}\d_{{\vec p}',{\vec p}''}\d_{{\vec q}',{\vec q}''}\d_{{\vec
    p}',-{\vec q}'}e^{i(p_l'-p_k')} \,.\nonumber
\ee
Contraction $2IV$:
\be
\frac{1}{L^6}\sum_{{\vec m}',{\vec m}''}\frac{1}{L^6}\d_{{\vec p}',{\vec
    p}''}\d_{{\vec q}',{\vec q}''}e^{i{\vec p}'(({\vec m}''+{\hat l})-{\vec
    m}')}e^{i{\vec q}'({\vec m}''-({\vec m}'+{\hat k}))}
= \frac{1}{L^6}\d_{{\vec p}',{\vec p}''}\d_{{\vec q}',{\vec q}''}\d_{{\vec
    p}',-{\vec q}'}e^{i(p_l'+p_k')} \,.\nonumber 
\ee
We can simplify the expressions if we define
\be
{\overline K}^{-1}((p_0',{\vec p}'),5,\a_1,\a_2)=
\sum_{p_5',p_5''}{\tilde \D}^{(N_5)}(p_5'){\tilde \D}^{(N_5)}(-p_5'')
K^{-1}(p'',5,\a_1; p',5,\a_2) \,.\label{Kbar}
\ee
The components of the propagator diagonal in the Euclidean index are even under
reflecting the spatial momenta, which means
\be
{\overline K}^{-1}((p_0',-{\vec p}'),5,\a_1,\a_2)
= {\overline K}^{-1}((p_0',{\vec p}'),5,\a_1,\a_2) \,.
\ee
If we define in addition
\be
{\overline {\overline K}}^{-1}(t,{\vec p}',\a_1,\a_2) = \sum_{p_0'}e^{ip_0't}
{\overline K}^{-1}((p_0',{\vec p}'),5,\a_1,\a_2) \,, \label{Kbbar}
\ee
and use the fact that $\S^{\a\dagger} = - \S^\a$ we can write
out the whole expression \eq{2der} as
\bea
&&\frac{6\cdot 16\cdot 4}{24}\frac{1}{{\cal N}^2} (P_0^{(0)})^4 
(\ov_0(0))^4 \sum_{{\vec p}'}\sum_{\a_1,\a_2,\a_3,\a_4}
{\rm tr}\Bigr\{ {\overline \s}^A \S^{\a_1}\S^{\a_2}\Bigl\}
{\rm tr}\Bigr\{ {\overline \s}^A \S^{\a_3}\S^{\a_4}\Bigl\} \nonumber \\
&\times& \Bigl[ \cos{(p_l'-p_k')}
{\overline {\overline K}}^{-1}(t,{\vec p}',\a_1,\a_3) 
{\overline {\overline K}}^{-1}(t,{\vec p}',\a_2,\a_4) \nonumber \\
&+& \cos{(p_l'+p_k')}
{\overline {\overline K}}^{-1}(t,{\vec p}',\a_2,\a_3) 
{\overline {\overline K}}^{-1}(t,{\vec p}',\a_1,\a_4)
\Bigr] \label{4der_v2}
\eea
Finally we simplify this expression in some special cases. If
the propagator is diagonal in the gauge index, \eq{4der_v2} becomes
\bea
&&\frac{16}{{\cal N}^2} 
(P_0^{(0)})^4 (\ov_0(0))^4 \sum_{{\vec p}'}\sum_{\a_1,\a_2}
{\rm tr}\Bigr\{ {\overline \s}^A \S^{\a_1}\S^{\a_2}\Bigl\}\cdot \nonumber\\
&& \Bigl[ {\rm tr}\Bigr\{ {\overline \s}^A \{\S^{\a_1}, \S^{\a_2}\}\Bigl\}
\cos{(p_l')}\cos{(p_k')}
+{\rm tr}\Bigr\{ {\overline \s}^A [\S^{\a_1}, \S^{\a_2}]\Bigl\}
\sin{(p_l')}\sin{(p_k')}  \Bigr]\cdot \nonumber\\
&&{\overline {\overline K}}^{-1}(t,{\vec p}',\a_1) {\overline {\overline
    K}}^{-1}(t,{\vec p}',\a_2) \,. \label{4der_v3}
\eea
On the torus, $\Sigma^0=0$ and $\Sigma^A=i\sigma^A$ imply 
that the term with the cosines
in \eq{4der_v3} is absent and so the toron does not contribute to the
correlator. Using ${\rm tr}\{\s^A\s^{A_1}\s^{A_2}\}=2i\e^{AA_1A_2}$ and
summing over the gauge index $A$ and the Euclidean indices with
$\d_{kl}$ gives the result in \eq{WT2mass}.

\end{appendix}


\end{document}